\documentclass[a4paper, 10pt, conference]{ieeeconf}
\IEEEoverridecommandlockouts
\usepackage{amsmath, amssymb}
\usepackage{graphicx}
\usepackage{subcaption}
\usepackage{color}
\usepackage{titlesec}
\usepackage{algorithm}
\usepackage{algpseudocode}

\begin{document}
\title{\LARGE \bf Advanced Algorithms for Autonomous Guidance of Solar-powered UAVs}
\author{Siyuan Li
\thanks{Siyuan Li is with the School of Electrical Engineering and Telecommunications, University of New South Wales, Sydney, Australia. (E-mail: siyuan.li4@student.unsw.edu.au).}
}

\maketitle
\begin{abstract}
Unmanned aerial vehicle (UAV) techniques have developed rapidly within the past few decades. Using UAVs provides benefits in numerous applications such as site surveying, communication systems, parcel delivery, target tracking, etc. The high manoeuvrability of the drone and its ability to replace a certain amount of labour cost are the reasons why it can be widely chosen. There will be more applications of UAVs if they can have longer flight time, which is a very challenging hurdle because of the energy constraint of the onboard battery. One promising solution is to equip UAVs with some lightweight solar panels to maximize flight time. Therefore, more research is needed for solar-powered UAVs (SUAVs) in different environments.

Firstly, this report introduces a privacy-aware navigation technique for a UAV as a fundamental of energy-aware navigation. A dynamic programming-based algorithm is developed to minimize the total privacy violation risk. This section serves as a preliminary work that inspired the author to focus on the navigation of SUAV.

Secondly, based on the inspiration of privacy-aware navigation, an energy-aware path planning algorithm for an SUAV in an urban environment is developed. Compared with the existing minimum distance path planning algorithms, the proposed method takes the cost during flight and the solar energy gain from the sunshine into account, which is suitable for an SUAV.

Finally, this report proposes a complex path planning strategy for an SUAV in a dynamic urban environment. Different from a static urban environment, some unknown moving obstacles make the problem difficult to solve. A hybrid mode path planning algorithm is proposed to adapt to this partially unknown environment. The algorithm is structured by a three-step progressive method, starting with an online energy-aware path planning algorithm, followed by a path following strategy and ending with the reactive obstacle avoidance method. The hybrid approach can handle the unknown uncertainties while maximizing the residual energy on the SUAVs.

The proposed methods are presented and compared with some existing works. Simulation results are presented to prove the effectiveness of the proposed methods.
\end{abstract}
\IEEEpeerreviewmaketitle
\section{Introduction}
\label{ch1}

\subsection{Background}
Nowadays, unmanned aerial vehicles (UAVs) have drawn increasing attention in the robotics field. The applications of UAVs can be traced back to the 1930s during the Second World War \cite{sullivan2006evolution}. With the rapid development of unmanned system technology, UAVs are becoming cheaper and more manoeuvrable. Thus, they are being used in more civilian applications. For example, wireless communication \cite{khan2021role, zeng2016wireless, merwaday2015uav}, environmental monitoring\cite{asadzadeh2022uav, green2019using, eskandari2020meta}, surveillance \cite{osco2021review, everaerts2008use, ramachandran2021review} and parcel delivery \cite{dissanayaka2023review, li2022traffic, wang2019routing}. While traditional UAVs do not need a human presence on board, they require remote control and navigation of humans. Thus, UAVs do not eliminate the need for human interaction. Developing and implementing autonomous systems on UAVs is essential to push the UAV technology to the next level \cite{quan2020survey}. In addition, motion planning, state estimation, perception system and dynamic controller are the key parts towards a fully autonomous system. Within these modules, the motion planning part is where humans can control the navigation of UAVs. Therefore, having a robust and reliable navigation system is essential for autonomous UAVs.

Given the progression of autonomous systems, more and more sensors are equipped on UAVs to fulfil various missions. However, a larger onboard battery cannot accommodate the higher electricity consumption as the increasing weight negatively affects the whole system. The endurance becomes a pain point that slows the evolution of mobile robots, especially for UAVs \cite{oettershagen2017design}. A long-endurance UAV can increase efficiency and help the sustainable development of the unmanned system. One solution to overcome this limitation is mounting solar panels on UAVs, allowing them to harvest solar energy \cite{el2021solar}. Global SUAV market 2020-2024 report states that the SUAV market is expected to grow by around \$500 million by 2024 and has the potential to reach \$3800 million by the end of 2033 \cite{data1}. In addition, numerous small private firms and large tech and aviation companies have been actively engaged in the SUAV market for a long time to develop state-of-the-art SUAVs. For example, Boeing, Google and Airbus \cite{engblom2012novel, harris2016project, data2}. Equipping proper solar panels on the UAVs provides the functionality to charge themselves while in the sunshine. In the meantime, the computational cost for path planning becomes higher due to the fact that the energy consumption and harvesting models need to be considered, which makes the navigation of SUAVs more complicated than traditional UAVs. Therefore, the navigation of solar-powered UAVs (SUAVs) is a promising research direction.

\subsection{Navigation Methods}
Existing navigation methods can be classified into three different categories: global path planning, local path planning and hybrid path planning. Whether a navigation system possesses the environment's information differs from the three categories. This section discusses some classic path planning techniques for mobile robots that are fundamental to the navigation of SUAVs.

\subsubsection{Global Path Planning}
Global path planning requires information about the environment to plan the path purely based on the map information \cite{patle2019review}. Obstacle and goal positions are necessary for global planners to generate a feasible path. If the map information is insufficient, the planner cannot avoid the unknown obstacles that make the result unrealistic. Global path planning can be further divided into search-based, sample-based, optimisation-based, and other mentionable methods.

\paragraph{Search-based Methods}
Search-based methods must discrete the map into several nodes that pivot the problem into graph searching. Each node has a different weight based on predefined rules, and the algorithm can explore the graph based on the weight information. Two fundamental algorithms are the baseline of most search-based algorithms: Breadth-first search (BFS) and Depth-first search (DFS) \cite{lavalle2006planning}. The first-in-first-out queue and last-in-first-out stack are used in BFS and DFS, respectively. The priority difference gives them different logic when exploring the graph tree. BFS can find more optimal solutions but requires a longer searching time compared to DFS.

Dijkstra algorithm is a typical BFS-based method to find the minimum distance path. However, unlike BFS, it selects the shortest path node from the remaining nodes and explores its neighbours. After that, the new weight is updated until the shortest path is found. The repeated process guarantees the optimal solution of the path. Another classic search-based method is the Bellman-Ford algorithm that uses a similar rationale to the Dijkstra algorithm \cite{goldberg1993heuristic}. However, the traversal speed of the Bellman-Ford algorithm is slower than that of the Dijkstra algorithm for a similar problem. The advantage of this method is that the edge weight can be set to a negative value, providing broader applications. The computational cost of both above mentioned methods is still high when dealing with a large map. In this way, the A* algorithm is developed with a higher computational efficiency. Based on Dijkstra, a heuristic function $h(n)$ is added to guide the exploring process. The cost function becomes $f(n) = g(n) + h(n)$, where $g(n)$ represents the actual cost of the current node. Even though the change is minimal, the positive effect dramatically increases the search speed. Because the heuristic function can guide the expanding process in a specific direction instead of evenly expanding. However, the design of the heuristic function $h(n)$ has to be admissible, or the effect might be negative \cite{quan2020survey}. The A* algorithm has extensive applications in the field of path planning. Jump point search (JPS)\cite{harabor2014improving} is a higher level algorithm based on A* algorithm. Instead of taking uniformed and symmetry grids, some longer jumps can be made to create optimised grids. The optimality feature of search-based method is preserved, while the computational cost is reduced dramatically. The shortcomings of JPS still exist, as it is not suitable for complex and dense environments where the line segment between each node may be obstructed. Choosing suitable algorithm in various environment is essential to the solve the problem.

There are some other novel search-based methods. For example, hybrid A* \cite{petereit2012application}, D* \cite{stentz1994d} and so on. For a more detailed explanation of search-based navigation, please refer to \cite{lavalle2006planning, dolgov2008practical, yang2014literature}

\paragraph{Sample-based Methods}
Sample-based path planning methods must also have complete map information like search-based methods. Unlike search-based methods, sample-based methods do not rely on a predetermined grid or graph. Instead, they randomly sample the state space and connect the sampled vertices. The main advantage is that this method suits high-dimensional or unstructured space. The shortcoming is that the generated path cannot guarantee optimality.

\begin{algorithm}
\caption{Rapidly-Exploring Random Trees (RRT)}
\label{RRT algo}
\begin{algorithmic}[1]
\State \textbf{Input:} Initial configuration $x_{init}$, step distance $\Delta x$, iterations $N$
\State \textbf{Output:} RRT Tree $T$
\State $T.\text{init}(x_{init})$
\For{$i = 1 \text{ to } N$}
    \State $x_{rand} \gets \text{RandomConfiguration}()$ 
    \State $x_{near} \gets \text{NearestVertex}(x_{rand}, T)$ 
    \State $x_{new} \gets \text{NewConfiguration}(x_{near}, x_{rand}, \Delta x)$ 
    \If{\text{CollisionFree}}
        \State $T.\text{addVertex}(x_{new})$ 
        \State $T.\text{addEdge}(x_{near}, x_{new})$
    \EndIf
\EndFor
\State \textbf{return} $T$
\end{algorithmic}
\end{algorithm}

LaValle \cite{lavalle1998rapidly} developed the Rapidly-exploring Random Tree (RRT) method in 1998. Algorithm \ref{RRT algo} shows a pseudocode for RRT. RRT starts from an initial node and expands several tree nodes iteration by iteration. The predefined map information can be divided into free space and obstacle regions. The random nodes are only generated in the free space. Additionally, the feasibility of new connections between the nodes can be verified by whether or not they have any intersection between the obstacle region. By doing this, the algorithm can be applied in a more complicated environment with much higher efficiency to find a feasible path. However, the optimality cannot be guaranteed due to the sampling feature itself. Some RRT-based algorithms are developed to improve the results to produce an asymptotically optimal path \cite{karaman2011sampling}. For example, they were Rapidly Exploring Random Graph (RRG), RRT*, and Probabilistic Road Map* (PRM*). Those methods can provide a close-to-optimal path if the sampling iteration is large enough to leverage the searching speed and optimality. To take RRT* as an example, the redundant edges not part of the shortest path are removed. This process is called "rewiring", the main distinction from the traditional RRT algorithm \cite{noreen2016comparison}. This process can optimize the shortest path to asymptotically optimal if sufficient iterations are given.

\paragraph{Other Mentionable Methods}
Many artificial techniques are inspired by some natural creatures' behaviours; the same goes for path planning. Some bio-inspired algorithms are practical to find the feasible path, such as Particle Swarm Optimization (PSO) \cite{kennedy1995particle}, Genetic Algorithm (GA) \cite{mitchell1998introduction}, Ant Colony Optimization (ACP) \cite{dorigo2006ant}, Neural Network \cite{glasius1995neural} and so on. PSO was inspired by the bird cluster activity, similar to RRT*; the optimal solution can be found through enough iterations. Position and velocity are two essential parameters that the particle is guided by the local best position and global best position. It is worth mentioning that PSO doesn't require the optimization problem to be differentiable.

Another group of algorithms is based on potential fields, such as Wavefront Expansion Algorithm \cite{dorst1989optimal}, Navigation Functions \cite{lavalle2006planning} and Artificial Potential Field (APF) \cite{xu2020path}. The simplest example is that each coordination has a potential cost that can guide the robot to move to the target node with the minimum possible cost. However, simply using the potential field navigation method may guide the robot stuck at some local minima nodes before reaching the target node. Sang \cite{sang2021hybrid} combined the potential field method with the A* method to escape local minima and ensure the optimality of the path.

With the rapid advancement of artificial intelligence technology, some machine learning and reinforcement learning techniques \cite{wang2020mobile, xiao2022motion, liu2019new} are also applied to the path planning phase. Those techniques are suitable for highly constrained or highly diverse environments. The optimal solution can be produced as long as the map information is sufficient and the learning time is adequate.

\subsubsection{Local Path Planning}
With the help of a global path planner, an initial feasible path can be generated as a reference for the UAVs. However, due to the lack of the latest map information, global path planners cannot detect unknown obstacles. In this way, the local path planner can take the role. Local path planning is also called sensor-based or reactive mode navigation, meaning sensor engagement is an essential requirement. Video cameras, LiDAR cameras and ultrasonic sensors are typical onboard sensors that sense the actual environment. With the information sensors provide, robots can take the obstacle avoidance action directly or after updating the global map \cite{hoy2015algorithms, wang2018strategy, savkin2013Reactive}. When the computational capability is limited by the onboard computer, controlling the robot directly without updating the map is a more secure option.

The artificial potential field (APF) method can be used not only for global path planning but also for local path planning. This can be achieved by assigning a high potential value to the obstacle to push the UAV further away from the obstacle \cite{sfeir2011improved}. The closer the UAV is to the obstacle, the higher the repulsive force. Rostami \cite{rostami2019obstacle} proposed an obstacle avoidance planning algorithm based on a modified APF algorithm. Both attraction function and repulsive function are designed to avoid the local minima. MATLAB simulations are presented to show the robot can reach the target without colliding with any unknown obstacles.

Some other local path planning methods were widely used in the previous research. Boundary following \cite{toibero2009stable, matveev2012real} can be seen as a straightforward technique to avoid dynamic obstacles. When detecting an unknown object, it follows the boundary of it until an escape criterion is fulfilled. However, the robot's behaviour is relatively uncontrollable because it may follow the obstacle in the wrong direction for a long time. Another group of methods is based on the minimum distance between the closest obstacle. Matveev et al. \cite{matveev2011method} presented a sliding-mode-control-based obstacle avoidance strategy, taking the minimum distance between obstacles as the control input. The proposed algorithm can be applied to border control applications. Some scholars also use motion-prediction techniques \cite{yang2017obstacle, dai2017distributed} to predict the future route of unknown obstacles, then replan the path to avoid them. Some other mentionable methods are dynamic window approach \cite{fox1997dynamic, ogren2005convergent}, curvature velocity method \cite{simmons1996curvature}, lane curvature method \cite{ko1998lane} and collision cone \cite{chakravarthy1998obstacle, xu2020dynamic}. The low-latency feature of local path planning makes it the best fit to handle unknown uncertainties. More obstacle-avoidance techniques can be developed with the advancement of sensor technologies.

\subsubsection{Hybrid Path Planning}
Global path planning has the advantage of generating optimal paths based on the given map information, and local path planning can avoid unknown obstacles to provide a collision-free path. The hybrid path planning approach combines the advantages of both methods, which is a higher-level navigation technique for real-time navigation. Hybrid approaches use the pre-generated path by the global planner as a reference while enabling the local planner to ensure safe navigation. A replan may needed if the actual path is far from the reference path during the obstacle-avoidance stage.

Zhu et al. \cite{zhu2012new} use the A* algorithm as the global planner and a bug-algorithm-based local planner to navigate in a partially unknown environment. In addition, Li et al. \cite{li2020path} combined an improved A* algorithm with a dynamic window approach to fulfil the mission in a large-scale dynamic environment. The hybrid approach provides fewer nodes and a smoother path for the mobile robots.

In conclusion, the path planning phase plays a crucial role in securing the collision-free navigation of mobile robots. Multiple path planning schemes are discussed in this section. Based on those navigation techniques for normal UAVs, the navigation algorithms for SUAVs have a solid fundamental. The additional energy model for SUAVs increases the difficulty of resolving the problem.

\section{Report Organization}
Eight main sections in this report are briefly outlined here. Section \ref{ch1} introduces the research problem's background and briefly summarises the navigation methods for mobile robots. Section \ref{ch2} reviews the application of SUAVs and their navigation methods. Section \ref{ch3} contains a case study that UAVs need to be sensitive to the privacy region in the environment, which is the technical knowledge fundamental for the navigation of an SUAV. Section \ref{ch4} presents a global energy-efficient 3D navigation method for SUAV in an urban environment. Two algorithms are proposed for two different needs in the specific problem. Section \ref{ch5} adds an obstacle avoidance strategy based on the previous algorithm that enables it to navigate in a dynamic urban environment with unknown obstacles. Section \ref{ch6} concludes the whole report and provides suggestions for future research.

\section{Literature Review}
\label{ch2}
This section provides an overview of the recent applications for single and multiple SUAVs. In addition, the energy harvesting models are presented in the second section. Finally, the related navigation techniques are outlined in an organised manner. A comprehensive review of the navigation and deployment of SUAV can be found in \cite{li2024navigation}.

\textbf{Some of the work is part of the paper}: \textbf{S Li}, Z Fang, SC Verma, J Wei, and AV Savkin. "Navigation and Deployment of Solar-Powered Unmanned Aerial Vehicles for Civilian Applications: A Comprehensive Review" Drones 8.2 (2024): 42.

\subsection{Applications of SUAVs}
\label{section2}
The development of UAVs is vast in various industries and domains. From military to commercial applications, SUAVs with longer operation times have a dramatic advantage over traditional UAVs. Most SUAVs are installed with electric motors driven by photovoltaic cells and rechargeable onboard batteries. In addition, the size of the SUAV can be customized according to the required applications \cite{rajendran2015implications}. Therefore, they are trendy among various commercial applications. Numerous prototype demonstrations have been developed in the past \cite{el2021solar}. Papers \cite{pal2020recent, zhu2014solar} briefly list the typical applications of SUAVs, but a more comprehensive review is needed. This section summarises and details the specific applications for single and multiple SUAVs. 

\subsubsection{Vision Based Applications}
With the rapid development of computer vision and vision-based technology, related applications are gradually being developed \cite{lu2018survey}. It is becoming common for UAVs to be fitted with cameras or sensors that allow them to be used in vision-based applications. Because fitting such hardware would significantly increase the aircraft's energy consumption, SUAVs with longer operation time could be more adaptable to these application scenarios \cite{chu2021development}. Some state-of-art vision-based applications of UAVs are discussed in \cite{li2021networked}, however similar review on SUAVs is lacking. This section reviewed and summarized vision-based applications for SUAVs.

\paragraph{Target Surveillance}
There are several papers about video surveillance applications of SUAVs \cite{huang2019method, huang2018energy, wu2018energy, huang2021energy,WU2017497, huang2016energy,  hosseinisianaki2013energy, Solar1} in a different environment. 	Most of them investigate the target surveillance problems. The determining factors influencing the complexity of the problem include the number of SUAVs, the number of targets, and the motion of the SUAVs, which are studied in these papers. A detailed comparison is listed in Table \ref{table:1}. 	
\begin{table}[ht]
\begin{tabular}{|c|c|c|c|}
\hline
\textbf{Ref} &  \textbf{Number of SUAVs} & \textbf{Number of Targets} & \textbf{Target State}\\
\hline
\cite{huang2019method} & Single & Single & Fixed \\
\cite{huang2018energy} & Multiple & Single & Fixed \\
\cite{wu2018energy} & Single & Single & Moving \\
\cite{WU2017497} & Single & Single & Moving \\
\cite{huang2016energy} & Single & Single & Moving \\
\cite{huang2021energy} & Multiple & Single & Moving \\
\cite{Solar1} & Multiple & Multiple & Moving \\
\hline
\end{tabular}
\caption{Vision-based Applications}
\label{table:1}
\end{table}

Authors \cite{huang2019method} focus on continuously surveilling a fixed ground target based on the limited 3D space and energy storage capability of SUAV. A target-centred virtual cylinder is constructed around the fixed target according to this specific mission, and a path planning framework is developed to maximize the overall energy efficiency. However, one shortcoming of the proposed methods is that the work does not consider the geometrical condition. Consequently, the virtual cylinder may not be feasible in a more complex environment. As a comparison, papers \cite{Solar1,  huang2016energy, wu2018energy, WU2017497,huang2021energy} investigate the target tracking problem for moving ground target instead of fixed target that elevates the level of complexity. Huang and Savkin \cite{Solar1} investigate a navigation problem for a team of SUAVs conducting video surveillance work for some moving targets. Multiple SUAVs are assigned to monitor the target periodically, maximizing the total monitoring time. Their proposed method is approved to be effective via computer simulation. However, one shortcoming of the navigation methods is that the trajectories of the moving targets are known information for the navigation system. Except for the target tracking mission, SUAVs can also do some general surveillance missions. Hosseinisianaki et al.\cite{hosseinisianaki2013energy} develop an energy-aware surveillance algorithm for SUAVs that leverages the energy storage and the surveillance coverage of the flight. The optimal coverage range of the sensor is challenging when facing the requirement of maintaining a high energy efficiency rate. Eventually, a comprehensive model is formulated for optimal coverage, path planning and power allocation for an SUAV. 

\paragraph{Environmental Applications}
With the help of onboard solar panels and vision-enabled technology, SUAVs also play an essential role in agriculture applications \cite{herwitz2002precision, herwitz2003solar}. Remote sensing technology allows SUAVs to conduct precision agriculture and smart farming missions. Herwitz et al. \cite{herwitz2002precision} discuss the potential of using SUAV for coffee cultivation because of the wireless communication and imaging capabilities. To take it a step further, author \cite{herwitz2003solar} demonstrates the safe and effective use of SUAVs for collecting commercial agricultural images, indicating the bright future to extend the usage in different agriculture sectors. Besides agriculture applications, SUAVs can also monitor environmental conditions \cite{malaver2015design, malaver2015development, thipyopas2019design}. Malaver et al.\cite{malaver2015design, malaver2015development} use SUAVs to monitor greenhouse gases with a wireless sensor network. The third paper \cite{thipyopas2019design} designs an SUAV system capable of conducting a 6-hour endurance mission at a 1000-meter altitude, further proving the promising future of using SUAV for environmental monitoring. Those successful prototypes can push the further vision-based applications for SUAVs.

\paragraph{3D Mapping}
Autonomous navigation is becoming more and more common to lower labour costs, which is essential to acknowledge sufficient map information to increase navigation accuracy. The map-based system has predefined map information to pre-plan the route to guide the SUAV to follow a feasible collision-free path. The 3D mapping \cite{oettershagen2016long, dwivedi2018maraal, karthik2021design} thus becomes a popular research direction for unmanned systems. The first paper \cite{oettershagen2016long} states the effectiveness of image capturing by SUAVs, which is crucial for some vision-based applications. In addition, Geographic Information Systems (GIS) is integrated into the SUAVs for advanced data analysis. MARAAL \cite{dwivedi2018maraal} is a mature SUAV developed by Unmanned Aerial Laboratory IIT Kanpur that can be used for 3D mapping applications. One highlight of this SUAV is that the coverage altitude is considerably high, varying from 1 km to 5 km above the ground. Lastly, Karthik et al. \cite{karthik2021design} focus on designing and developing an SUAV for surveying, mapping and disaster relief applications. In summary, integrating solar energy in SUAVs has revolutionized the application of 3D mapping, offering extended flight durations and enhanced 3D  mapping capabilities.

\subsubsection{Wireless Applications}
Wireless communication systems have become an important part of the big data era. A number of novel technologies need the support of a high data transfer rate \cite{chettri2019comprehensive}. However, the deployment of the traditional communication base stations is not cost-efficient and cannot meet current technology's needs. Therefore, a UAV-based aerial communication system is a promising solution, especially in a complex environment \cite{jawhar2017communication}. A further point worth mentioning is that SUAVs with much longer flight times are more suitable to be deployed as an aerial communication base. Moreover, the multi-SUAV communication system can be viewed as an example of networked control system \cite{matveev2009estimation, savkin2006analysis, matveev2003problem, gupta2009networked, yang2006networked, liang2018control, zhang2019networked}. In this section, some popular wireless system applications \cite{padilla2020flight, song2021energy, huang2020autonomous, huang2020energy, sun2018resource, sun2019optimal, romeo2004heliplat1, romeo2004heliplat2, wozniak2021selection, luo2023maximizing} are presented.

\paragraph{Communication System}
Wireless transceivers can be equipped on SUAVs to provide communication service while hovering at a specific position. In addition, the high maneuverability allows UAVs to secure communication services while flying. Paper \cite{padilla2020flight, song2021energy} studies the optimization problem for SUAV communication systems with multiple ground users. Both signal path loss and shadowing are considered to maximize energy efficiency and sustainability. Similarly, author \cite{huang2020autonomous} considers a problem in that one SUAV is used to secure communication with one ground node in the presence of eavesdroppers. The developed algorithm is focused on optimizing the remaining energy of the SUAV and the overall period during which the SUAV can communicate securely with the fixed ground node. The algorithm differs from traditional distance minimization, providing a larger time frame to secure the 
communication service. Furthermore, the presence of eavesdroppers increases the difficulty of solving this problem. To take it a step further, Sun et al. \cite{sun2019optimal} not only designed the optimal 3D trajectory but also developed a resource allocation strategy for multicarrier SUAV communication systems by using the sole information of channel states. The optimal design provides a stable and sustainable communication service for the SUAV communication system. In 2004, the first generation of European very long-endurance high-altitude solar-powered UAVs was designed and prototyped, called HELIPLAT \cite{romeo2004heliplat1, romeo2004heliplat2}. This HELIPLAT platform has several advantages, including cost efficiency by increasing flight endurance, decreased maintenance and spare costs, and enhanced flight safety by flying above traffic and weather conditions, with less interference with aviation traffic. It also offers extensive area coverage that requires fewer SUAVs per area. However, it does have limitations, notably in payload weight (less than 1500 N) and available power for the payload (1000–1200 W). Additionally, paper \cite{romeo2004heliplat1} shows the possibility of HELIPLAT being used for some communication applications. All in all, with the evidence support of simulation results and prototype testing, SUAVs can be used for various communication applications in different environments.

\paragraph{Data Acquisition}
Similar to communication applications, this section mainly focuses on the data acquisition applications \cite{wozniak2021selection, luo2023maximizing}. SUAVs are used to acquire specific information instead of an information base station. Wozniak and Jessa \cite{wozniak2021selection} develop a long-range data acquisition chain using SUAVs that specifically focuses on areas with poor electricity and communication infrastructures. The data acquired by SUAVs during the process can also be used for environmental monitoring applications. From their evaluation of more than fifty UAVs, it can be concluded that high cost and large size are not primary factors for the selection of UAVs. Additionally, the payload mass significantly influences the efficiency of different SUAVs. Building upon the insights and experiments provided by Wozniak and Jessa, Luo et al. \cite{luo2023maximizing} further explore other data acquisition applications. The authors address the problem of data acquisition from wireless sensor networks using SUAVs in an urban environment. Three algorithms are proposed to optimize the trajectory to maximize sensor data collection. Both of these papers explore the scenarios to collect data from remote or inaccessible areas using SUAVs, marking a significant advancement in environmental monitoring and connectivity solutions.

\subsubsection{Delivery Applications}
UAVs can potentially reduce labour costs for deliveries, save time for last-step deliveries, and respond more quickly to emergencies \cite{dorling2016vehicle}. In the near future, the delivery of UAVs will have some effect on the traditional transportation delivery system. The high maneuverability makes UAVs suitable for healthcare delivery for emergency surgery \cite{scott2017drone}. Paper \cite{tian2022routing} introduces the routing strategy for SUAV delivery systems in a complex urban environment. A pruning-based globally optimal algorithm is developed for this problem, giving a better simulation result than a traditional routing algorithm for normal UAVs. Unlike normal UAVs, SUAVs are not significantly limited by the capacity of the onboard batteries. In addition, the cost of deploying denser charging stations limits the development of UAV delivery systems. The research can not only focus on path planning for delivery applications \cite{huang2020reliable} but also on the deployment of charging stations for recharging purposes \cite{huang2020method}. Using SUAVs for longer-range delivery can thus be seen as a promising solution. More research is needed on how to land safely in a very complex environment to push the implementation of this technology.

\subsection{Energy Models of SUAVs}
\label{section3}
The main difference between a normal UAV and an SUAV is that the SUAV can harvest energy from the sunlight. Papers \cite{rajendran2015review, morton2015solar} conclude the energy system and solar cell selection for SUAVs. On-board solar panels can convert solar energy to electrical energy, which makes SUAVs have longer endurance in exposure to sunlight \cite{khan2023solar}. However, SUAVs cannot always harvest energy due to weather conditions or obstructions caused by obstacles. In this section, two main scenarios are considered: first, the energy model in cloudy conditions, and second, the energy model in cloud-free conditions.

\subsubsection{Line of Sight}
Before introducing the energy model, it is essential to find a way to determine whether obstacles are blocking sunlight. This is more applicable for multi-rotor SUAVs as they fit better in complex urban environments. For example, buildings taller than the SUAV in an urban environment could block direct sunlight, creating a shadowy area. Line of Sight (LoS) is commonly used to address this problem. In addition, LoS is commonly used in target tracking \cite{rysdyk2003uav, kim2008moving} and communication \cite{mohammed2021line} fields. Huang and Savkin \cite{Solar1} implement LoS for an application that contains energy harvesting and target surveillance problems. The buildings in an urban environment can be treated as obstacles, blocking not only the movement of SUAVs but also the sunlight and the LoS between SUAVs and the target. For the LoS, the straight line segment between two nodes $p_{\text{sun}}$ and $p$ are considered, where $p_{\text{sun}} = (x_s, y_s, z_s)$ represents the position of the sun and $p = (x, y, z)$ is the current coordinates of SUAV. In this way, it can be described as \cite{Solar1}
\begin{eqnarray}
	\begin{cases}
	x = x_s + \alpha_x\gamma\\
	y = y_s + \alpha_y\gamma\\
	z = z_s + \alpha_z\gamma\\
	\text{min}\{ x_p,x_s\} \leq x_s+\alpha_x\gamma\leq \text{max}\{x_p,x_s\}.
	\end{cases}
\end{eqnarray}
Therefore,  
\begin{eqnarray}
	(\alpha_x,\alpha_y,\alpha_z)=\frac{\vec {p_{\text{sun}}p}}{\|\vec {p_{\text{sun}}p}\|},
\end{eqnarray}
where $\gamma$ is a scalar variable. The LoS can be determined by checking the intersection points between $p_{\text{sun}}$ and $p$ and any buildings. The assumption is that the full map information is given so the LoS can be determined.

\subsubsection{Energy Harvesting Model}
The LoS check can provide information on whether the SUAV is under the sunlight. With this information, the effects of clouds and obstacles can be determined. There are two main aspects of the energy models of SUAV: energy consumption and harvest. The only difference in the energy consumption model is the increased weight due to additional solar panels. Thus, the energy consumption model is not being discussed in this section. One crucial decision factor for the energy harvesting model is the cloud condition. Clouds lead to higher attenuation of solar power, as solar power is mainly centred at wavelengths smaller than 1 mm \cite{zhang2020power}. Therefore, two types of energy harvesting models are discussed based on whether or not the cloud condition is considered.

\paragraph{Cloud-Integrated Energy Model}
This scenario needs to integrate the effect of the cloud into the solar energy model. Additionally, it is applicable when the altitude of the SUAV varies dramatically, potentially covering the altitude below, in, and over the clouds. The attenuation of the sunlight through a cloud can be modelled as \cite{kokhanovsky2004optical}:
\begin{eqnarray}
\label{solar cost 1}
	\varphi(d_{\text{\text{cloud}}})=e^{-\beta_cd_{\text{\text{cloud}}}},
\end{eqnarray}
Where $d_{cloud}$ represents the distance over which solar light travels through the cloud, and $\beta_c$ is the absorption coefficient that models the optical properties of the cloud. In addition, $\beta_c$ has to be a non-negative value. Now define the cloud's upper boundary and lower boundary to be $H_{up}$ and $H_{down}$ respectively. Thus, the solar panel output power can be described as \cite{sun2019optimal}:
\begin{eqnarray}
\label{solar cost 2}
	P = 
	\begin{cases}
	\eta S G, \quad  ~~~~~~~~~~~~~~~~~~~z \geq H_{\text{up}},\\
	\eta S G e^{-\beta_c(H_{\text{up}}-z)}, ~~~~~~~~ H_{\text{down}} \leq z \leq H_{\text{up}},\\
	\eta S G e^{-\beta_c(H_{\text{up}}-H_{\text{down}})},~~~~ z < H_{\text{down}},
	\end{cases}
\end{eqnarray}
where $z$ is the altitude of the SUAV. The given function describes a scenario where the $P$ grows exponentially as SUAVs ascend through the cloud but stabilizes at a constant value when below or above the cloud. 

\paragraph{Cloud-Free Energy Model}
One typical scenario for a cloud-free environment is when SUAVs fly high enough to eliminate the cloud influence. Some fixed-wing SUAVs can achieve this altitude. Another scenario is when SUAVs are flying at a low altitude, so the obstructions by the obstacles are the main factor. Furthermore, SUAVs are assumed to be always under the cloud, and it is not realistic to optimize the path to avoid the shadow of the cloud. Because the position of the cloud is not a proper pre-defined information, in this case, many scholars use the ASHRAE clear sky model \cite{sayigh2012solar} to address the problem adequately \cite{baldock2006study, le2022coverage, wei2020comprehensive}. It assumes that the SUAV always flies in a clear sky without any clouds. In addition, the model also provides data on the average solar irradiance. When the sunlight shines directly on the solar cells, the output electrical power $P(\theta)$ can be described as \cite{klesh2009solar}:
\begin{eqnarray}
\label{solar cost 3}
	P(\theta)=\eta G S \cos (\theta) \quad \text { if } \cos (\theta) \geq 0,
\end{eqnarray}
where $\eta$ stands for the efficiency of the solar cell, S represents the solar panels' total area, and $G$ is the solar spectral density. If only part of the solar panel is exposed to the sunlight, $\eta$ can be changed accordingly to address the problem. The incident angle $\theta$ depends on the azimuth angle $a$ and elevation angle $e$, which have the following relation:
\begin{eqnarray}
	\cos (\theta)=\cos (\phi) \sin (e)-\cos (e) \sin (a-\psi) \sin (\phi).
\end{eqnarray}
Thus, the total solar energy harvested during time frame $[t_{o}, t_{f}]$ is
\begin{eqnarray}
\label{solar cost 4}
	E=\int_{t_{o}}^{t_{f}} P(\theta) \mathrm{d} t.
\end{eqnarray}
This model is the classic solar energy harvest model and is the foundation for many other models. Equation (\ref{solar cost 2}) is also based on it.

Another widely used model is modified based on equation (\ref{solar cost 2}) and can illustrate the effect of flight height better \cite{fu2021joint, luo2023maximizing}.
\begin{equation}
\label{solar cost 5}
	\eta S G e^{\alpha_c-\beta_c e^{\frac{-z} {\delta_c}}},
\end{equation}
where $\alpha_c$ represents the maximum value of atmospheric transmittance and $\delta_c$ stands for the scale height of the earth. It can be seen that the SUAV at higher altitudes can harvest more energy than that at lower altitudes.

\subsection{Navigation Methods}
\label{section4}
UAVs can take off and land in various complex environments, making them widely used in multiple applications. A collision-free path is essential to the whole system when executing the missions. Some traditional and widely used algorithms, for example, Dijkstra algorithm \cite{bondy1982graph}, A-star algorithm \cite{Hart1968} and Rapidly-exploring Random Tree (RRT) algorithm \cite{lavalle2001rapidly} can provide an initial path based on the map information. However, for SUAVs, energy consumption and harvest also need to be considered, making the path planning problem much more challenging. This section summarizes commonly used path planning strategies for SUAVs. Table \ref{table:2} provides the list of references, the corresponding navigation strategy and the number of SUAVs investigated.
\begin{table}[ht!]
\centering
\begin{tabular}{|c|c|c|}
\hline
\textbf{Ref} &  \textbf{Navigation Strategy} & \textbf{Number of SUAVs}\\
\hline
\cite{huang2019method} & Optimization & Single\\
\cite{huang2018energy} & Optimization & Single\\
\cite{WU2017497} & Distributed Model Predictive Control & Multiple\\
\cite{huang2016energy} & Particle Swarm Optimization & Multiple\\
\cite{huang2021energy} & RRT & Multiple \\
\cite{Solar1} & RRT & Multiple \\
\cite{hosseinisianaki2013energy} & Optimization & Single \\
\cite{padilla2020flight} & Optimization & Single\\
\cite{huang2020autonomous} & RRT & Single \\
\cite{huang2020energy} & RRT & Single\\
\cite{sun2019optimal} & Monotonic Optimization & Single\\
\cite{fu2021joint} & Optimization & Single\\
\cite{huang2021path} & RRT & Single \\
\cite{wu2018path} & Whale Optimization Algorithm & Single\\
\cite{wirth2015} & Dynamic Programming & Single \\
\cite{tuan2021mpc} & Model Predictive Control &  Single\\
\cite{7859311} & Optimization & Single\\
\cite{dai2012optimal} & Optimization & Single\\
\cite{spangelo2013power} & Optimization &  Single\\
\cite{hosseini2013optimal} & Optimization & Single\\
\cite{lun2022target} & Optimization & Multiple\\
\cite{kim2019flight} & Optimization & Single\\
\hline
\end{tabular}
\caption{Navigation Strategy}
\label{table:2}
\end{table}

It can be seen from the table that the majority of previous works address the path planning problem of SUAVs as an optimization problem. The total energy model must be defined and formulated as the cost function. In the meantime, some of them use RRT or MPC-based algorithms to plan the path. The generated path is sub-optimal compared to optimization-based methods, but the results are still feasible and reasonable. Similarly, most researchers focus on single SUAV path planning rather than multiple SUAVs. For multiple SUAV systems, the cost functions are more complex than a single SUAV system in the same scenario. Some work addresses the coverage path planning problem instead of traditional minimum-cost path planning. Therefore, this section focuses on three parts: sample-based methods, optimization-based methods and coverage navigation methods.

\subsubsection{Sample-based Methods}
Sample-based path planning algorithms always require certain environmental information, and then the map is sampled as a set of nodes. Based on the newly sampled map, the algorithm can find the feasible path by searching and tuning the path. Rapidly-exploring Random Tree (RRT) is one of the effective methods to address the energy-aware path planning problem \cite{huang2021energy, Solar1, huang2020autonomous, huang2020energy, huang2021path}. However, one shortcoming of RRT is that the generated path is not optimal. The path planning part of paper \cite{huang2020energy} is straightforward for SUAVs. Firstly, some feasible paths are obtained using RRT* while satisfying the communication constraints and energy requirements at the same time. Secondly, select the trajectory with the minimum energy consumption. The simplicity is the advantage of this method so long as it has energy harvesting and consumption models. However, the algorithm itself is not involved with the energy models. In this case, authors \cite{huang2021path} further modify the RRT algorithm to make it sensitive to energy cost. They research the application of inspecting mountain sites using an SUAV. The modified RRT* algorithm first finds a feasible path and then rewires the nodes if the residual energy is higher than the threshold energy level. The feasible path is broken into multiple sub-tours, and the algorithm aims to reduce the length of each sub-tour by replacing longer paths with more energy-efficient ones while maintaining the UAV's energy requirements. This approach balances energy efficiency to reduce the overall completion time of the inspection tour, which can provide a relatively more optimal path than the previous one.
\subsubsection{Optimization-based Navigation}
Unlike sample-based navigation, optimization-based methods can find the optimal solution based on the derived objective function. Fu et al. \cite{fu2021joint} consider a 3D trajectory planning problem while collecting data from the ground devices using one SUAV. Both the energy model and communication model are considered in the issue. In addition, the SUAV is also used to charge ground communication devices through laser charging, making the problem difficult to solve. Multiple objective functions are built to represent the mathematical models in the system. A convex optimization solver based on the interior-point method is used to solve this problem. Authors \cite{7859311} solve the optimal path planning problem based on the gravitation potential energy. The objective function is defined by gravitational and battery residual energy while setting the yaw and pitch angles as control inputs. On the other hand, paper \cite{dai2012optimal} studies a similar problem but uses the branch and bound (BNB) method for optimization. Moreover, Euler angles and unit quaternions address the complexities in 3D flight models. Both papers aim to maximize the energy efficiency and effectiveness of SUAVs by exploring different methodologies in optimal path planning. In contrast to previous algorithms, Lunn et al. \cite{lun2022target} focus on the target searching problem in a dynamic environment. This path planning problem differs from others because the target node is unknown during the planning phase. The grey wolf optimization algorithm has also been chosen as the solver for this problem. In contrast with previous research,  multiple SUAVs cooperation is considered, and the repulsive force field in the artificial potential field method is used to design the collaboration cost function. However, environmental effects, such as the effect of the wind, are not considered in any of the papers above. To fill the gap, paper \cite{wirth2015, kim2019flight} considers the wind effects during the path planning stage. Wirth et al. \cite{wirth2015} use a particular software, "Meteorology-aware Trajectory Planning and Analysis Software for Solar-powered UAVs" (METPASS), for the simulation part. The model considers environmental hazards, as well as wind and solar radiation, making the result more realistic. It is worth mentioning that dynamic programming technique is used to plan the path more efficiently. Similar to this paper, authors \cite{kim2019flight} consider the SUAV path planning problem in wind fields using direct collocation. Solar radiation and wind effects are evaluated using the numerical forecast model. 
\subsubsection{Coverage Navigation}
Traditional path planning algorithms focus more on obtaining a feasible path from a start node to a target node while satisfying some soft or hard constraints \cite{aggarwal2020path}. Sufficient map information is essential to planning the path effectively. Then, the coverage path planning algorithm is designed to fulfil the task. Papers \cite{7101619, le2022coverage, vasisht2015trajectory} research on the specific coverage path planning algorithm for SUAVs. Franco et al. \cite{7101619} develop a back-and-forth algorithm for convex and concave areas to find the optimal coverage path. Likewise, the coverage path optimization model in the second paper \cite{le2022coverage} is developed based on the undirected graph search method. In the meantime, another mixed integer linear programming model is used to minimize the total flight time. The authors provide a large number of simulations to prove the effectiveness and adaptability of the proposed method. In contrast to Franco's work, the second paper restricts the flying zone. It uses multiple SUAVs to complete the mission corporately, which makes the algorithm able to be applied in more scenarios and be more adaptable to complex environments. Unlike the previous two pieces of literature, paper \cite{vasisht2015trajectory} adds a hotspot surveillance strategy to the coverage planning algorithm. Priority distribution is visible to SUAVs, making the path denser near higher-priority nodes. 

In conclusion, the coverage navigation for SUAVs is a relatively immature area that few researchers have considered. The solar energy models make it more complicated than regular UAV coverage navigation algorithms. More research is needed to fill the gap.

\subsection{Research Challenges}
In the previous sections, the navigation and deployment techniques for civilian applications of SUAVs were introduced. However, there are still several open problems that merit further investigation. Because of the immature nature of the investigated research area, many issues are left to be fully resolved before integrating SUAV technologies to real life.

Primarily, while there are several small-scale SUAV navigation techniques for intricate settings, the majority of these researches fail to consider environmental aspects. Accordingly, exploring the impact of environments' variations in wind and solar angles of incidence is challenging that could serve as an exciting field for further direction. In \cite{wang2019overview}, the author discuss the effects of different wind patterns on UAVs, providing an in-depth explanation of the mechanisms that interfere the UAV's orientation, location and velocity. Integrating wind-based features and mathematical models to categorize wind quantitatively is an innovative concept. The factors taken into account in the overall classification are similarly extensive. Ultimately, the results are verified in terms of energy transmission and the influence of unmanned UAV trajectory under varying wind conditions. The wind can not only affect the speed of the vehicle, but also influence the gesture of the SUAV. In the present of solar panels, the multi-rotor UAV may therefore increase the area affected by the wind, as well as the angle of light it receives. These influences are difficult to fully represent in simulation, and the impact is even greater in real-world application scenarios.

Based on the previous challenge, another perspective can be drawn out. The variation in the angle of sunlight incident on the solar panel can dramatically influence the energy absorption efficiency. This variation is not only due to the variability of sunlight itself but also due to the changing orientation of the SUAVs. The efficiency may also vary during a long flight, as the vehicle may travel through areas with low sunlight intensity, potentially leading to energy shortages for the SUAV. Furthermore, since the vehicle may not maintain a fixed altitude, variations in altitude and latitude are other factors that need to be considered. In addition, since the sun moves during the flight, the incident angle may change significantly if the mission is of sufficient duration. Multiple aforementioned factors can potentially influence the solar cell efficiency on the SUAV, making the investigated problem require very careful consideration. How to comprehensively consider all factors before deploying a drone, or to create a model that is closest to reality during simulation, is a challenging aspect.

In addition to the research discussed above, it is important to note that none of the studies have focused on the navigation of SUAVs in dynamic environments. However, the ability to operate in such environments is crucial, as it more closely mirrors real-life applications. Dynamic environments are characterized by various uncertainties that can obstruct the SUAV's movement, making predetermined paths infeasible. Traditional path planning methodologies alone are insufficient for addressing these challenges. Instead, a more advanced algorithm is required to navigate effectively through unpredictable conditions and ensure the SUAV's successful operation. Developing an algorithm for SUAVs in dynamic environments represents a research area of significant importance.

Last but not least, the hardware considerations for SUAVs present a significant challenge. The advantage of equipping solar panel on UAVs is obvious, but there are some constraints need to be discussed. The increase of the vehicle weight may compromise the energy consumption during the flight. More energy are drawn in the same given amount of time when the total weight in increased. Additionally, the payload of the vehicle can be influence because of that. Compromises must be considered during the initial stages of building SUAVs. Furthermore, the manufacturing costs of SUAVs are higher than those of traditional UAVs. Although investing more in the early stages can be profitable in the long run, this factor should not be overlooked. These issues need to be fully considered when implementing SUAV. It is because of these challenges that the implementation of SUAV has been relatively slow for civilian applications.

\section[Privacy-Aware Navigation of a UAV]{Privacy-Aware Navigation of a UAV} 
\label{ch3} 

This section considers a privacy-aware navigation strategy for a UAV. The objective is to find a minimum privacy violation risk path from a start node to a target node. A dynamic programming-based algorithm is developed to fulfil the goal. In addition, the nature of the algorithm is an information-aware algorithm, which is similar to the energy-aware navigation algorithm for an SUAV. The work discussed in this section not only inspired the author to conduct further research on navigation for SUAVs but also serves as a technical foundation for the following sections.

\subsection{Introduction}
The development of unmanned aerial vehicles has rapidly been applied to various applications, from defence-related to civilian applications. However, the privacy and safety issue is attracting more and more attention at the same time \cite{zhi2020security, blank2018privacy, ly2021cybersecurity}. Leakage of private information exists not only through video surveillance but also through the interception of data transmission. Because the features that UAVs can be controlled remotely and deployed in a covert way, their privacy concern is unavoidable. On the other hand, the hackers may attack the UAV to steal some information as well \cite{westerlund2019drone}. Thus, developing algorithms that enable UAVs to avoid certain privacy-sensitive areas can address the issue. One commonly used method for privacy-aware navigation is to treat privacy regions as no-fly zones. In this way, the problem is transformed into an obstacle avoidance problem. However, the limitation is that when privacy-sensitive regions occupy a significant amount of space, a feasible path that can avoid all such areas may not exist. \cite{luo2020privacy}

In this section, the privacy region is not treated as a no-fly zone; instead, a combination of soft and hard constraints represents the privacy region. Only some high privacy violation risk regions are set to no-fly zones, and enough space is left for the UAV to find a feasible path. Furthermore, some quantitative measurements are made to measure the privacy risk so the UAV can see the path with minimum privacy risk. A dynamic programming-based navigation strategy has been developed to address the problem.

The rest of the section is structured as follows. The problem being investigated is described in section \ref{PS}. Section \ref{NL} introduce the dynamic programming-based navigation law. Section \ref{CS} shows the computer simulation of the proposed methods with a comparison with other path planning methods. Section \ref{C} concludes the whole section.

\subsection{Problem Statement}
\label{PS}
In this section, the UAV flies in a bounded 3D space ${\cal A}\subset {\bf R}^3$. $P(t)= (x(t),y(t),z(t))$ represents the position of the UAV at time $t$. Then, the UAV flight model can be described as:
\begin{eqnarray}
\label{eq1}
	\dot{P}(t) = vQ(t),
\end{eqnarray}
where $v$ is the speed and $Q(t)$ is the heading of the UAV. It can also be described as:
\begin{eqnarray}
	P(t) = P(0)+v\int_{0}^{t} Q(\tau)d\tau.
\end{eqnarray}
In addition, the flying height have a hard constraint that hold at all times.
\begin{eqnarray}
\label{eq2}
	z^{min}\leq z(t)\leq z^{max}.
\end{eqnarray}
Here $z^{min}$ and $z^{max}$ are some constant values. 

In the 3D space, some high privacy violation risk areas are set as no-fly zones, and the UAV is never allowed to enter those areas. Those no-fly zones are already excluded from $\cal A$. Therefore, the coordinates of the UAV $P(t)= (x(t),y(t),z(t))\in {\cal A} \forall t$.

The assumption is made that there are $N$ numbers of private regions, and the privacy violation intensity is assigned to each region $j=1,\ldots, N$ that is modelled by
\begin{eqnarray}
	\begin{cases}
	F_j(x,y,z) = 0, 		\quad \quad \quad \text{no violation}\\
	0 < F_j(x,y,z) < 1, 	\quad \text{acceptable violation}\\
	F_j(x,y,z) = 1, 		\quad \quad \quad \text{high violation},
	\end{cases}
\end{eqnarray}
where the no-fly zone is the third case that $F_j(x,y,z) = 1$. In this section, privacy regions are modelled as spheres, a realistic representation in the field of telecommunications. Thus, the center of the privacy regions are represented as $C_j=(x_j,y_j,z_j)$ for privacy region $j$. To ensure safe navigation, another hard constraint can be introduced
\begin{equation}
\label{eq3}
	F_j(x,y,z) < 1.
\end{equation}
This section aims to safely navigate a UAV from a start node to a target node in a given amount of time while minimizing the total privacy violation risk. To measure the total privacy violation risk, the following equation is made
\begin{equation}
\label{sum}
	\sum_{j=1}^N\int_0^{t_f} F_j(x(t),y(t),z(t))dt.
\end{equation}
This equation will be explained further in the following section.

{\bf Problem 1: } Consider a UAV whose motion is described by (\ref{eq1}) flying in a 3D space ${\cal A}$, design control laws for $v$ and $Q(t)$ to achieve a collision-free navigation from $P_0$ to $P_{target}$. In the meantime, try to achieve a minimum cost function (\ref{sum}) and satisfy the safety requirements in (\ref{eq2}) and (\ref{eq3}).

A typical application of the investigated problem might arise during a parcel delivery task when residential properties, national defence territories or critical infrastructures exist along the route. Regulations dictate a minimum distance between the UAV and these locations, and the flight becomes riskier the closer the UAV is to them. Our algorithm can produce a proper path in this scenario.

\subsection{Navigation Law}
\label{NL}
This section uses a dynamic programming-based algorithm to find the optimal path.

Define $M>0$ as a given integer and $\delta:= \frac{T_max}{M}$, where $T_{max}$ is the upper bound of the time. For all $i=0,1,\ldots,M$, introduce sets ${\cal S}(i)\subset {\cal A}$ as follows. The set ${\cal S}(M)$ consists of one point $P_f$, i.e., ${\cal S}(N):=\{P_f\}$. For any $i<M$, the set ${\cal S}(i)$ consists of all points $\hat{P}\in {\cal A}$ such that there exists a solution of (\ref{eq1}) over the interval $[0,\delta]$ with $v(t)=v^{max}$ and constant $Q(t)$ for all $t\in [0,\delta]$ such that $P(0)=\hat{P}$, $P(\delta)\in {\cal S}(i+1)$ and $P(t)\in {\cal A}~~\forall t\in [0,\delta]$. Introduce also functions $V(i,\hat{P})$ defined on ${\cal S}(i)$ for $i=0,1,\ldots, M$ as follows:
\begin{eqnarray}
\label{V}
	V(M,P_f):=0, \nonumber V(i,\hat{P}):=\\
	 \min \left[ V(i+1,P(\delta))+ \sum_{j=1}^N\int_0^{\delta} F_j(x(t),y(t),z(t))dt\right],
\end{eqnarray}

where the minimum is taken over all solutions $P(t)=(x(t),y(t),z(t))$ of (\ref{eq1}) on  the interval $[0,\delta]$  with $v(t)=v^{max}$ and constant $Q(t)$ for all $t\in [0,\delta]$ such that $P(0)=\hat{P}$, $P(\delta)\in {\cal S}(i+1)$ and $P(t)\in {\cal A}~~\forall t\in [0,\delta]$. Moreover, let $\hat{Q}(i,\hat{P})$ be the vector such that the minimum of (\ref{V}) with $P(0)=\hat{P}$ is achieved with $Q(t)\equiv \hat{Q}(i,\hat{P})$. Introduce the set ${\cal I}=\{0,1,\ldots,M-1\}$ of indexes $i\in {\cal I}$ such that $P_0\in {\cal S}(i)$. If ${\cal I}$ is not empty, let $i_0\in {\cal I}$ be the index such that the minimum in $\min_{i\in {\cal I}}V(i,P_0)$ is achieved at $i=i_0$, i.e. 
$V(i_0,P_0)\leq V(i,P_0)$ for all $i\in {\cal I}$.

The UAV trajectory with an initial condition $P(0)=P_0$ and the following $t_f$ and the control inputs:
\begin{eqnarray}
\label{Tr}
	t_f:=\frac{(M-i_0)T_g}{M}=(M-i_0)\delta,\nonumber\\
	v(t)=v^{max}~~\forall t\in [0,t_f],\nonumber\\
	W(t)=\hat{Q}(i,P((i-i_0)\delta))
\end{eqnarray}
for all $t\in [(i-i_0)\delta,(i-i_0+1)\delta)$ and for all $i=i_0,i_0+1,\ldots,M-1$.
\subsubsection{Computer Simulation}
\label{CS}
In this section, the computer simulation of the proposed method is presented below. Both 2D and 3D environments are considered to validate the algorithm better. In addition, MATLAB is used to implement the algorithm and visualize the results. As a global path planner, the map information is fully known to the system, and there are no dynamic or unknown obstacles. To better illustrate the privacy region, the assumption is made that any non-convex privacy regions can fit in a finite radius sphere. Let $E((x_j,y_j,z_j),(x,y,z))$ denote the 3D Euclidean distance between privacy region $j$ and the UAV in the 3D, and $0<c_1<c_2$ be given constants. The design of $F_j(x(t),y(t),z(t))$ is shown as follow
\begin{eqnarray}
	\begin{cases}
	F_j(x,y,z) = 0, 		\quad \quad \quad \text{if} ~ E\geq c_2\\
	0 < F_j(x,y,z) < 1, 	\quad \text{if} ~ c_1 < E < c_2\\
	F_j(x,y,z) = 1, 		\quad \quad \quad \text{if} ~ E\leq cd_1.
	\end{cases}
\end{eqnarray}
To be more specific, $F_j(x,y,z) := E((x_j,y_j,z_j),(x,y,z))\times (\frac{1}{c_1-c_2})-\frac{c_2}{c_1-c_2}$. This cost function will be used in the dynamic programming algorithm to find the minimum privacy violation path. Equations (\ref{eq2}) and (\ref{eq3}) are the hard constraints.

\subsubsection{Simulations in 2D Environment}
\begin{figure}[ht]
\centerline{\includegraphics[width=7 cm]{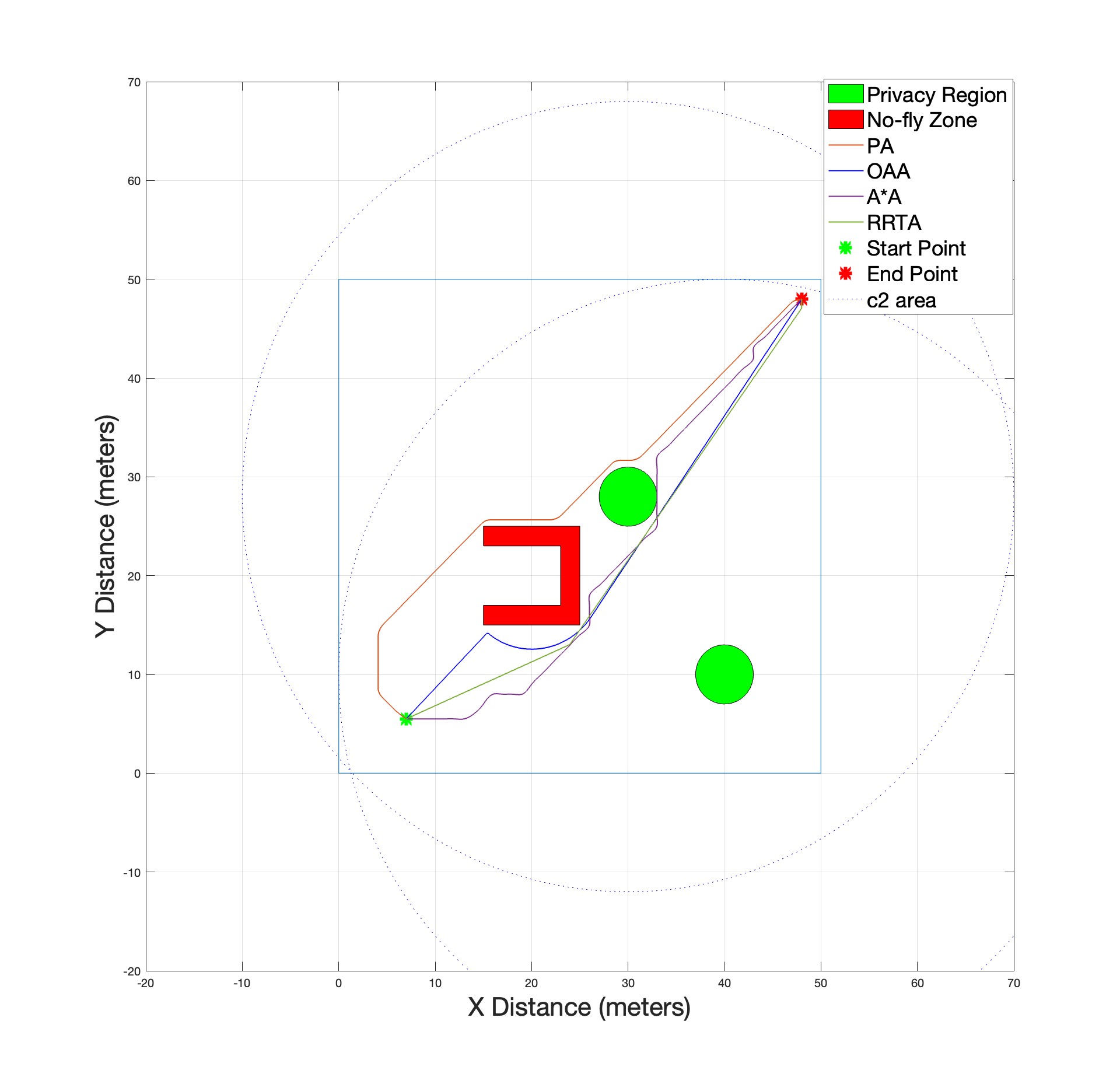}}
\caption{2D: UAV trajectories for PA, OAA, A*A and RRTA}
\label{f1}
\end{figure}

\begin{figure}[ht]
\centerline{\includegraphics[width=7 cm]{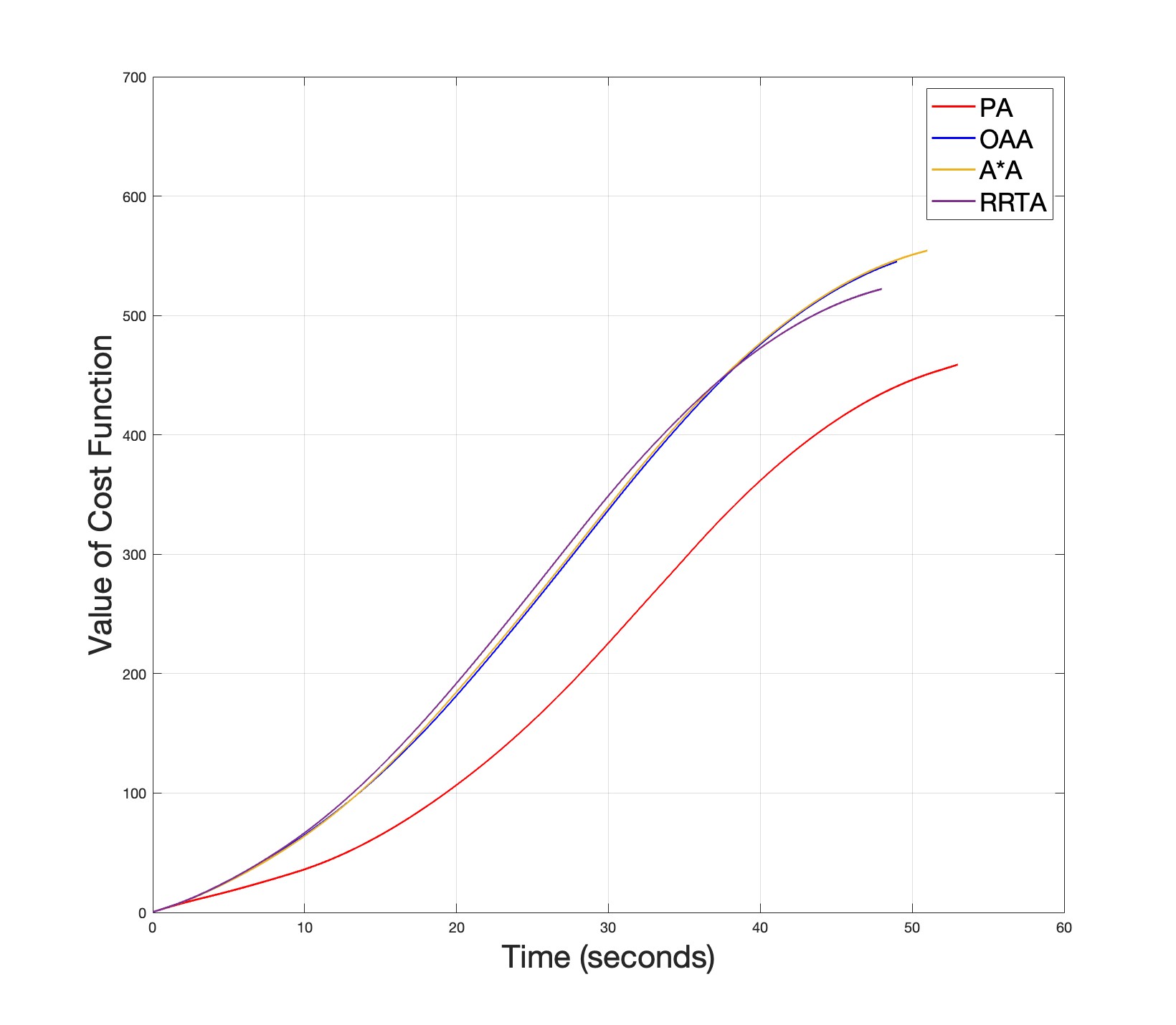}}
\caption{2D: Cost Function for PA, OAA, A*A and RRTA}
\label{f2}
\vspace{1 cm}
\centerline{\includegraphics[width=7 cm]{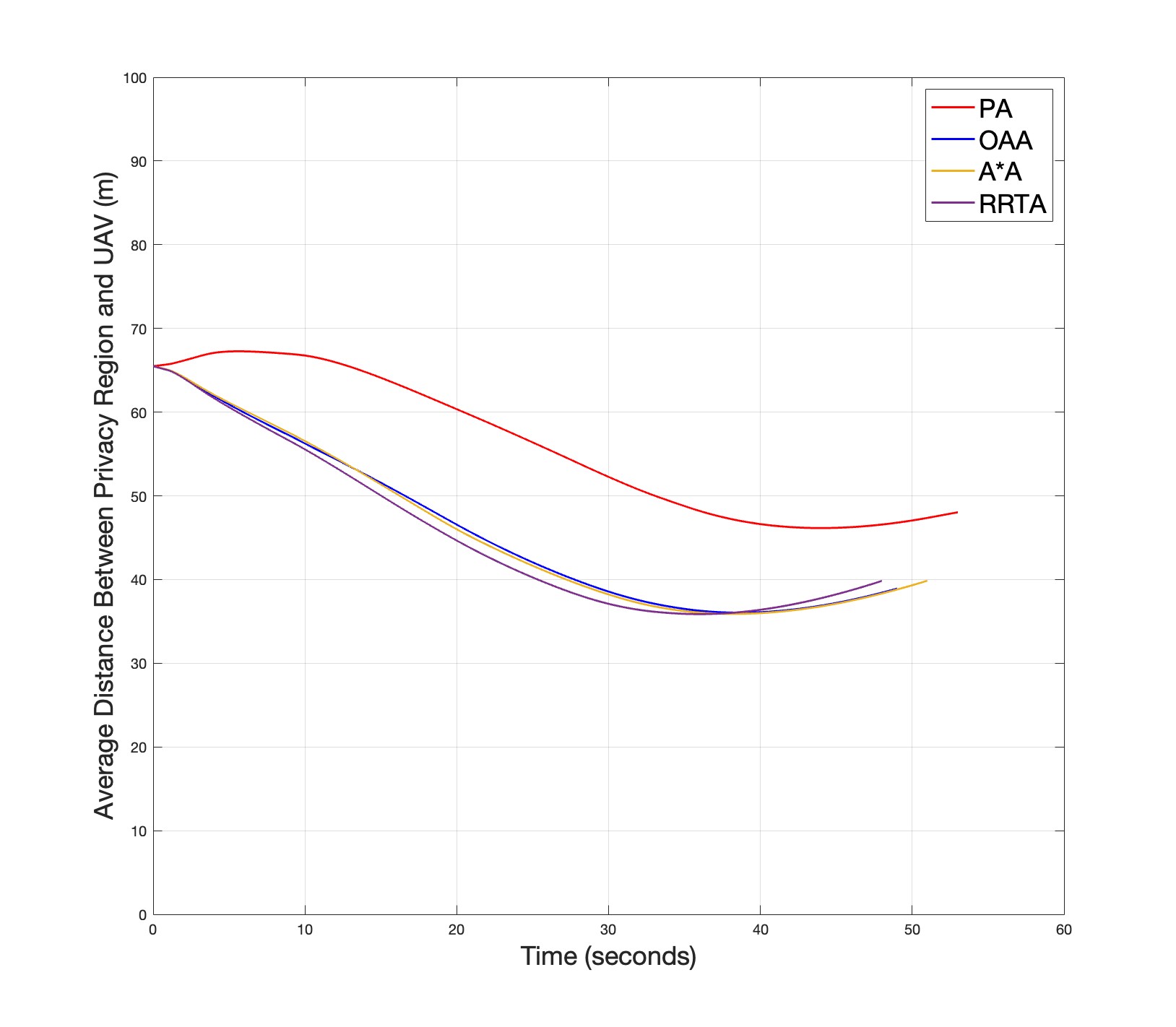}}
\caption{2D: Average minimum distance for PA, OAA, A*A and RRTA}
\label{f3}
\end{figure}

Figure \ref{f1}-\ref{f3} shows the 2D simulation of the proposed method. It can be seen that green circles represent privacy zones, and a non-convex no-fly zone has been added to assess the feasibility of the proposed methods within a complex environment. Furthermore, the blue dashed line shows the region of $c_2$. 

Three different methods are used to compare with the proposed methods. The high-privacy regions and no-fly zones are set as obstacles for the following methods. A sliding mode-based obstacle avoidance algorithm (OAA) \cite{savkin2013simple} is used as the first comparison method. The second method is the well-known A* algorithm (A*A) \cite{gunawan2019smoothed}. The third method is the Rapidly-exploring Random Tree algorithm (RRTA) \cite{sun2017two}. 

It can be seen in figure \ref{f1} that all four methods can avoid the high privacy regions and non-convex no-fly zones. The proposed privacy-aware method (PA) can sense the privacy intensity that takes a different route than the other three methods. Figure \ref{f2} presents the total privacy violation risk for all four methods. It clearly shows a lower cost for the PA method, which is about 16\% less than that of the other three methods. For figure \ref{f3}, the average distance for the PA method is further than the other three methods because the algorithm is trying to avoid the privacy regions. In conclusion, in a 2D environment, the PA method outperforms all other three methods in the total cost.

\subsubsection{Simulations in 3D Environment}

\begin{figure}[ht]
\centerline{\includegraphics[width=7 cm]{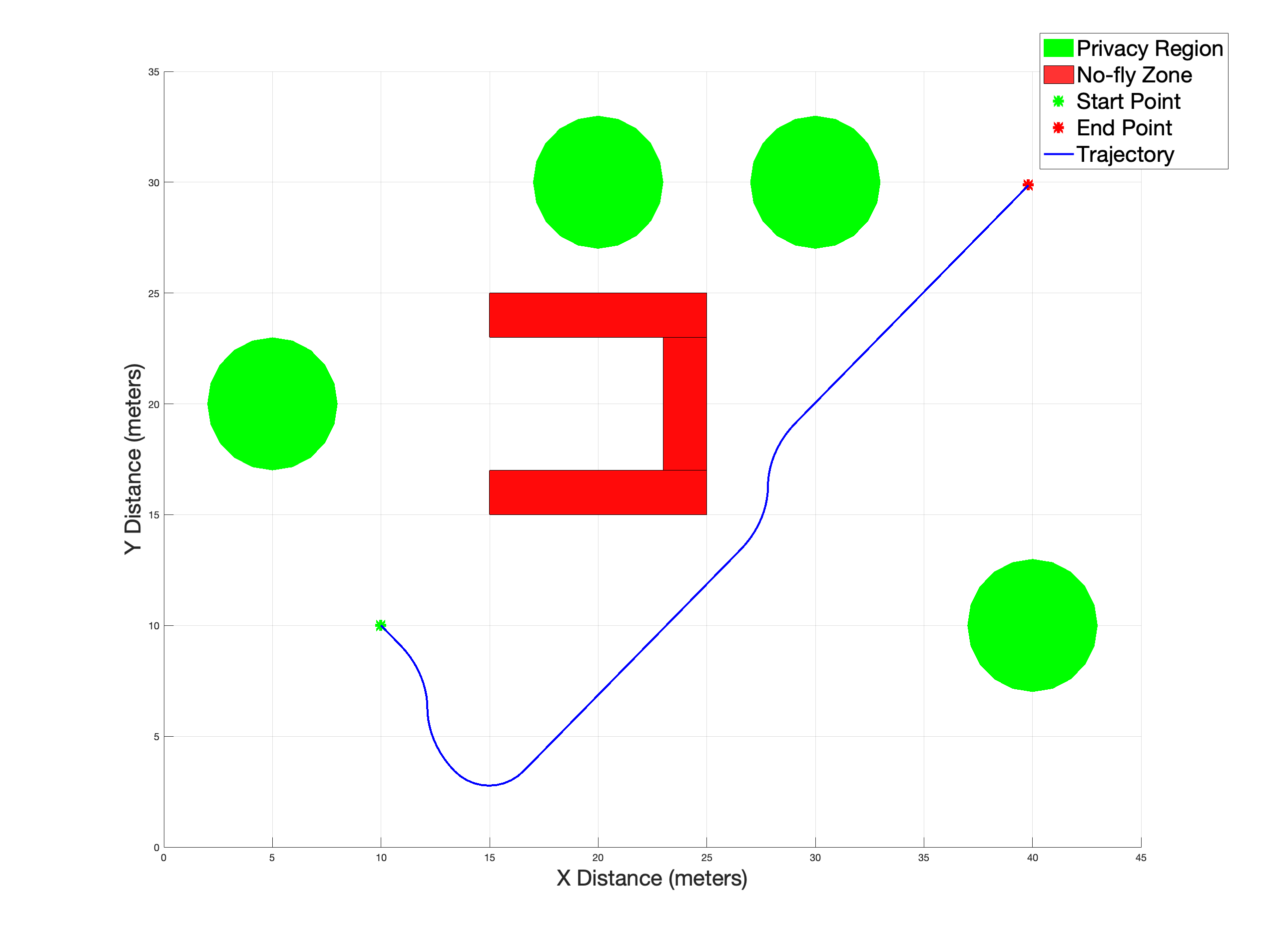}}
\caption{3D: Top View}
\label{f4}
\vspace{1 cm}
\centerline{\includegraphics[width=7 cm]{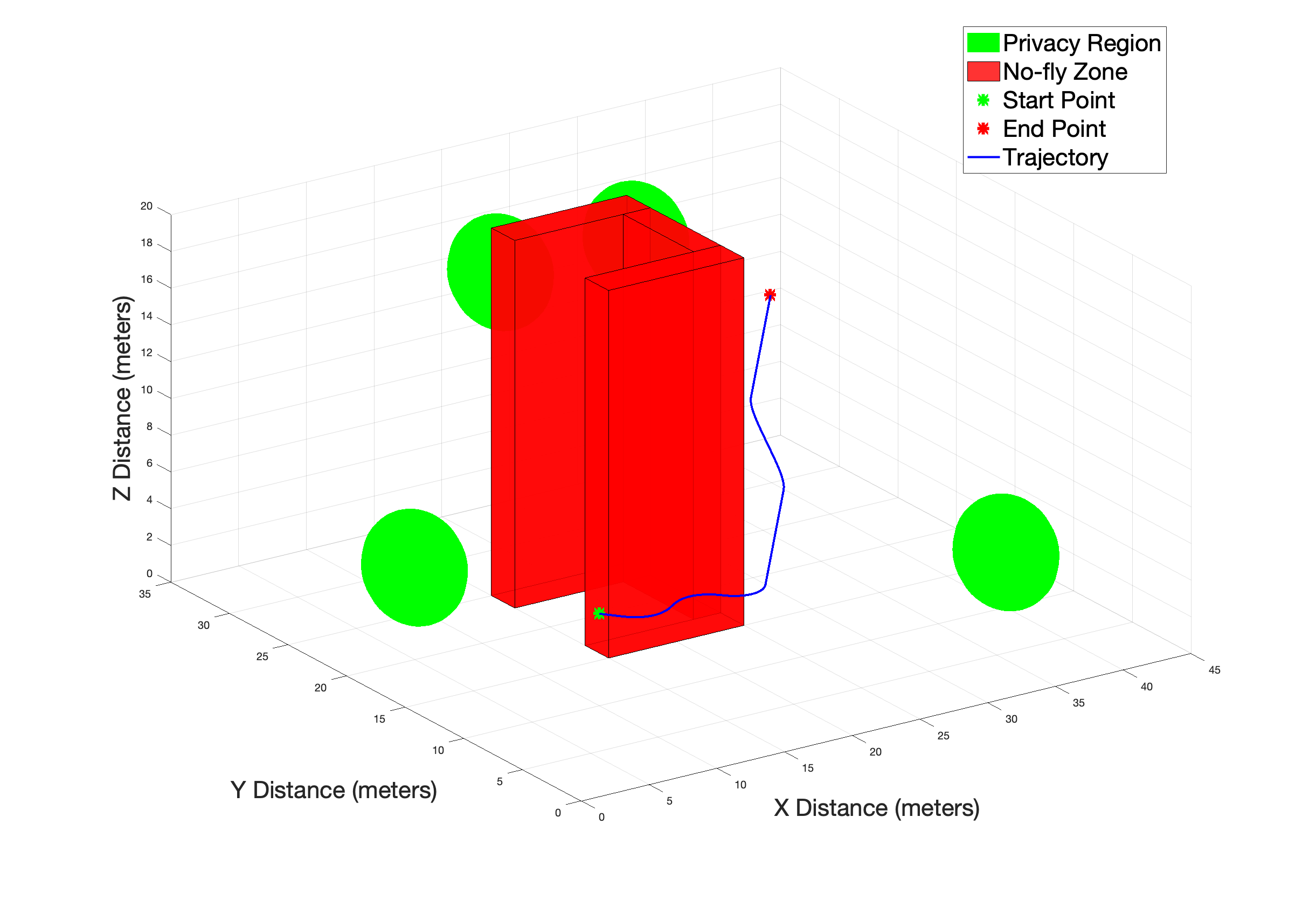}}
\caption{3D: Side View}
\label{f5}
\end{figure}

Figure \ref{f4} and \ref{f5} show the simulation in a 3D environment. Similar to the 2D environment, it contains a non-convex no-fly zone and multiple privacy regions. $z^{min}$ and $z^{max}$ are set at 3m and 20m respectively. It can be seen that the generated route is not the minimum-distance path, and the PA method can find the feasible path in a 3D environment while minimizing the total privacy cost.

\subsection{Conclusion}
\label{C}
In this section, a dynamic programming-based privacy-aware 3D navigation algorithm was developed to find the minimum cost path. A non-convex no-fly zone was added to address the feasibility of the proposed method. In addition, some comparisons were made to prove the usefulness of the algorithm that was developed. The optimization problem is convex that the global optimum can be achieved. The nature of dynamic programming is to break down the problem into some sub problems. The developed approach ensures that each sub problem is solved optimally, leading to an overall optimal solution for the entire problem. The computational cost is reasonable that can be used for real-time execution. However, it's more suitable for off-line planning because of the need for adequate map information. The privacy-aware algorithm can be viewed as an information-aware algorithm with broader applications. Inspired by this section, energy-aware navigation for a solar-powered UAV could be an extension of this application. Thus, the following sections will focus on the navigation of a solar-powered UAV.

\section[Energy-Aware 3D Navigation of a Solar-powered UAV in an Urban Environment]{Energy-Aware 3D Navigation of a Solar-powered UAV in an Urban Environment} 
\label{ch4} 

This section considers a scenario in which a solar-powered unmanned aerial vehicle (SUAV) tries to find an energy-efficient path in a complex 3D urban environment and return to the charging station as soon as possible while maintaining positive residual energy. In the presence of solar energy, the maximum capacity of the on-board battery is considered. Two path planning algorithms are developed to find the energy-efficient and time-efficient path for this situation. Computer simulations are conducted to prove the effectiveness of the proposed methods. Compared to some benchmark methods, the proposed methods show better results.

\subsection{Introduction}
With the rapid development of unmanned system technology, the applications of unmanned aerial vehicles (UAVs) have gradually entered the public's vision, which has the potential to revolutionize the labour landscape by replacing manual labours and enhance work efficiency and productivity. The applications of UAVs include mountain site surveying, agriculture inspection, and remote eavesdropping \cite{fan2020review, tsouros2019review, ye2018secure}. Additionally, UAVs can be flexible solutions for diverse needs in urban settings such as target surveillance, communication systems, parcel delivery and 3D mapping \cite{9580756, 10005002, 7317490, nex2014uav}. The navigation algorithm in such a complicated urban environment is more challenging. Thus, this section will focus on the applications of UAVs in urban environments. 

UAVs can be classified into three types commonly based on their rotor type: single-rotor, multi-rotor, and fixed-wing \cite{s21186223}. The main advantage of fixed-wing UAVs is that they carry higher onboard weight for more extended-distance flights, but they cannot hover over one place to conduct long-time surveillance. Single-rotor UAVs have the capacity of longer flight durations compared to multi-rotor UAVs of similar size equipped with comparable battery capacity but lack manoeuvrability due to the control design \cite{arjomandi2006classification}. Most applications in urban environments require high manoeuvrability, so multi-rotor UAVs are the best fit in this setting.
\begin{figure}[ht]
\centerline{\includegraphics[width=6 cm]{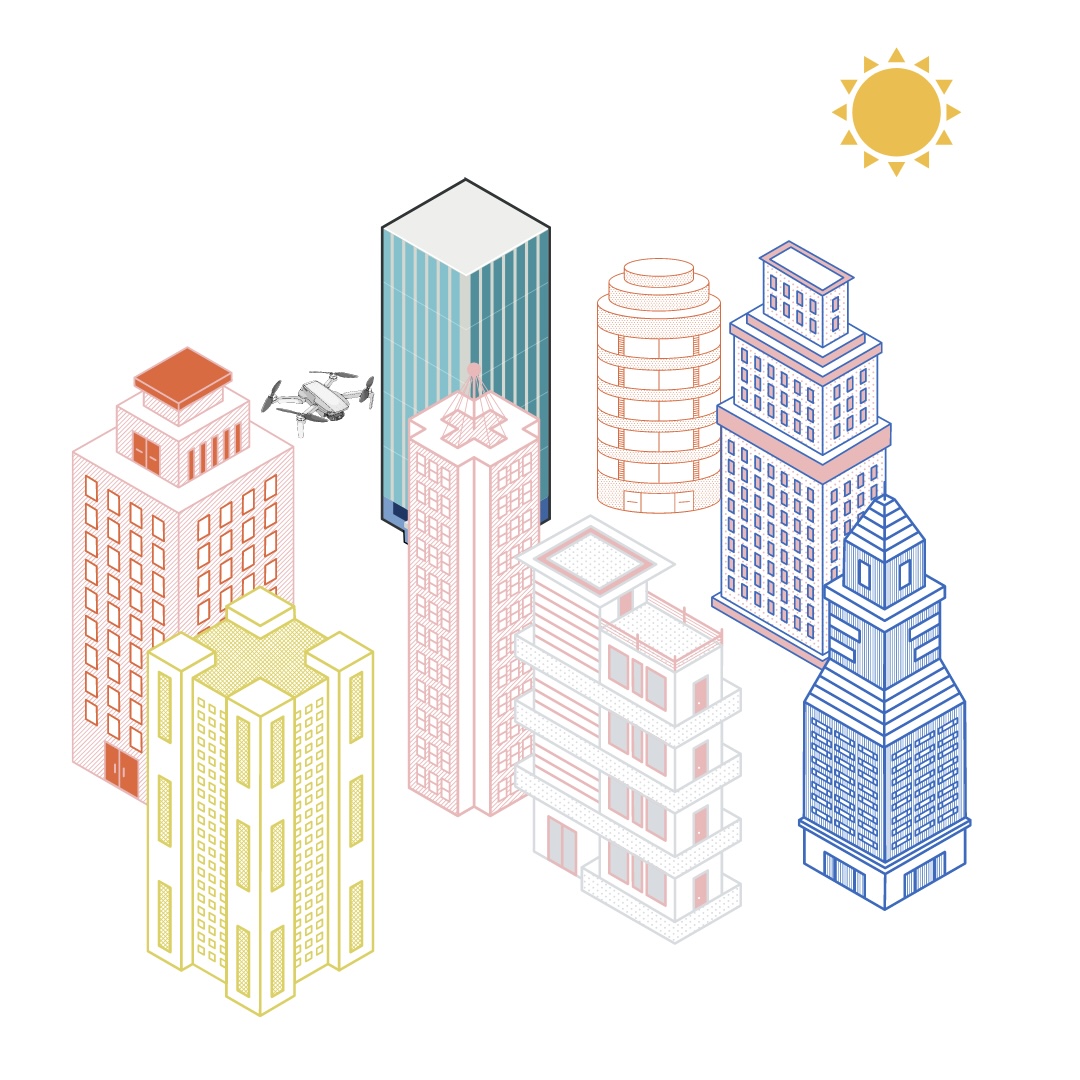}}
\caption{Illustration of Urban Environment}
\label{ch3f1}
\end{figure}
Most applications of UAVs require long flight distances and durations, and the standard methods to minimize the flight consumption per unit time are lowering the overall weight of UAVs or optimizing the deployment of the UAVs. However, these solutions have reached a bottleneck  \cite{morton2015solar}. Finding a straightforward and efficient way to extend the duration of drone flights is a trending topic. 

Renewable energy has been developed rapidly to reduce the use of fossil fuels, which is beneficial to the environmental reservation. Solar energy is one of the earliest forms of energy source and is widely used till now, and solar-powered products have become increasingly common in our lives \cite{hayat2019solar}. For instance, it is commonly used in solar-powered electric cars, solar-powered charging stations and solar-powered UAVs (SUAVs). Since multi-rotor UAVs have a strict weight limitation, it is not feasible to fit a large-sized onboard battery. Especially in a complicated urban environment, running out of battery could lead to severe consequences for pedestrians and society. The advancement by using solar energy on SUAVs allows extending the flight time to over 24 hours \cite{7353711}. Therefore, powering UAVs with solar energy is a promising solution to extend their flight time without sacrificing manoeuvrability. 

Some related works have been published focusing on the navigation of solar-powered UAVs. Authors \cite{Solar1} consider a navigation problem for a group of SUAVs to track the moving target in an urban environment. An improved rapidly exploring random tree (RRT) algorithm is applied to find the energy-efficient path. However, the environment can be considered as 2.5D because the flying altitude of the UAV is set to be fixed. Furthermore, the RRT-based algorithm finds multiple paths and chooses the one with the minimum cost. The energy cost is not part of the path planning algorithm. Additionally, authors \cite{wu2018path} comprehensively evaluate the navigation problem in urban environments for SUAVs, but the flying altitude is still at a fixed level instead of various ranges. Further research on the pure 3D environment is needed in those scenarios. In addition, SUAVs also have some valuable applications in the wildlife environment. Inspecting mountain sites \cite{huang2021path} is an appropriate example, as solar power can provide a more extended endurance capability for the UAV. Authors use an RRT-based path planning algorithm to find a feasible path first, then an optimized path if there is excess residual energy at the end of the flight. This technique can address practical concerns, such as the duration required to finish the inspection and the amount of initial energy the onboard battery uses. Lee and Yu \cite{7859311} use gravitational potential energy to optimize cost during flight for SUAVs. Utilizing gravitational potential energy during the night can decrease reliance on batteries, which in turn can reduce the aircraft's weight and increase its flight endurance.

This section considers a path planning problem: an SUAV flies in a complex urban environment with constraints. Several tall constructions will block the movement of the SUAV, which means the constructions will be treated as obstacles in the pre-defined map. In addition, the constructions taller than the SUAV will create shadow areas, preventing it from absorbing solar energy from the sun. These two factors make the problem difficult to solve. The first objective of this section is to navigate an SUAV from the start node to the end node along an energy-efficient path with minimum energy consumption. The second objective is to find a time-efficient path considering the constraint of on-board battery capacity. Moreover, the algorithms are proposed for a pure 3D environment instead of a fixed flight height. Compared with the RRT-based algorithm developed by previous scholars \cite{Solar1}, the energy-efficient algorithm considers the solar energy model, providing a lower energy cost path.

The remainder of the section is organized in the following manner. Section \ref{ch3model} introduces the SUAV system model, energy models and the problem statement. Section \ref{ch3law} explains the 3D path planning method in the presence of solar energy and battery capacity. The following section \ref{ch3simulations} presents the 3D computer simulation in an urban environment, showing the improvement of the proposed methods. The last section \ref{ch3conclusion} concludes the overall section.

\subsection{System Model and Problem Statement}
\label{ch3model}
\subsubsection{UAV System Model}
Assuming an SUAV is moving in the 3D free space, the position of the SUAV is defined as $p(t)=(x(t), y(t), z(t))$ at each time step $t$. The dynamic system of the SUAV can be modelled using the same methodology as traditional UAVs, given that the reliance on solar energy is the sole distinguishing feature between UAVs and SUAVs. Therefore, it can be described as: 
\begin{eqnarray}
\label{ch3UAV}
	\begin{cases}
	\dot{x}(t) = v(t)cos(\theta(t))\\
	\dot{y}(t) = v(t)sin(\theta(t))\\
	\dot{z}(t) = u(t)\\
	\dot{\theta}(t) = \omega(t),
	\end{cases}
\end{eqnarray}
where $\theta$ is the horizontal heading angle with respect to the x-axis, and $v(t)$ denotes the flying speed of the SUAV. It is bounded by an interval $0 < v(t) \leq v^{max}$ where $v(t)\in {\bf R}$ and the upper limit $v^{max}$ is given as a constant positive number in this section. This mathematical model is widely used to describe the motion of UAVs, missiles and autonomous underwater vehicles (AUVs) \cite{shao2021new, manchester2006circular, savkin2022optimal, wu2019coordinated, luo2021distributed}. SUAV is set to fly within a range of altitude rather than at a fixed altitude; the vertical coordinate $z(t)$ thus lies within a reasonable range of the minimum height $z^{min}$ to the maximum height $z^{max}$. Moreover, SUAVs are considered non-holonomic vehicles in this section in that they are subject to motion constraints due to restricted manoeuvrability in certain directions, the angular velocity $\omega(t)$ of SUAV is limited to be less than $\omega^{max}$. Therefore, the following requirements are stated,
\begin{eqnarray}
\label{ch3Z}
	\begin{cases}
	z^{min}\leq z(t)\leq z^{max}\\
	-\omega^{max}\leq \omega(t)\leq \omega^{max}.
	\end{cases}
\end{eqnarray}

\subsubsection{Shadow Area}
The shadow area needs to be defined properly in the 3D environment. For SUAVs, one concern is capturing direct sunlight as a power source. However, depending on the height, location, and orientation of tall buildings, shading from high-rise buildings can obstruct a certain amount of direct sunlight reaching the solar panel, decreasing the overall power output. Consequently, it is important to accurately define the shadow area of tall buildings in this section. Let $V(p(t))$ be an indicator to determine whether the SUAV is under the sunlight at time $t$. If $V(p(t)) = 1$, the solar panels on the SUAV are shaded, which means the SUAV enters the shadow area. If $V(p(t)) = 0$, SUAV is exposed to the sunshine and can absorb solar energy from the sun. Thus, 
\begin{align*}
	\begin{cases}
	V(p(t)) = 1, \text{shadow}\\
	V(p(t)) = 0, \text{not shadow}.
	\end{cases}
\end{align*}

To identify the shadow area, let the position of the sun be $p(s) = (x(s), y(s), z(s))$ and take the straight line segment between $p(s)$ and the current position of SUAV $p(t)$ to check whether there is an intersection between the constructions or not, which can be represented as follow \cite{Solar1}:
\begin{align*}
	\begin{cases}
	x = x_s + \alpha_x\gamma\\
	y = y_s + \alpha_y\gamma\\
	z = z_s + \alpha_z\gamma\\
	min\{ x_p,x_s\} \leq x_s+\alpha_x\gamma\leq max\{x_p,x_s\}.
	\end{cases}
\end{align*}
Since the distance between the sun and the ground is considerably far, $z(s)$ is set to a relatively large value. $x(s)$, $y(s)$ and $z(s)$ vary with time to simulate the movement of the sun. Additionally, $p(t)=(x_p,y_p,z_p)$, $(\alpha_x,\alpha_y,\alpha_z)=\frac{\vec {p(s)p(t)}}{\|\vec {p(s)p(t)}\|}$, and $\gamma$ is a scalar variable. 
\begin{figure}[ht]
\centerline{\includegraphics[width=7 cm]{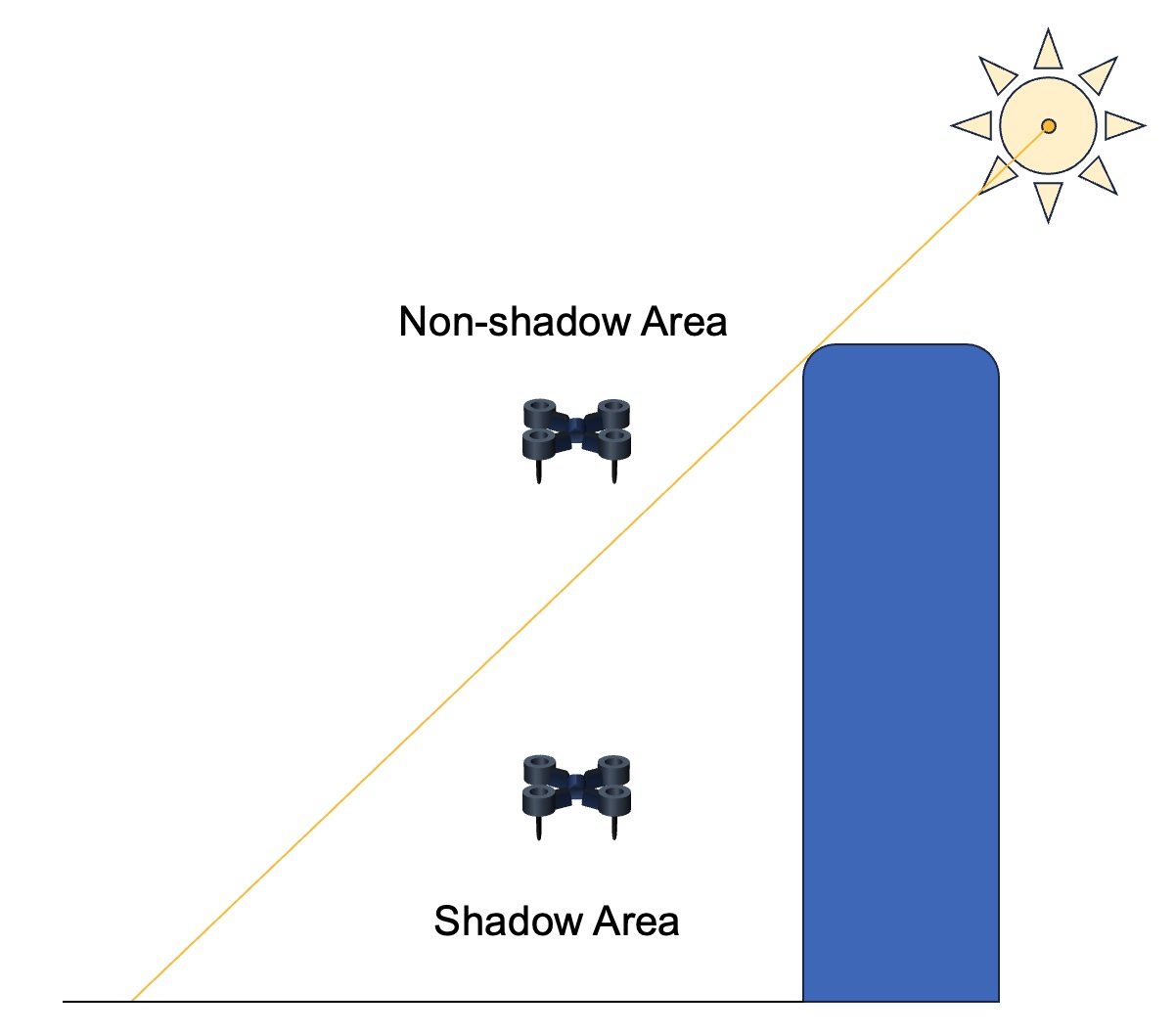}}
\caption{Illustration of Shadow Area}
\label{ch3f2}
\end{figure}

Fig.\ref{ch3f2} provides a visual representation of a typical situation where the height of a tall building causes a shadow area. This figure indicates clearly that it is easy to determine whether an SUAV is exposed to the sun by identifying whether there is an intersection between an SUAV and the sun. If the intersection exists, the SUAV is in the shadow area and $V(p(t)) = 1$. 

\subsubsection{Energy Consumption Model}
SUAVs need to consume energy while flying; various factors will influence the overall energy needed in a given amount of time, including weight, fly height, fly speed, etc. Authors \cite{7101619} conduct comprehensive research on the energy consumption during flight for a quad-rotor UAV. When a constant acceleration rate is given, the energy consumed to increase the speed from $v_{1}$ to $v_{2}$ can be represented by
\begin{align*}
	E_{a}=\int_{t_{1}: v=v_{1}}^{t_{2}: v=v_{2}} P_{a}(t)dt.
\end{align*}
When the speed $v$ is given as a constant, the energy consumed in a flight cover distance $d$ can be computed as
\begin{eqnarray}
\label{ch3UAV Cost2}
	E_{v}=\int_{0}^{d / v} P(v) d t=P(v) \frac{d}{v}.
\end{eqnarray}
When an SUAV is trying to climb up or descend for a height of $\Delta h$, the energy consumption can be computed as
\begin{eqnarray}
\label{ch3UAV Cost3}
	E_{up}=\int_{h_{1} / \hat{v}_{up}}^{h_{2} / \hat{v}_{up}} P_{up} dt=P_{up} \frac{\Delta h}{v_{up}},
\end{eqnarray}
\begin{eqnarray}
\label{ch3UAV Cost4}
	E_{down}=\int_{h_{2} / \hat{v}_{down}}^{h_{1} / \hat{v}_{down}} P_{down} dt=P_{down} \frac{\Delta h}{v_{down}}.
\end{eqnarray}
For the research problem being investigated, SUAV will be flying in a 3D environment. Therefore, the consumption power $P_{\text{out}}$ and energy cost $E_{\text{out}}$ depend on the movement of the SUAV.

\begin{figure}[ht]
\centerline{\includegraphics[width=7 cm]{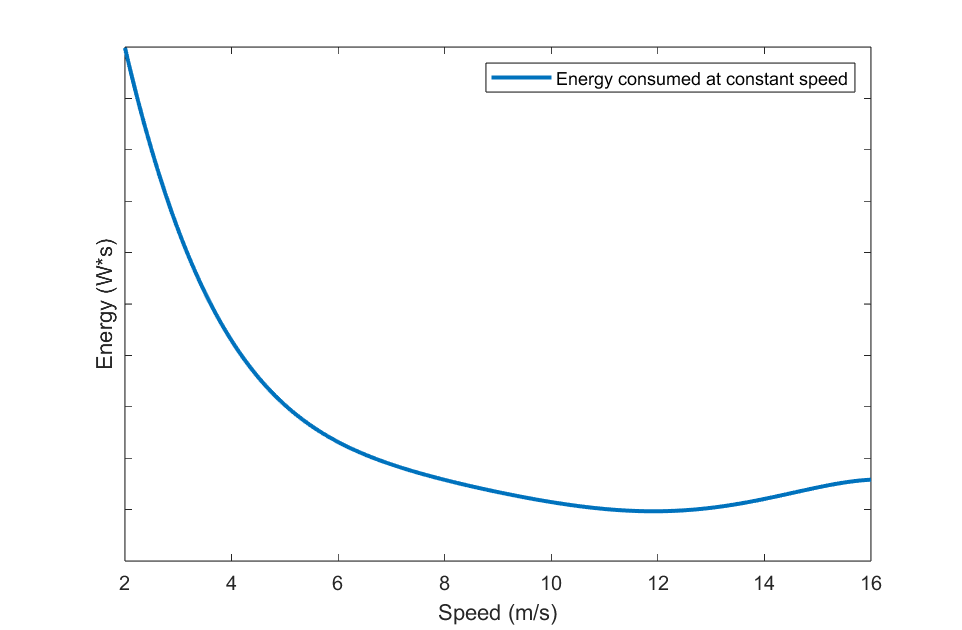}}
\caption{Energy Consumption at Constant Speed $v$ for a Distance $d$}
\label{ch3f3}
\end{figure}

Fig.\ref{ch3f3} illustrates the relationship between SUAV speed $v$ and the energy consumed during a flight of distance $d$ \cite{7101619}. It can be seen that SUAV consumes the most energy when flying at a low speed, and it becomes more energy-efficient when speeding up until it reaches a speed around $12 \mathrm{~m} / \mathrm{s}$. Thus, this section assumes that the SUAV is flying at a constant speed $v \simeq 12 \mathrm{~m} / \mathrm{s}$ to get close to optimal cost.

\subsubsection{Energy Harvesting Model}
A clear sky model is used for the investigated problem. Therefore, the assumption is made that the effect of wind and clouds is ignored. Only constructions in the urban environment can result in shadow areas for the SUAV. In addition, solar cells are equipped on the SUAV to collect solar power from direct sunlight. They are placed on top of the SUAV to gain maximum radiation from the sun. The incidence angle decides how much energy the SUAV can get from the sun. Then the relationship can be defined by \cite{klesh2009solar}:
\begin{align*}
	\cos (\theta)=\cos (\phi) \sin (e)-\cos (e) \sin (a-\psi) \sin (\phi),
\end{align*}
where $a$ and $e$ stand for azimuth and elevation angles, respectively. $\phi$ is the bank angle and $\psi$ is the heading angle. $\theta$ is the incidence angle of the sun rays upon the solar cells. Moreover, the solar energy power gained from the solar cell can also be defined by
\begin{eqnarray}
\label{ch3Solar Cost2}
	P_{\text {solar }}(\theta)=\eta P_{\mathrm{sd}} S \cos (\theta) \quad \text { if } \cos (\theta) \geq 0,
\end{eqnarray}
where $S$ represents the total area of the solar penal on SUAV and $\eta$ stands for the solar cell efficiency. $P_{\mathrm{sd}}$ is a constant value representing the solar spectral density.
It is straightforward to calculate the total energy gained within the time $\left[t_{o}, t_{n}\right]$ as
\begin{eqnarray}
\label{ch3Solar Cost3}
	E_{\text {gain }}=\int_{t_{o}}^{t_{n}} P_{\text {solar }}(\theta) \mathrm{d} t.
\end{eqnarray}
Therefore, $P_{\text {solar }}$ and $P_{\text {out }}$ are two determinants of the cost at any given time.

\subsubsection{Problem Statement}
In this section, an SUAV is flying in a bounded 3D environment, and ${\cal A}\subset {\bf R}^3$ is introduced to be a given bounded area in 3D that SUAV can fly freely without collision within this area. The constructions can be treated as a no-fly zone and excluded from $\cal A$ in the offline map. Therefore, no obstacle avoidance algorithm is needed in this research problem. In addition, the coordinates of the SUAV $p(t)=(x(t),y(t),z(t))\in {\cal A}~\forall t$.

In the presence of a shadow area, the solar power gained from the sunlight depends on the value of $V(p(t))$,
\begin{align*}
	\begin{cases}
	P_{\text {solar}}(\theta)=\eta P_{\mathrm{sd}} S \cos (\theta) \quad \text { if } V(p(t)) = 1\\
	P_{\text {solar}}(\theta) = 0 ~~~~~~~~~~~~~~~~~~ \text{if } V(p(t)) = 0.
	\end{cases}
\end{align*}
In this manner, SUAV cannot harvest solar energy from the sun if it is in the shadow area. Conversely, SUAV can effectively capture solar energy at a power output of $P_{\text {solar}}$.

From the previous section, energy models are derived for consuming and harvesting the energy. Equation (\ref{ch3UAV Cost2}) will be used when SUAV is flying at a steady altitude level, equation (\ref{ch3UAV Cost3}) and (\ref{ch3UAV Cost4}) will be used in climbing up and descending situations respectively. In the meantime, the solar panel equipped on SUAV can gain energy from the sunlight. The capacity of battery can be defined by $E_{\text {Batt}}$,
\begin{align*}
	E_{\text {total}} = E_{\text {Batt}} +  E_{\text {gain}} - E_{\text {out}},
\end{align*}

For the purpose of safety, SUAV cannot drain the onboard battery to fully empty. A threshold value $E_{\text{min}}$ is introduced to be set as a hard constraint for this problem. 
\begin{eqnarray}
\label{ch3Total Cost3}
	E_{\text {total }} \geq E_{\text{min}}.
\end{eqnarray}

In this section, two different scenarios are being investigated:

{\bf Scenario 1:} In an urban environment, an SUAV is trying to find the best route from $p(0)$ to $p(n)$ while maximizing the residual battery energy, given that other constraints are satisfied.

This is a typical situation in which an SUAV wants to find an energy-efficient route. The objective is to maximize the $E_{\text {total }}$ while satisfying (\ref{ch3UAV}), (\ref{ch3Z}), and (\ref{ch3Total Cost3}).

{\bf Scenario 2:} In an urban environment, an SUAV is trying to find the time-efficient route from $p(0)$ to $p(n)$ while ensuring the residual energy stored in the battery cannot be lower than the threshold value $E_{\text{min}}$, given that other constraints are satisfied.

This is the situation where the SUAV runs out of battery and aims to return to the charging station as soon as possible. the objective is to minimize $t$ while satisfying equations (\ref{ch3UAV}), (\ref{ch3Z}), and (\ref{ch3Total Cost3}).

\subsection{Navigation Law}
\label{ch3law}

A modified search-based algorithm \cite{Hart1968} is used in this section to find the energy-efficient path. This technique balances heuristic information and actual path cost data to guide the search process. It begins with a priority queue containing the initial node, indicated by a cost function $f(n)$. This function is defined as the sum of the heuristic estimate $h(n)$ and the actual cost from the starting node to the current node, $g(n) = E_{\text{gain}} - E_{\text{out}}$. While searching the minimum-cost nodes, the algorithm assigns priority to those with the lowest $f(n)$ value and updates $g(n)$ and $f(n)$ for all adjacent nodes simultaneously. In the meantime, equation (\ref{ch3Total Cost3}) holds at all times. The algorithm iteratively expands nodes until it reaches the target node or determines that a feasible solution is impossible. In this way, it guarantees the discovery of an energy-efficient and feasible solution, provided that the heuristic function is admissible and consistent. The proposed algorithm is illustrated in Algorithm \ref{ch3algo1}. More specificity, $FindMinimumF$ is used to find the node that has a minimum $f(n)$ value within the $\text{SET}_{open}$, and this node $N_{\text{current}}$ is the node to be evaluated. To check the battery level at all times, the $CheckBattery$ function is used to finish the task. In addition, due to the limitation of onboard batteries, the energy stored cannot exceed the maximum capacity.
\begin{algorithm}
\caption{Finding the energy-efficient path}
\label{ch3algo1}
\begin{algorithmic}[1]
\State initialize $SET_{open}$ and $SET_{closed}$
\While{$SET_{open}$ is not empty}
    \State $N_{current} = FindMinimumF(SET_{open})$
    \If{$N_{current}$ reach the target}
        \State Return
    \EndIf
    \State Remove $N_{\text{current}}$ from the $SET_{open}$, and add it to $SET_{closed}$
    \For{all the neighbour of $N_{\text{current}}$}
        \If{neighbour is in $SET_{closed}$}
        	\State Continue
        \Else
            \If{$CheckBattery = \text{False}$ }
            		\State Continue
            \EndIf
            \State $g(n) = g(c) + E_{\text {gain }} - E_{\text {out }}$
            \If{$\nu(i) \notin SET_{open}$}
            	\State add neighbour to $SET_{open}$
            \ElsIf{$g(n) \geq g(c)$}
            	\State Continue
            \EndIf
            \State EnergyLevel = EnergyLevel - $E_{\text {out }} + E_{\text {gain }}$
            \If{EnergyLevel $>$ $E_{Batt}$}
            	\State EnergyLevel = $E_{Batt}$
            \EndIf
            \State Update $g(n), h(n), f(n)$
            \State Update parent node
        \EndIf
    \EndFor
\EndWhile
\end{algorithmic}
\end{algorithm}

Algorithm \ref{ch3algo2} is developed based on the energy-efficient algorithm. A time-efficient path can be obtained by changing the objective cost function. The new objective cost function is $g(n) = g(c) + Time(current, neighbour)$. The $Time$ function calculates the time taken from the current node to the neighbour nodes, providing the objective cost function for this situation. The residual energy of the battery needs to be checked to ensure it will not exceed the battery capacity. It also cannot drop below the threshold value $E_{min}$. 

\begin{algorithm}
\caption{Finding the time-efficient path}
\label{ch3algo2}
\begin{algorithmic}[1]
\State initialize $SET_{open}$ and $SET_{closed}$
\While{$SET_{open}$ is not empty}
    \State $N_{current} = FindMinimumF(SET_{open})$
    \If{$N_{current}$ reach the target}
        \State Return
    \EndIf
    \State Remove $N_{current}$ from the $SET_{open}$, and add it to $SET_{closed}$
    \For{all the neighbour of $N_{current}$}
        \If{neighbour is in $SET_{closed}$}
        	\State Continue
        \Else
            \If{$CheckBattery = False$}
            	\State Continue
            \EndIf
            \State $g(n) = g(c) + Time(current,neighbour)$
            \If{$\nu(i) \notin SET_{open}$}
            	\State add neighbour to $SET_{open}$
            \ElsIf{$g(n) \geq g(c)$}
            	\State Continue
            \EndIf
            \State EnergyLevel = EnergyLevel - $E_{\text {out }} + E_{\text {gain }}$
            \If{EnergyLevel $>$ $E_{Batt}$}
            	\State EnergyLevel = $E_{Batt}$
            \EndIf
            \State Update $g(n), h(n), f(n)$
            \State Update parent node
        \EndIf
    \EndFor
\EndWhile
\end{algorithmic}
\end{algorithm}

\subsection{Computer Simulations}
\label{ch3simulations}
This section analyses the performance of proposed path planning algorithms. MATLAB is used as a software tool to simulate the urban environment and generate the trajectories to verify the effectiveness of the proposed methods. A smoothed A-star method \cite{8938467} is used to compare as the benchmark method. The essential parameters of the simulation are listed in the table \ref{ch3table:1}.

\begin{figure}[ht]
\centerline{\includegraphics[width=6 cm]{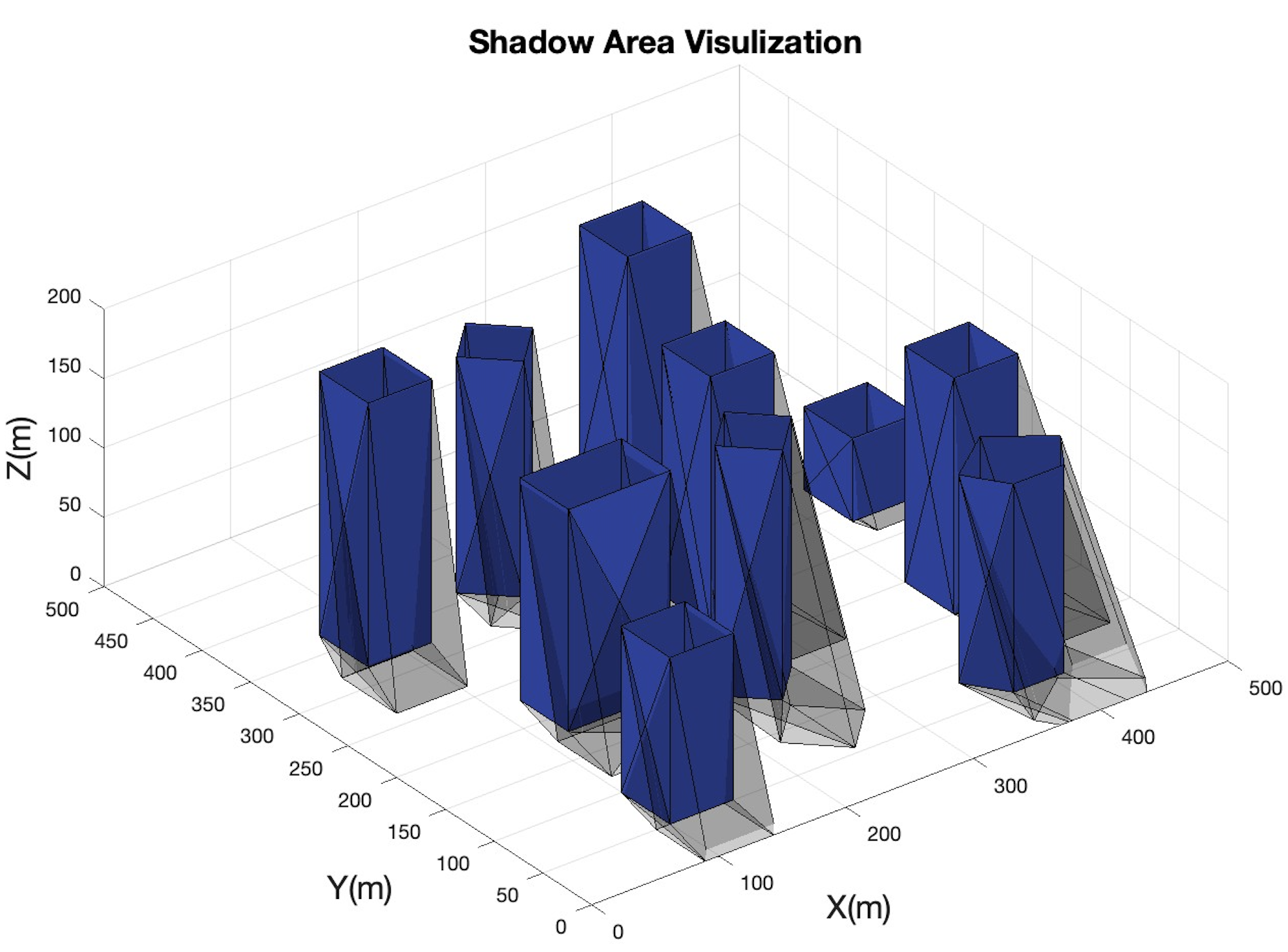}}
\caption{Shadow Area Simulation in Section \ref{ch4}}
\label{ch3f4}
\end{figure}

\begin{figure}[ht]
\centerline{\includegraphics[width=7 cm]{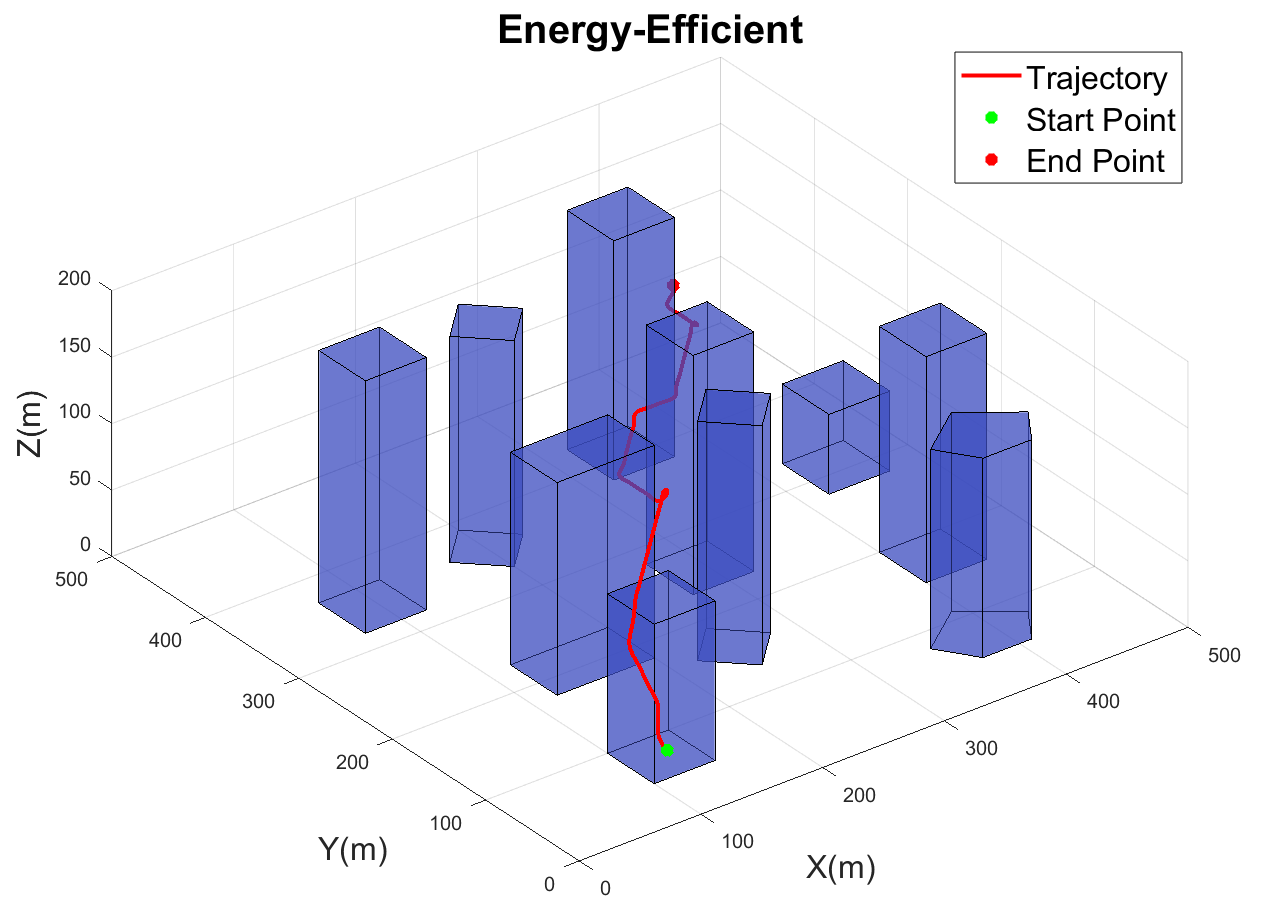}}
\caption{Energy-Efficient Algorithm Result in Section \ref{ch4}}
\label{ch3f5}
\end{figure}

\begin{figure}[ht]
\centerline{\includegraphics[width=7 cm]{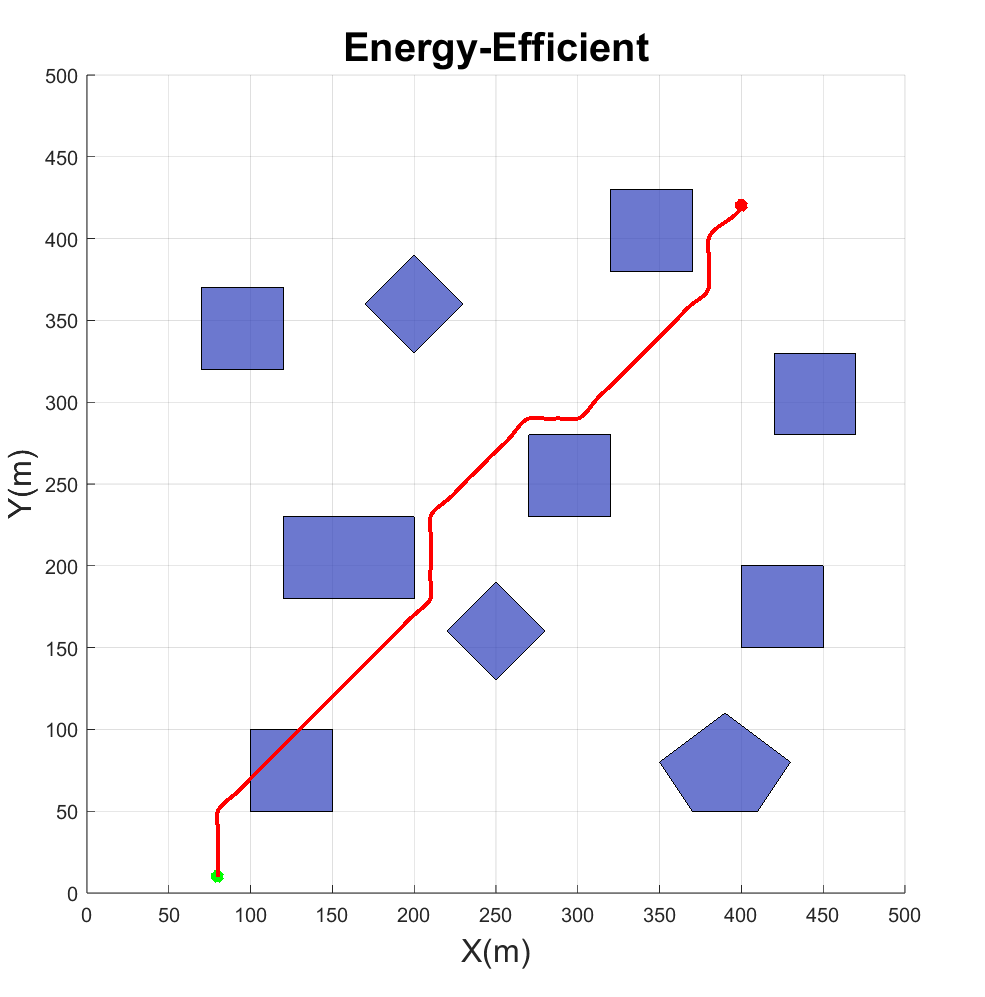}}
\caption{Energy-Efficient Algorithm Top View in Section \ref{ch4}}
\label{ch3f6}
\end{figure}

\begin{table}[h!]
\centering
\begin{tabular}{|c|c|c|}
\hline
\textbf{Parameter} & \textbf{Value}\\
\hline
Flying speed, $v$ & 12 $m/s$ \\
Minimum height, $z^{min}$ & 40 $m$ \\
Maximum height, $z^{max}$ & 200 $m$ \\
Power at constant speed, $P(v)$ & 30 $W$ \\
Power when climbing up, $P_{up}$ & 34 $W$ \\
Power when descending down, $P_{down}$ & 26 $W$ \\
Solar spectral density, $P_{\mathrm{sd}}$ & 380 $W/m^2$ \\
Solar cell efficiency, $\eta$ & 20$\%$ \\
Solar panel area, $S$ & 0.3 $m^2$ \\
Battery Capacity, $E_{Batt}$ & 670 $J$ \\
\hline
\end{tabular}
\caption{Parameters Used in the Simulation of Section 4}
\label{ch3table:1}
\end{table}
The urban environment is modelled by using different shapes of prisms. Other height parameters are assigned to illustrate the complex urban constructions better. Figure \ref{ch3f4} shows the shadow area when the sun is in front of the environment with the sun position of $p(s) = (250,800,1800)$. The sun's height is set at a relatively high value to reasonably illustrate the situation. Even though 1800 $m$ is not as high as the exact height of the sun, considering that the sunlight is illuminated in parallel, the actual angle of incidence of the sun is equivalent to that at 1800 $m$. Furthermore, because of the incident angle of the sun, higher constructions have larger shadow areas that block the SUAV from getting more sunlight. Fig. \ref{ch3f5} shows the trajectory using the energy-efficient method in the urban environment. It can be seen that the SUAV is trying to avoid going into the shadow area that leads to a lift-up at the beginning of the flight. Please note that even though some of the path seems to be in the shadow area in Fig. \ref{ch3f6}, it is not actually in the shadow prism in the 3D environment. 

\begin{figure}[ht]
\centerline{\includegraphics[width=7 cm]{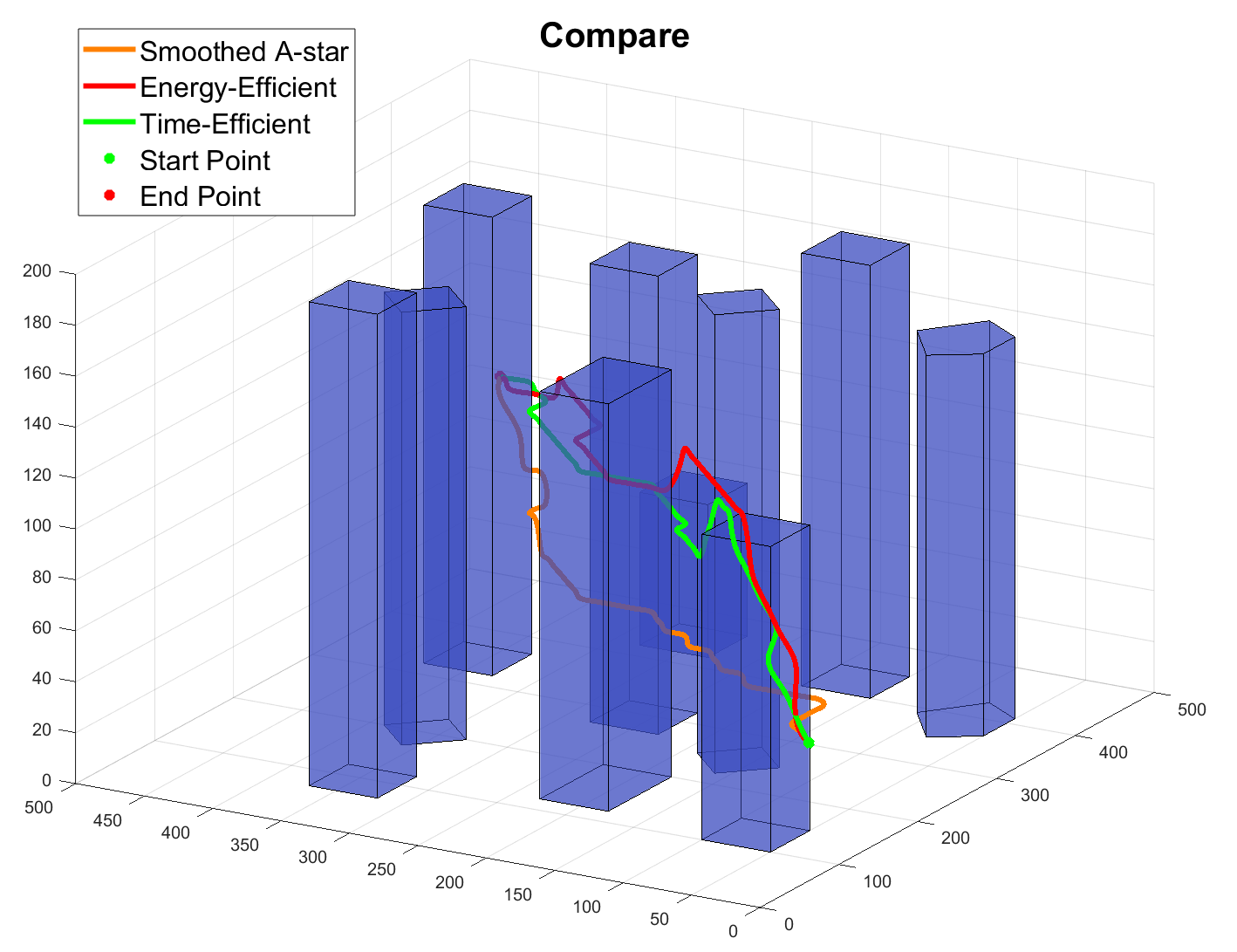}}
\caption{Comparison Between Three Methods in Section\ref{ch4}}
\label{ch3f7}
\end{figure}

\begin{figure}[ht]
\centerline{\includegraphics[width=7 cm]{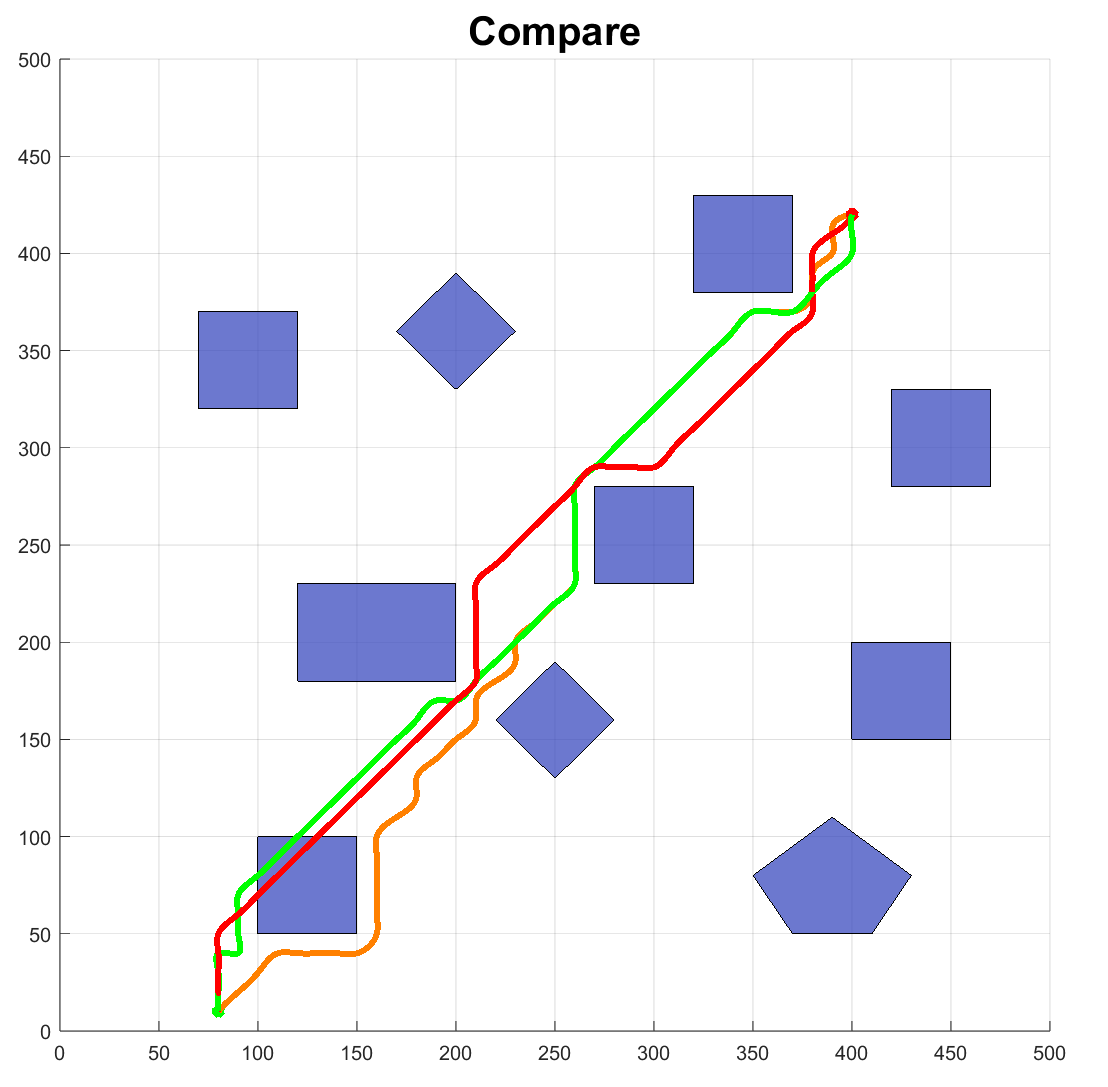}}
\caption{Comparison Between Three Methods Top View in Section \ref{ch4}}
\label{ch3f8}
\end{figure}

\begin{figure}[ht]
\centerline{\includegraphics[width=9 cm]{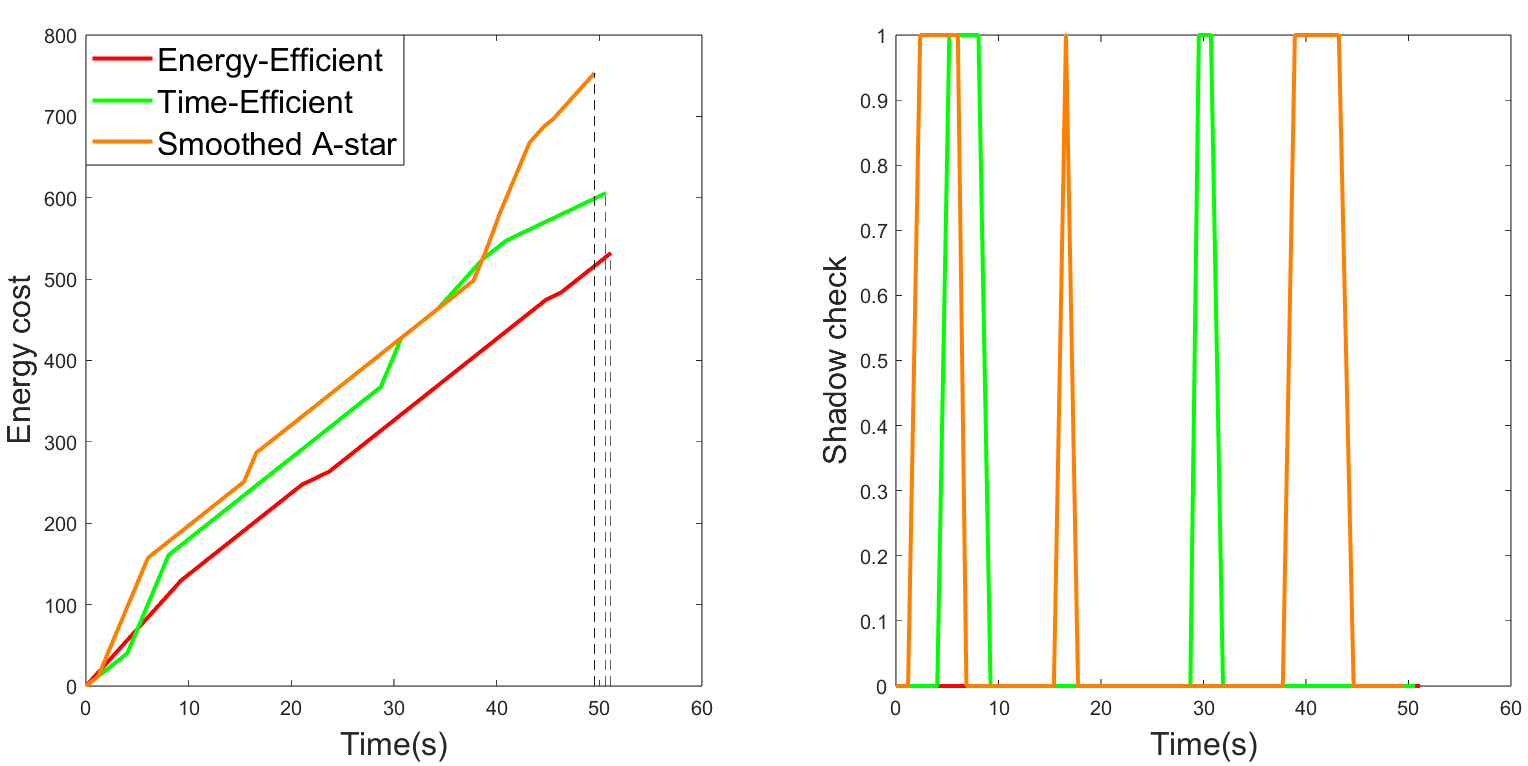}}
\caption{Result Comparison in Section \ref{ch4}}
\label{ch3f9}
\end{figure}

The smoothed A-star algorithm is used as a benchmark to compare with the proposed methods. Figure \ref{ch3f7} and \ref{ch3f8} compare energy-efficient, time-efficient, and A-star algorithms. It can be seen that the time-efficient method has some overlaps with the A-star method from the top view, but they are different at the end of the trajectory. Because the time-efficient method takes a lift to gain more solar energy, as shown in figure \ref{ch3f7}. Fig. \ref{ch3f9} and TABLE \ref{table:2} show the total energy cost for all those three methods. The energy-efficient method has a minimum cost of 532.1$J$, far less than the battery capacity of 670$J$. The threshold energy $E_{min}$ is set at 50$J$, which means the total energy cost of SUAV cannot exceed 620$J$. In this way, the time-efficient method has a total cost of 605.6$J$ that precisely meets the energy requirement. Then, the A-star method provides a shorter trajectory. Still, the total cost is well above the battery capacity, making the actual application's trajectory infeasible. On the right-hand side of Fig. \ref{ch3f9}, the shadow condition check $V(p(t))$ is presented. It can be observed that the energy-efficient trajectory never crosses the shadow area, and the A-star method spends a more extended time under shadow than the other two methods. The energy-efficient method provides much lower cost, and the time-efficient method balances both time and cost. 
\begin{table}[ht]
\centering
\begin{tabular}{|c|c|c|c|c|}
\hline
\textbf{Method} & \textbf{Time(s)} & \textbf{Cost(J)}\\
\hline
Energy-Efficient		& 51.13 & 532.1 \\
Time-Efficient 		& 50.64 & 605.6 \\
Smoothed A-Star 		& 49.55 & 753.3 \\
\hline
\end{tabular}
\caption{Result Comparison in Section \ref{ch4}}
\label{ch3table:2}
\end{table}

\subsection{Conclusion}
\label{ch3conclusion}
In conclusion, a path planning algorithm is proposed to generate an energy-efficient path in a complex urban environment. With further modification, a time-efficient algorithm is used to plan the path that can return to the charging station as soon as possible without violating the battery energy limitation. Computer simulation is completed to demonstrate the expected performance of the proposed method. Both algorithms can be used in a 3D environment with an extensive range of altitudes. Compared with the A-star method, the proposed method shows a much lower energy cost for the energy-efficient algorithm, and the time-efficient method can find a feasible path considering the limitation of onboard batteries. The optimization problem is convex that the global optimum can be achieved. The design of the heuristic function $h$ is key to achieving the global optimum. For this, $h$ must be both admissible and consistent with the problem. In addition, improving the computational efficiency can be a future research direction. The relative high computational cost making it non ideal for real-time execution.

\section[A Hybrid Approach for Navigation of a Solar-powered UAV in a Dynamic Urban Environment]{A Hybrid Approach for Navigation of a Solar-powered UAV in a Dynamic Urban Environment} 
\label{ch5} 

This section considers a scenario in which a solar-powered UAV travels in a dynamic urban environment with unknown static and moving obstacles. A framework is proposed, including a three-phase hybrid approach for this problem. Firstly, an energy-aware path planning algorithm is proposed based on the limited information of the environment. Secondly, a pure pursuit controller is applied to follow the pre-generated online path. Finally, an energy-aware reactive obstacle avoidance algorithm is used to avoid collision with unknown obstacles. The effectiveness of the proposed framework is verified based on computer simulation. This section is an extension of the previous section, in which a more complicated scenario is considered.

\subsection{Introduction}

The origin of unmanned Aerial Vehicles (UAVs) can be traced back to the initial efforts to construct a pilotless aircraft in 1916 \cite{4161584}. Although UAVs were widely intended for military utilisation in the early stages, UAVs are becoming a more compelling area of interest because they are capable of being equipped with various advanced sensors (e.g. visible light sensors, multispectral sensors, thermal sensors) \cite{DENG2018124} to collect data with high precision, articulated arms to inspect infrastructure such as the ceilings of bridges and powerlines \cite{ramon2019planning}, and an Automated External Defibrillator (AED) rapidly aiding in cardiac arrest cases exposed to traffic \cite{saboor2110elevating}. In the upcoming decades, UAVs are expected to play a more critical role in Smart Cities with the advancement of wireless sensors, networked unmanned systems and the Internet of Things (IoTs) \cite{drones6040095}. However, the precision flying of UAVs is challenging in an urban environment in terms of various flight altitudes and paths and long flight periods under diverse unpredictable conditions, including both daytime and night-time operations. If these difficulties are unsolved, they may cause concerns about drone accidents. The fall of UAVs may pose a risk to people on the ground, especially in urban environments with a high population concentration \cite{9165709}. This section will focus on the applications of SUAVs in an urban environment, which require stringent endurance and energy management strategies. 

In light of the promotion of renewable energy and the more matured solar energy-related technologies, the propellers of UAVs in this section are driven by electric motors not only powered by the onboard battery but also powered by the solar energy collected via the solar cells equipped on the surfaces of UAVs. The solar-powered unmanned aerial vehicle (SUAV) considered in this section qualifies and estimates the energy required for the total flight from the starting point to the destination, which is also necessary to monitor the energy loss simultaneously. Considering the size of photovoltaic cells equipped on small-sized SUAVs is generally restricted and the corresponding conversion efficiency of solar panels, it is essential to optimize the path planning of SUAVs, thereby adjusting the instantaneous incidence angle between the direction of the sunlight and solar panels to maximise the solar energy harvesting \cite{wu2018path}. Besides the sun position and the spatial orientation of SUAVs, buildings in urban environments also affect the amount of solar radiation incident on the photovoltaic panels because they may cause sunlight occlusions \cite{WU2017497}. Therefore, it is essential to consider the shadow region when optimising the path planning of SUAVs in urban environments. Wu et al. \cite{wu2018path} integrates the Improved Whale Optimization Algorithm (IWOA) with a Reactive Influence Field Differential System (RIFDS) to generate an optimised path planning of SUAVs in urban environments. In their cost function, three factors include collision avoidance for static and dynamic obstacles, energy efficiency and path length. Similarly, for the path planning of SUAVs in 2D space, Huang et al. \cite{huang2016energy} combine Receding Horizon Control (RHC) with Particle Swarm Optimization (PSO). The authors of \cite{Solar1} propose a navigation strategy for a multi-SUAV system in urban environments to conduct a mobile target surveillance mission. However, the shadow area is treated as an obstacle in the environment, simplifying the energy problem to be solved.

Autonomous exploration in an unknown environment requires accurate time mapping, reactive control and online trajectory replanning to ensure navigation free of collisions \cite{shim2005autonomous, verma2023hybrid,wang2018strategy, savkin2013Reactive}. Predicting and obtaining comprehensive real-time map information is difficult in real urban environments. Various uncertainties may block their movements, such as flying birds, construction materials, balloons, and other UAVs. Online path planning can use map information to provide a close-to-optimal path based on limited information. However, online path planning cannot handle the uncertainties in dynamic environments. Local planners or methods that use sensors apply the knowledge gained from these sensors to quickly respond to obstacles close by, which is suitable in a dynamic environment \cite{elmokadem2021hybrid}. This is based on what they can see in their immediate surroundings at any given time. On the other hand, simply relying on reactive navigation may lead to some dead loop and unpredictable motions \cite{8484144}. A hybrid approach can thus leverage the advantage of both online path planning and reactive navigation methods, which is a better option in a partially unknown environment. Authors \cite{gia2020real} use the modified A-star algorithm as the global planner, a weighted-sum model was employed to choose a temporary target and a smoothed sub-trajectory was found that could lead the robot to avoid the unknown obstacles. A hybrid path planning algorithm based on the safe A-star algorithm and adaptive window approach \cite{zhong2020hybrid} was proposed to handle the uncertainties in a large-scale dynamic environment. The authors of \cite{li2016hybrid} develop a local rolling optimization algorithm to constantly optimize the results of the pre-generated online path while sensing the information of the environment.

This section aims to propose a hybrid navigation approach to guide a solar-powered UAV to move in a dynamic urban environment containing multiple unknown obstacles. The proposed hybrid approach includes three phases: firstly, an online energy-aware path planning phase; secondly, a path tracking phase; and lastly, a reactive control obstacle avoidance phase. The online path planning algorithm generates a collision-free path based on the limited information about the environment. Then, a path-tracking controller is applied to ensure the SUAV can follow the pre-generated path properly. While following the path, if there is some unknown obstacle blocking, a reactive obstacle avoidance strategy will be used to handle the uncertainty, providing a safe and collision-free trajectory. In the meantime, the sun's position significantly influences the control input that can guide the SUAV to gain more solar energy from the sun. The online path planning algorithm is modified based on the A-star algorithm \cite{Hart1968}, a pure pursuit controller \cite{pure} is used to track the path, and the reactive obstacle avoidance is inspired by \cite{savkin2013simple}. Compared with some other work, the proposed method provides a more improved result in the dynamic environment for the SUAV. Additionally, it is worth mentioning that no previous work has been implemented in this scenario.

The structure of this section is outlined as follows. Section \ref{model} introduces the SUAV system model and its energy models. The problem being tackled in this study is discussed in Section \ref{problem}. Section \ref{law} explains the implementation of the proposed hybrid strategy. The following section \ref{simulations} presents how the proposed method enhances operations through computer simulation in a dynamic urban environment. The final Section \ref{conclusion} summarizes the entire section.

\subsection{System Model}
\label{model}
\subsubsection{UAV System Model}
Assuming a non-holonomic solar-powered UAV moving in the 3D space, the absolute position of the SUAV is defined as $p(t)=(x(t), y(t), z)$ at each time step $t$. The SUAV is considered to be flying at a fixed altitude $z$ in the environment and is assumed to employ the same motion control strategy as a normal UAV. A well-known model for the motion of UAV is represented as follows: 
\begin{eqnarray}
\label{UAV}
	\begin{cases}
	\dot{x}(t) = v(t)cos(\theta(t))\\
	\dot{y}(t) = v(t)sin(\theta(t))\\
	\dot{\theta}(t) = u(t),\\
	\end{cases}
\end{eqnarray}
where $\theta(t)$ is the heading of the SUAV; angular speed and horizontal speed are represented by $u(t)$ and $v(t)$ respectively. In real life, this kinematic model is often supplemented by a dynamic model of a UAV with advanced controller and state estimators such as H-infinity controllers \cite{petersen2000robust, bansal2013design, azar2020backstepping} and robust state estimators \cite{petersen1999robust, pathirana2005node, savkin1998robust, marantos2015uav, hajiyev2012robust}. Furthermore, some hard constraints must be stated for the flight's speed and height. $v(t)$ and $u(t)$ have lower and upper boundaries at all times. Because of the physical limitations, the SUAV model has the following constraints on the control inputs: the effect of wind is ignored,
\begin{eqnarray}
\label{constraint}
	\begin{cases}
	v^{min}\leq v(t)\leq v^{max}\\
	u^{min}\leq u(t)\leq u^{max}.\\
	\end{cases}
\end{eqnarray}

\subsubsection{Urban Constructions Model}
Many constructions of different shapes and heights exist in a complex urban environment. They will be treated as obstacles in the path-planning phase, requiring an appropriate methodology. This section models each urban construction as the smallest prism capable of enclosing the structure, which can be formulated as \cite{wu2018path}:
\begin{align*}
	\Gamma{(p)}=\\ \left(\frac{x-obs_{x_{0}}}{a}\right)^{2d}+\left(\frac{y-obs_{y_{0}}}{b}\right)^{2e}+\left(\frac{z-obs_{z_{0}}}{c}\right)^{2f}
\end{align*}
$p(t)$ is the position of the UAV at time $t$, while the obstacle center is denoted by $(obs_{x_{0}},obs_{y_{0}},obs_{z_{0}})$, and the obstacle has axes lengths $a$, $b$ and $c$. Additionally, $d$, $e$ and $f$ are the parameters that can define the shape of the construction. When $d = 1$, $e = 1$ and $f > 1$, the construction can be treated as a cylinder. While if $d > 1$, $e > 1$ and $f > 1$, the obstacle is a rectangle. After evaluating the function, it can be concluded that the SUAV will have collisions with constructions if $\Gamma{(p(t))} \geq 1$; Now a hard constraint is introduced:
\begin{eqnarray}
\label{obstacle}
	\Gamma{(p(t))} < 1,
\end{eqnarray}
which can obtain a collision-free path in the 3D space during the online path planning phase. Equation (\ref{obstacle}) must hold at all times to ensure the safe operation of the SUAV.

\subsubsection{Energy Model}
In this section, the SUAV has on-board solar cells that can gain solar energy from the sunlight. In the meantime, the SUAV needs to consume a certain amount of energy while flying. A clear sky model \cite{ASHRAE} is used to estimate the solar radiation rate correctly without considering the cloud condition in the sky. The solar energy gained from the direct sunlight can be computed as \cite{klesh2009solar}:

\begin{eqnarray}
\label{Solar Cost}
	E_{\text {gain }}=\int_{t_{o}}^{t_{n}} \eta P_{\mathrm{sd}} S \cos (\theta) \mathrm{d} t,
\end{eqnarray}
where $\eta$ stands for the solar cell efficiency and $S$ represents the size of the onboard solar panel. The incidence angle size $\theta$ depends on the azimuth angle and the sun's elevation angle, where both angles vary with respect to the time of the day. Then $E_{\text {gain }}$ is the total solar energy gained from the solar panel within the time frame $\left[t_{o}, t_{n}\right]$.

Different models of UAVs consume different levels of energy within a given amount of time. Flying speed is the main factor that affects the rate at which energy is consumed. A common energy model \cite{7101619} for a quadrotor UAV can be used when the speed of the UAV is constant at a fixed height.
\begin{eqnarray}
\label{UAV Cost}
	E_{out}=\int_{0}^{d / v} P(v) d t=P(v) \frac{d}{v}.
\end{eqnarray}
All in all, the energy model of the SUAV is presented in the above equations.

\subsection{Problem Statement}
\label{problem}
In this section, a scenario is considered that an SUAV is flying in a bounded 3D urban environment ${\cal A}\subset {\bf R}^3$ with some static and known obstacles ${\cal O}_k \subset {\cal A}$ as well as some moving or static unknown obstacles ${\cal O}_u = \{{\cal O}_1, {\cal O}_2, \cdots, {\cal O}_n \} \subset {\cal A}$. The statically known obstacles can be the urban constructions from the given map information. Furthermore, the unknown obstacles can be the dated map information or some birds in the environment that cannot be predicted. Thus, the primary set of the obstacles can be defined as ${\cal O} = {\cal O}_k \cup {\cal O}_u$.
 
Line of Sight (LoS) is one of the best and easiest ways to determine whether the SUAV is exposed to the sun in the urban environment. Given the position of the sun as $p_{sun} = ({x_s,y_s,z_s})$, a line segment is constructed between $p_{sun}$ and $p_{t}$. By evaluating the intersection of this line segment with any potential obstacles ${\cal O}_n$, it can be conclusively determined whether the SUAV is situated within a shadow area, which can be defined as follows:
\begin{align*}
	\begin{cases}
	LoS(p(t)) = 1, \text{shadow}\\
	LoS(p(t)) = 0, \text{not shadow}.
	\end{cases}
\end{align*}
In the presence of LoS, the solar power gained from the direct sunlight depends on the $LoS(p(t))$
\begin{align*}
	\begin{cases}
	P_{\text {solar }}(\theta)=\eta P_{\mathrm{sd}} S \cos (\theta) \quad \; \; \, \text { if } LoS(p(t)) = 1\\
	P_{\text {solar }}(\theta) = 0 \quad \quad \quad \quad \quad \quad \quad \text{if } LoS(p(t)) = 0.
	\end{cases}
\end{align*}
Considering the power consumption during the flight, the residual on-board energy can be defined as:
\begin{eqnarray}
\label{Cost3}
 	E_{\text {total }} = E_{\text {batt }} +  E_{\text {gain }} - E_{\text {out }} \geq E_{min},
\end{eqnarray}
where $E_{min}$ is the lowest threshold for the battery level, and $E_{\text {batt}}$ is the initial energy stored in the battery.

In addition, the following assumptions are made.

{\bf Assumption 1:} The shapes of the unknown obstacles are various, but they can fit in a sphere of radius $R_n$.

{\bf Assumption 2:} The moving speed of each of the unknown obstacles ${\cal O}_n$ is at a constant speed of $v_o(n)$, and the maximum speed of the SUAV is greater than the full speed of the moving obstacles as $v_o^{max}< V < v^{max}$.

{\bf Assumption 3:} The speed $v_o(n)$ and the mass center of the obstacle $O_n$ is known to the robot.

The following statement summarises the problem being considered for this section.

{\bf Problem 1: } Consider an SUAV whose motion is described by (\ref{UAV}) flying in a 3D space ${\cal A}$. Under assumptions 1, 2 and 3, design control laws for $v(t)$ and $u(t)$ to achieve collision-free navigation from $p_0$ to $p_{target}$. In the mean time, try to achieve a minimum $E_{\text {total}}$ and satisfy the safety requirements in (\ref{constraint}), (\ref{obstacle}) and (\ref{Cost3}).

\subsection{Navigation Law}
\label{law}
\subsubsection{Online Path Planning}
The first phase of the hybrid approach is to generate an online path as a reference path for the SUAV. The proposed online path planning algorithm is inspired by the famous A-star algorithm \cite{Hart1968}. The A-star algorithm, distinguished for its robustness and adaptability across many systems beyond robotics, has been widely adopted by researchers. 

In general, the basic A-star algorithm designs some cost functions and minimizes the cost function
\begin{eqnarray}
\label{costf}
	f(n) = g(n) + h(n)
\end{eqnarray}
where $n$ stands for the next position of the robot. $g(n)$ can be calculated based on the actual cost from the initial node $p_{0}$ to the next node $p_n$, and $h(n)$ is a heuristic function that can estimate the lowest cost from $p_n$ to the $p_{target}$. The heuristic function is problem-specific and ideally estimates the actual cost to reach the goal, thus enabling the A-star algorithm to prioritize paths that appear to lead to a faster solution. The result is not optimal since this technique uses both heuristic information and actual path cost data in the search process. However, properly designing the heuristic function can guarantee a close-to-optimal solution.

This section aims to find an energy-aware path that can provide a minimum total energy cost. According to the energy model (\ref{Solar Cost}) and (\ref{UAV Cost}) in section \ref{model}, the current cost $g(n)$ can be represented by $E_{out} - E_{gain}$ from $p_{0}$ to $p_n$, and the heuristic function $h(n)$ is designed to be the total energy cost associated with a line segment from node $p_n$ to $p_{target}$. In addition, due to the physical feature of the battery, the total energy stored cannot exceed a maximum threshold value $E_{max}$. In the meantime, equation (\ref{costf}) holds at all times. Then, it completes the online path planning phase, and an energy-aware path can be computed.

\subsubsection{Path Tracking}
\begin{figure}[ht]
\centerline{\includegraphics[width=7 cm]{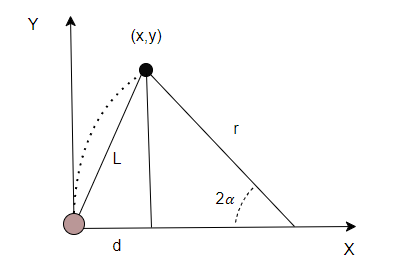}}
\caption{Pure pursuit}
\label{ch5f0}
\end{figure}

With the information provided in the previous section, an energy-aware path is given as a matrix ${\cal W} = \{{\cal W}_1, {\cal W}_2, \cdots, {\cal W}_n \}$. A path-following algorithm is needed to direct the SUAV to follow the path properly as the second phase of the hybrid approach. The pure pursuit tracking \cite{pure} algorithm is chosen due to its reliability and simplicity. Rather than employing a complicated polynomial, the displacements along the $x$ and $y$ axes are utilized to match a circular arc to the target point. Hence, the trajectory is pursued by consecutively adapting various arcs to different target points while the SUAV moves forward. While following the path, the pure pursuit controller will choose a lookahead point $p* \subset \cal W$ as the next virtual target, which is $L$ distance ahead of the closest node. The pure pursuit controller will adjust its steering angle to guide the SUAV towards the virtual lookahead point. Figure \ref{ch5f0} shows the geometric relationship of the pure pursuit controller. Brown circle is the SUAV under investigation, and the black circle is the next lookahead point. 

The curvature $k$ can be computed as
\begin{eqnarray}
\label{pure1}
	\frac{L}{sin(2\alpha)} = \frac{r}{sin(\frac{\pi}{2}-\alpha)}\\
	\frac{L}{2sin\alpha cos\alpha} = \frac{r}{cos\alpha}\\
        \frac{L}{sin\alpha} = 2r\\
        k=\frac{L}{r}=\frac{2sin\alpha}{L}
\end{eqnarray}

Therefore, the steering angle $\delta$ can be calculated as:
\begin{eqnarray}
\label{pure2}
	\delta = v(t)*k
\end{eqnarray}

\subsubsection{Local Obstacle Avoidance Algorithm}
While executing the path-following phase, there will be some unknown obstacles in the environment that may block the movement of the SUAV. The third phase of the hybrid approach is to use a reactive obstacle avoidance algorithm to handle this problem. The algorithm is developed based on a simple biologically inspired algorithm \cite{savkin2013simple} that is used to navigate a wheeled mobile robot in a dynamic, cluttered environment with some unknown moving obstacles.
\begin{figure}[ht]
\centerline{\includegraphics[width=6 cm]{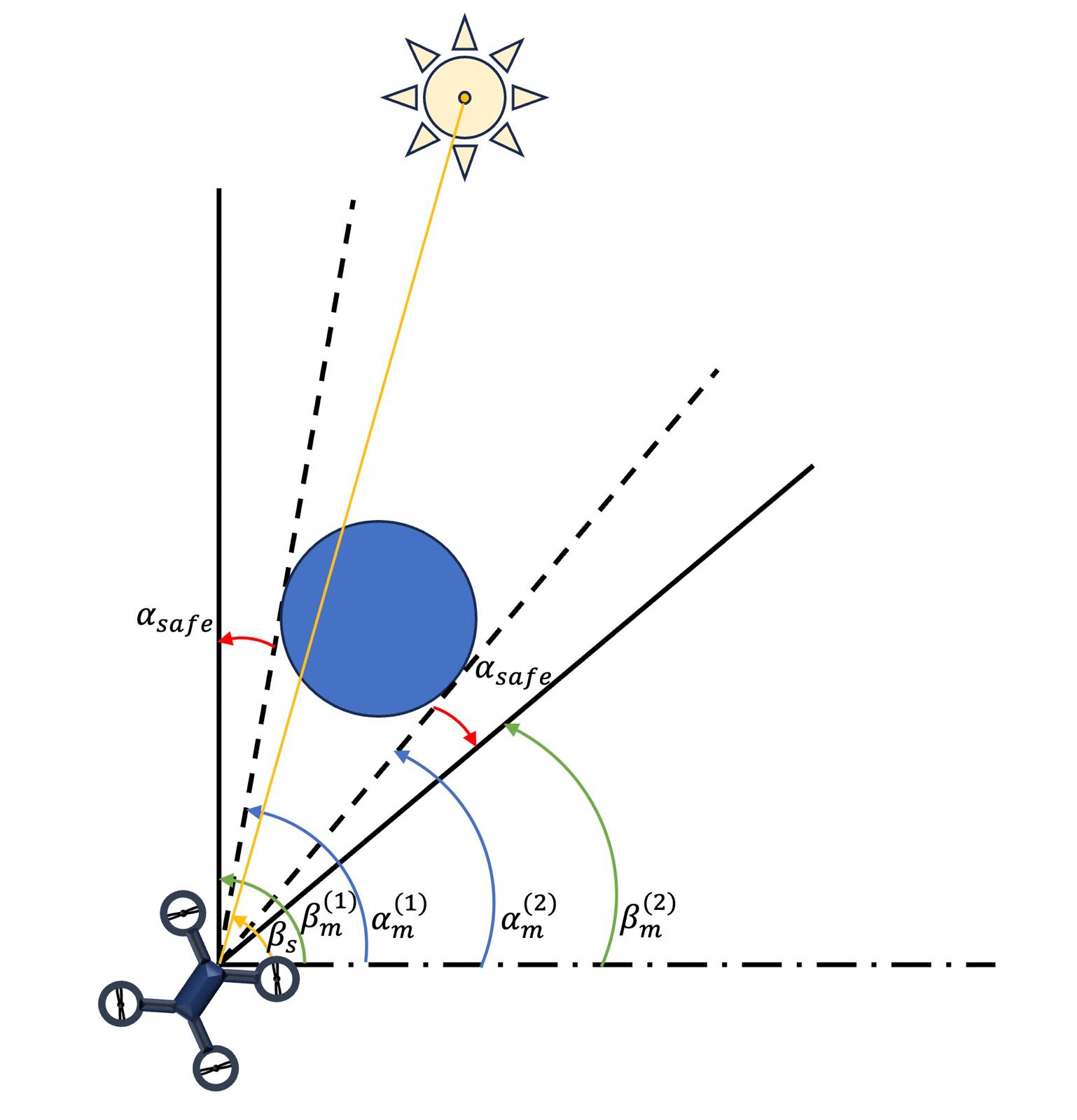}}
\caption{Vision Cone that Represents the Sensing Area of the SUAV}
\label{ch5f1}
\end{figure}

A front-facing sensor is assumed to be equipped on the SUAV with a sensing range of $R_{\text{sensor}}$ that can sense the occurrence of obstacles. This field of view is typically treated as an enlarged cone. When an obstacle enters the field of view, its outline intersects with the boundaries of the camera view. Within those intersections, two intersections giving the largest and smallest absolute angles can be defined as key intersections. These two angles can be defined as $\alpha_m^{(1)}$ and $\alpha_m^{(2)}$. Now let $\alpha_{\text{safe}}$ be a constant angle, $0 < \alpha_{\text{safe}} < \pi/2$. So
\begin{eqnarray}
\label{betaf}
	\beta_m^{(1)} = \alpha_m^{(1)} + \alpha_{\text{safe}}\\
	\beta_m^{(2)} = \alpha_m^{(2)} - \alpha_{\text{safe}},
\end{eqnarray}
Then the vector of $\beta_m$ can be represented by $\gamma_n^{(m)}, m = 1,2$
\begin{eqnarray}
\label{gammaf}
	\gamma_n^{(m)}(t) := (v^{max}-V)\left[cos((\beta_m^{(n)}t)),sin((\beta_m^{(n)}t))\right],
\end{eqnarray}
which defines the two boundaries of an enlarged vision cone of the obstacle. Now let $\gamma_1$, $\gamma_2$ be the vectors from the current position and $\alpha(\gamma_1, \gamma_2)$ be the angle between $\gamma_1$ and $\gamma_2$ in a counter-clockwise direction as shown in figure \ref{ch5f1}. Then
\begin{eqnarray}
\label{tauf}
\tau(\gamma_1,\gamma_2):=
	\begin{cases}
	0 \quad \quad \alpha(\gamma_1, \gamma_2) = 0\\
	1 \quad \quad 0 < \alpha(\gamma_1, \gamma_2) \leq \pi\\
	-1\quad -\pi < \alpha(\gamma_1, \gamma_2) < 0.
	\end{cases}
\end{eqnarray}

There are two vectors in equation (\ref{gammaf}), two possible directions for the SUAV to move towards in the future. To make this decision, $m\in\{1, 2\}$ value needs to be determined. For the navigation of the SUAV, the sun's position can be considered in this problem. Let $\beta_s$ be the angle formed by the sun's projection point on the ground and the horizontal coordinate. Let $\gamma_s$ be the vector in that direction. Then
\begin{eqnarray}
\label{epsilonf}
	\epsilon_m &= \alpha(v_o(n)+\gamma_n^{(m)}(t),v(t))
\end{eqnarray}
\begin{eqnarray}
\label{mf}
	m_0:= 
	\begin{cases}
	\min |\epsilon_1, \epsilon_2| \quad \quad \; \; \, |\epsilon_1 - \epsilon_2 | \geq \Theta\\
	\min |\alpha (\epsilon_m, \gamma_s)| \quad  |\epsilon_1 - \epsilon_2 | < \Theta
	\end{cases}
\end{eqnarray}
where $\Theta$ is a relatively small angle that can be turned to affect the algorithm's performance. The meaning of the above equation is that among two vectors $\alpha(v_o(n)+\gamma_n^{(1)}(t)$ and $\alpha(v_o(n)+\gamma_n^{(2)}(t)$, the vector that has a smaller angle between itself and the heading of the SUAV is chosen when the difference between those two vectors are smaller than the threshold value $\Theta$. If the difference between two vectors is small, the vector closer to $\gamma_s$ will be chosen. Therefore, the SUAV has a trend to gain more direct sunlight. 

A sliding-mode-control law-based navigation law can be proposed:
\begin{align}
\label{uf}
	u(t) &= -u^{max}\tau(v_o(n)+\gamma_n^{(m)}(t),v(t))\\
	v(t) &= \| v_o(n) + \gamma_n^{(m_0)}(t)\|
\end{align}
It can be seen that $u(t)$ is defined by the value of $\tau$, which means the angular velocity is consistently at the maximum value either in a positive or negative direction. Additionally, the control law leads the robot to avoid the obstacle at a constant avoiding angle $\alpha_{\text{safe}}$.

To further explain the rationale of the control law, the close-loop system (\ref{UAV}) and (\ref{uf}) are ordinary differential equations with discontinuous right-hand side. The control law will steer the heading of the SUAV to a sliding surface that is outside of the avoiding object. A fixed avoiding angle $\alpha_{\text{safe}}$ was set between the robot's current direction of motion and one of the two boundary rays within the obstacle's vision cone as seen from the robot. In a nutshell, the control law involves navigating towards the nearest of the two boundaries of the expanded vision cone of the obstacle, as specified by equation (\ref{gammaf}).

\subsubsection{Switching Law}
A switching mechanism is used to switch between path tracking of the global path mode and obstacle avoidance reactive control mode. The main factor is the distance between the obstacle and the SUAV. The sensing range of the onboard sensor is $R_{\text{sensor}}$. Now introduce a threshold value $D$ as the trigger distance and $d_i$ as the minimum distance between obstacles. The proposed switching law belongs to the class of switched controllers \cite{utkin2017sliding, utkin1978sliding, savkin2002hybrid, skafidas1999stability}.

{\bf Switching Law 1:} Switching from path tracking mode to reactive control mode when the minimum distance between obstacles is less than the threshold value. i.e., $d_i \leq D$.

{\bf Switching Law 2:} Switching from reactive control mode to path tracking mode when SUAV is oriented towards the next virtual target point $p*$ and the minimum distance between obstacles is larger than threshold value $D$.

\subsection{Computer Simulations}
\label{simulations}
This section presents the simulation results for the proposed navigation algorithm. The primary objective of this simulation is to find a collision-free trajectory with lower energy cost compared with the purely reactive control method. MATLAB was chosen as the simulation software for this problem. The environment contains four static prisms as the urban buildings, and a combination of moving and static obstacles makes the problem difficult to solve. The essential parameters of this simulation are listed in the table \ref{ch5table:1}. 

\begin{table}[ht!]
\centering
\begin{tabular}{|c|c|c|}
\hline
\textbf{Parameter} & \textbf{Value}\\
\hline
Flying speed, $v$ & 12 $m/s$ \\
Maximum flying speed, $v^{max}$ & 20 $m/s$\\
Flying height, $z$ & 100 $m$ \\
Safe angle, $\alpha_{\text{safe}}$ & 40 \\
Lookahead distance, $L$ & 20 $m$ \\
Power at constant speed, $P(v)$ & 30 $W$ \\
Solar spectral density, $P_{\mathrm{sd}}$ & 380 $W/m^2$ \\
Solar cell efficiency, $\eta$ & 20$\%$ \\
Solar panel area, $S$ & 0.3 $m^2$ \\
Battery Capacity, $E_{\text{Batt}}$ & 750 $J$ \\
\hline
\end{tabular}
\caption{Parameters Used in the Simulation of Section \ref{ch5}}
\label{ch5table:1}
\end{table}

\begin{figure}[ht]
\centering
\includegraphics[width=6 cm]{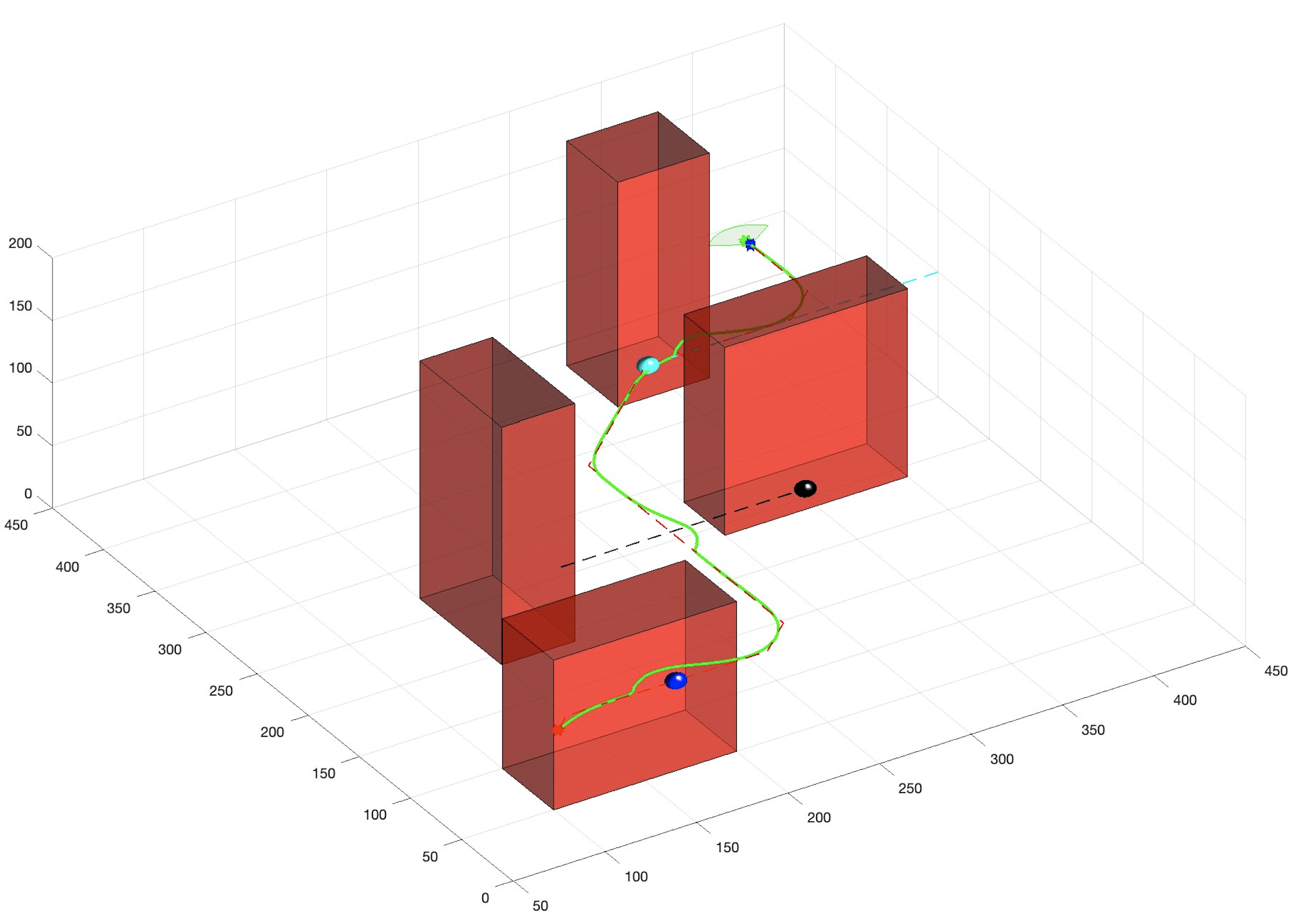}
\caption{3D Trajectory in a Dynamic Urban Environment}
\label{ch5f2}
\end{figure}

\begin{figure}
		\centering
        \begin{subfigure}{0.5\textwidth}
        	\centering
            \includegraphics[width=5 cm]{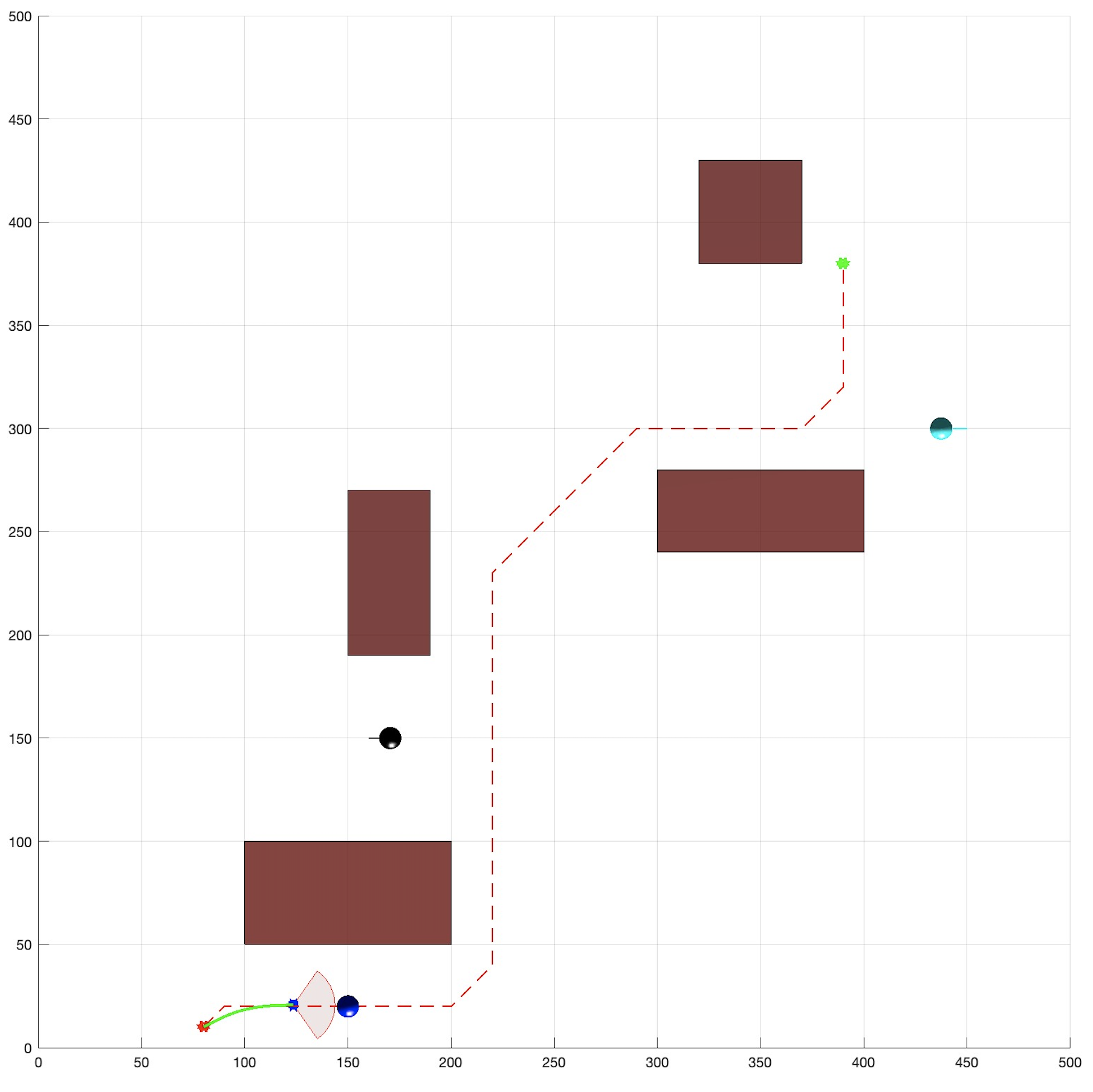}
            \caption{Top View 1}
            \label{fig:sub1}
        \end{subfigure}
        
        \begin{subfigure}{0.5\textwidth}
        	\centering
            \includegraphics[width=5 cm]{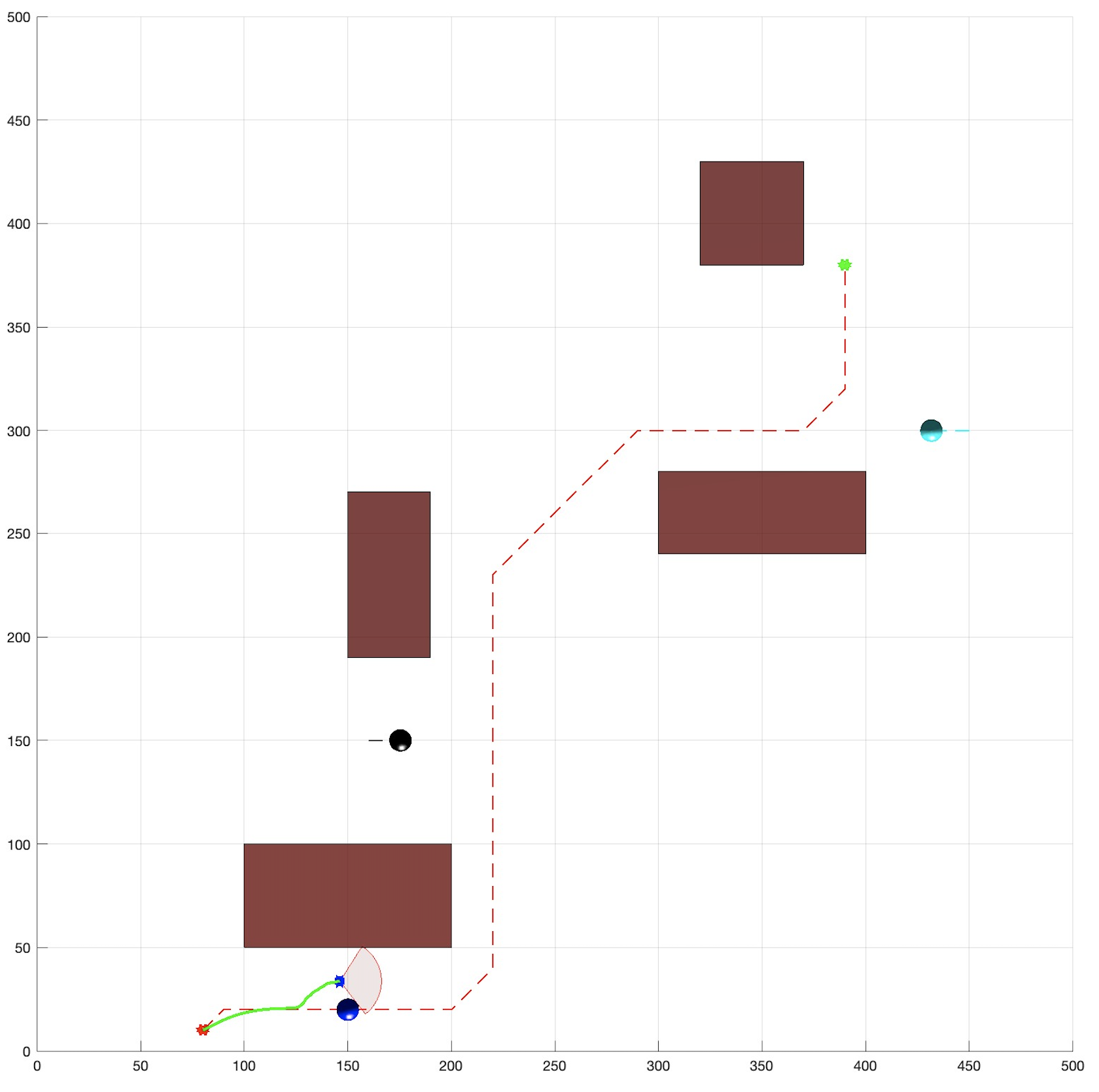}
            \caption{Top View 2}
            \label{fig:sub2}
        \end{subfigure}

        \begin{subfigure}{0.5\textwidth}
        	\centering
            \includegraphics[width=5 cm]{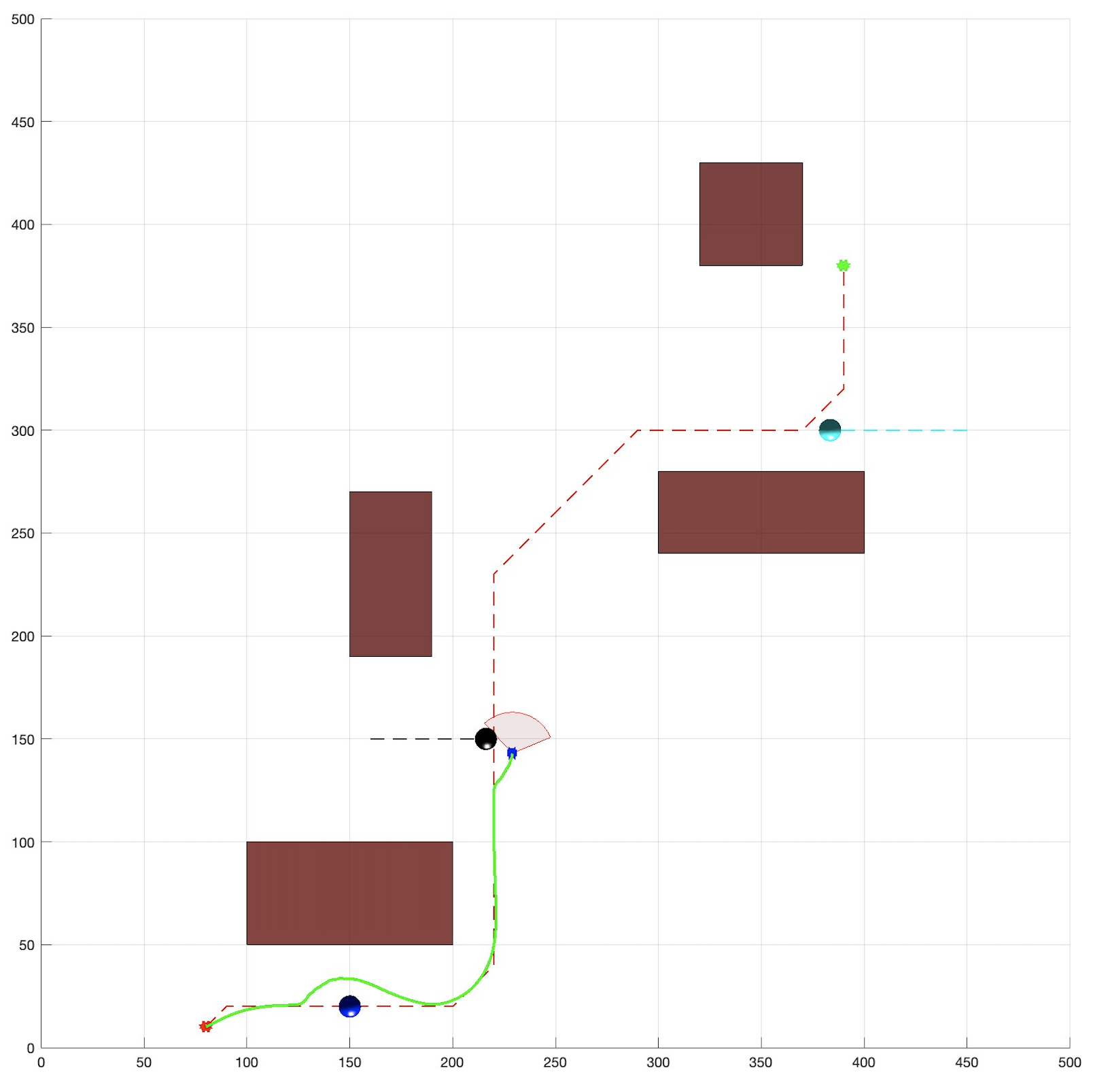}
            \caption{Top View 3}
            \label{fig:sub3}
        \end{subfigure}

        \begin{subfigure}{0.5\textwidth}
        	\centering
            \includegraphics[width=5 cm]{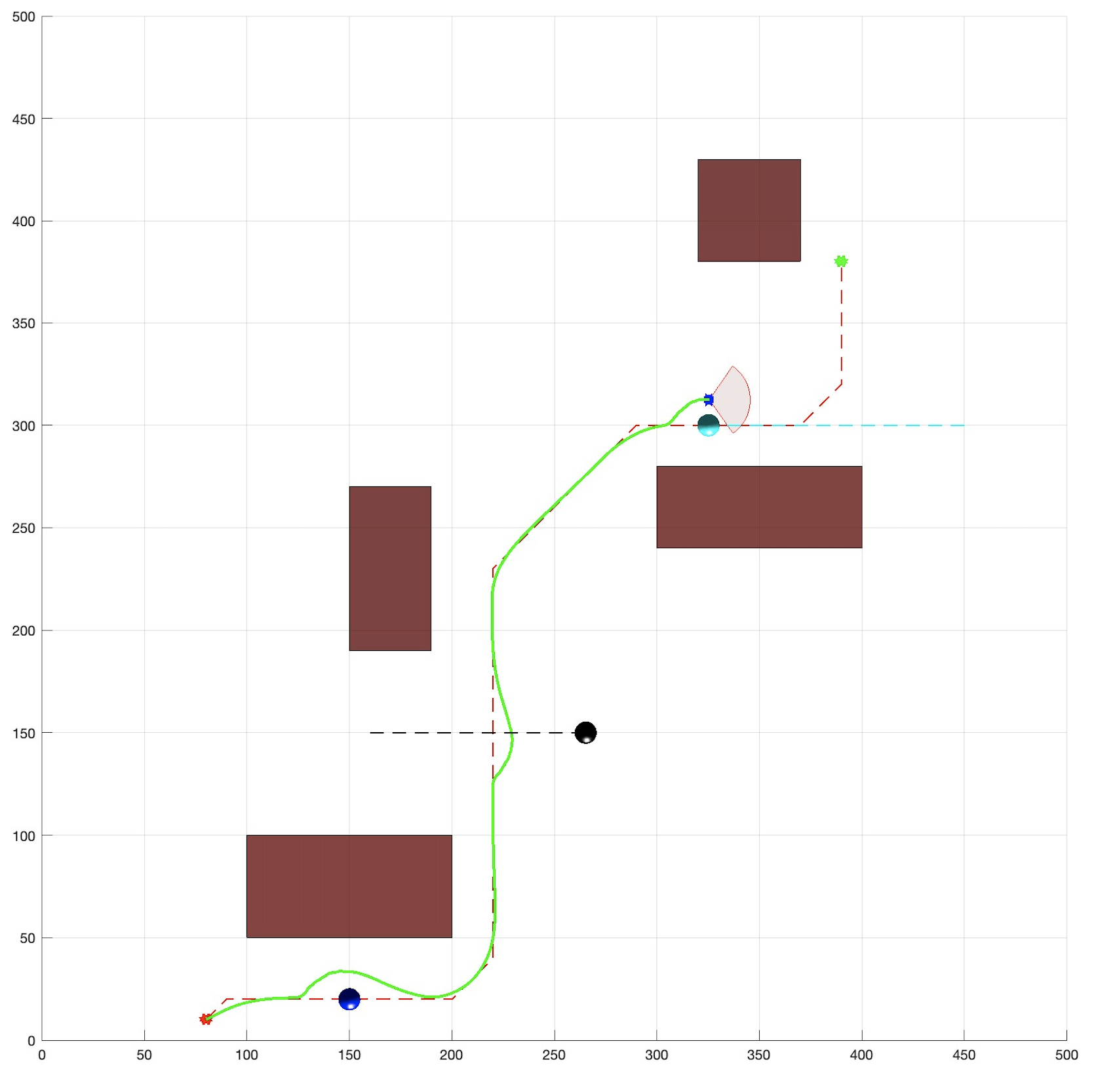}
            \caption{Top View 4}
            \label{fig:sub4}
        \end{subfigure}
    \caption{Top View of 3D Trajectory in a Dynamic Urban Environment}
    \label{ch5f3}
\end{figure}

\begin{figure}[ht]
\centerline{\includegraphics[width=8 cm]{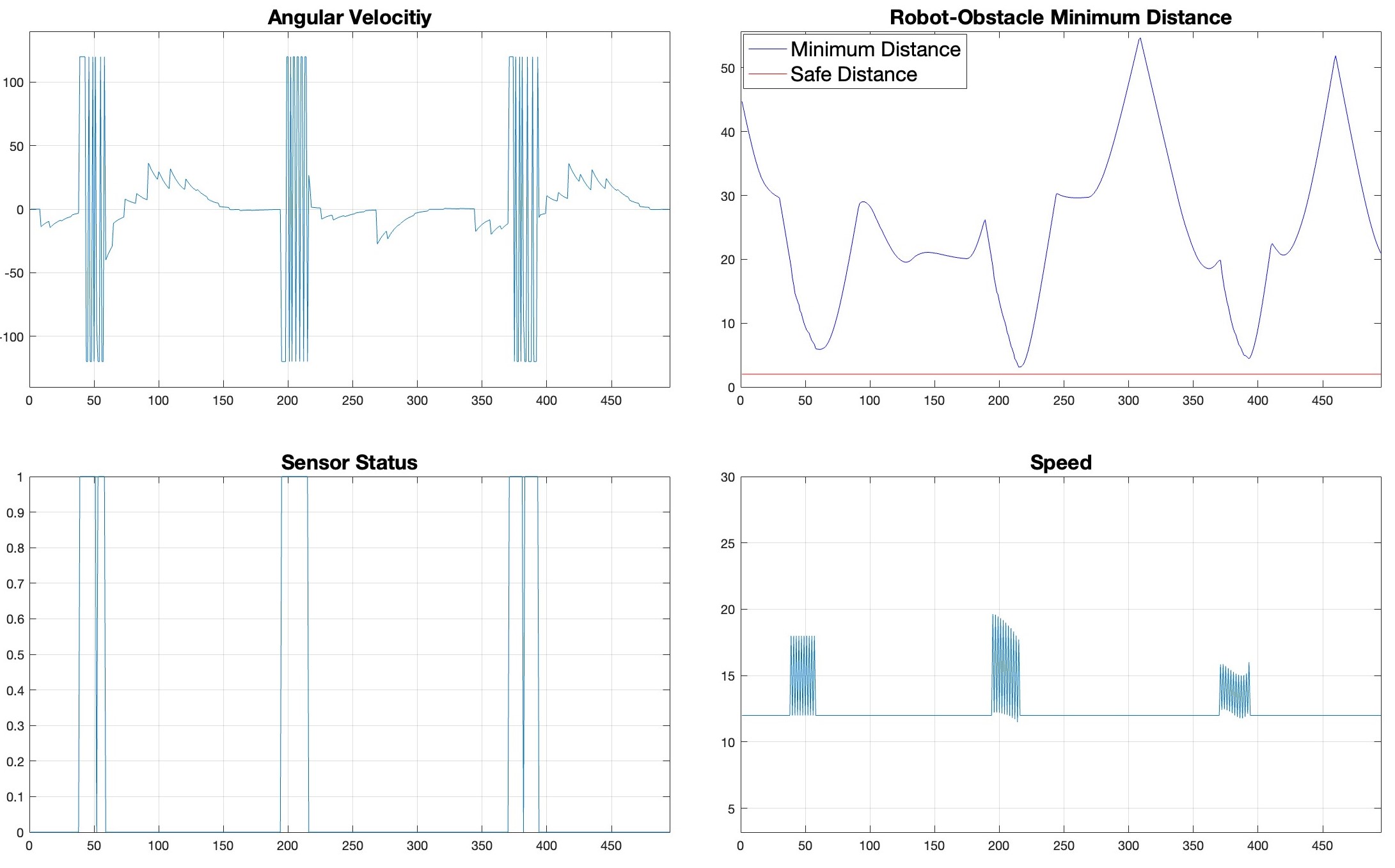}}
\caption{Simulation Data in a Dynamic Urban Environment}
\label{ch5f4}
\end{figure}

Figure \ref{ch5f2} shows the 3D trajectory in the urban environment; four prism-shaped blocks represent the buildings in the environment, which are the known static obstacles. One unknown blue static obstacle is added to the environment, along with two unknown moving obstacles that are black and green. When approaching the first unknown stationary blue colour obstacle, because $|\epsilon_1 - \epsilon_2 | < \Theta$, the motion vector is oriented along the positive y-axis instead of the negative y-axis. In this way, the SUAV can potentially avoid flying into the shadow area created by the unknown obstacle. Figure \ref{ch5f3} shows the top view of the whole navigation process. The angular velocity information and sensor status are shown in figure \ref{ch5f4}; it is indicated that when an obstacle is detected, the sensor status is set to $1$, and the $u$ values jump between $+120$ to $-120$. Additionally, the speed of the SUAV is set at a constant value of $12 m/s$ during the second phase until an obstacle is detected. The minimum distance between the SUAV and the closet obstacle is shown in figure \ref{ch5f4} as well; it can be seen that the safe distance requirement is consistently met. The proposed method can generate a collision-free trajectory while minimizing energy costs.

\begin{figure}[ht]
\centerline{\includegraphics[width=9 cm]{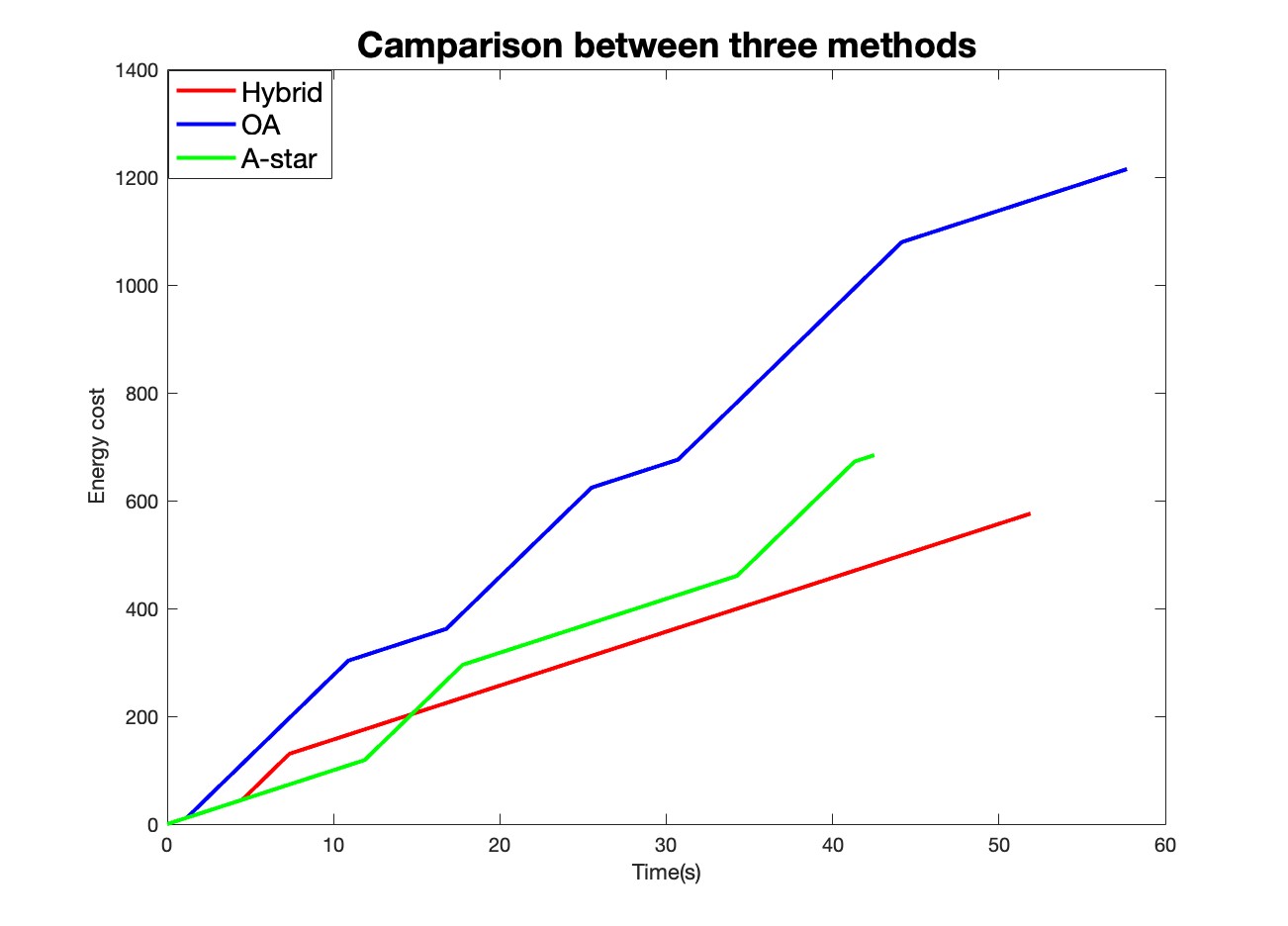}}
\caption{Cost Comparison of Three Methods}
\label{ch5f7}
\end{figure}

\begin{figure}[ht]
\centerline{\includegraphics[width=7 cm]{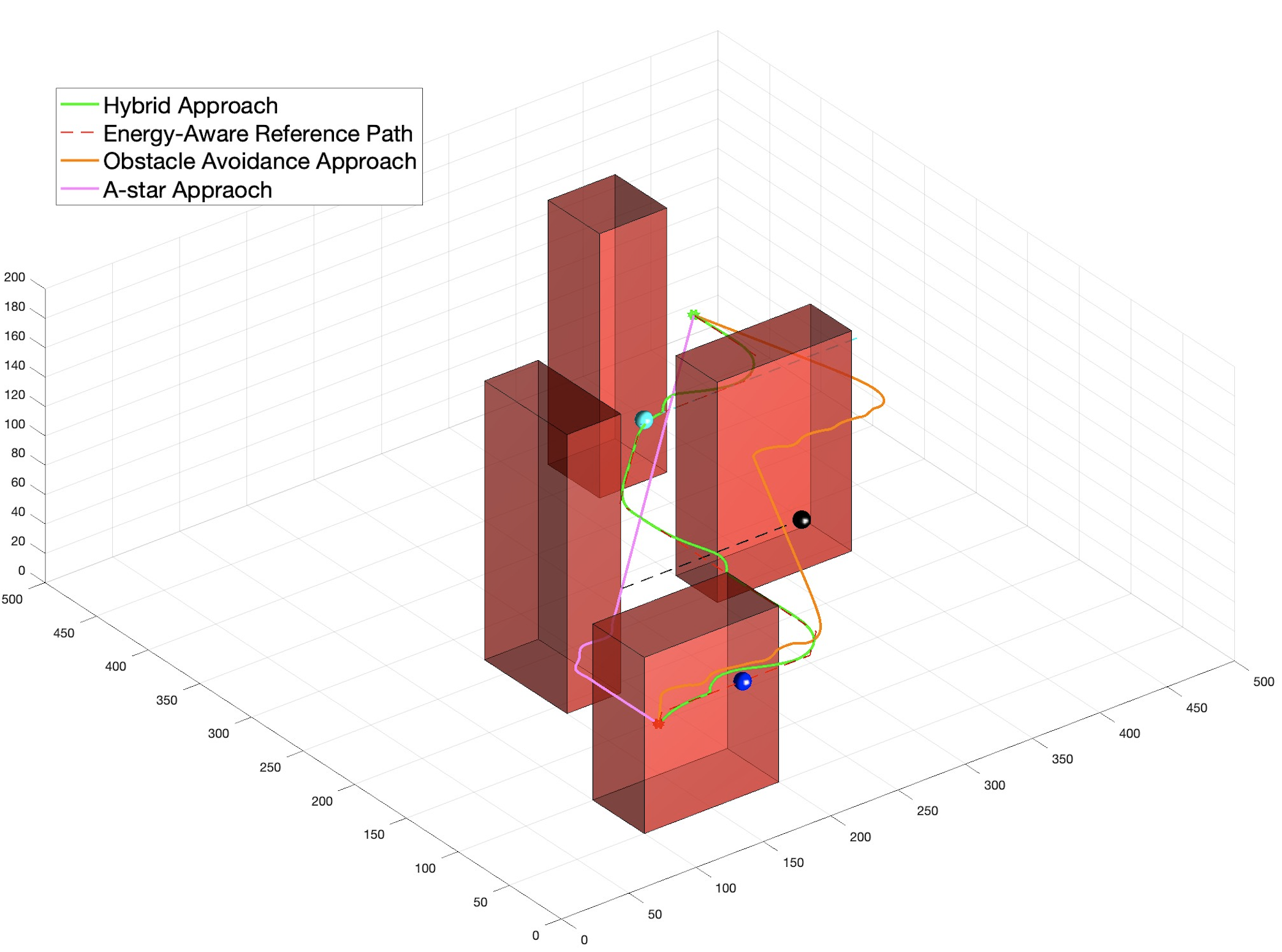}}
\caption{Comparison Result in 3D}
\label{ch5f5}
\vspace{1 cm}
\centerline{\includegraphics[width=6 cm]{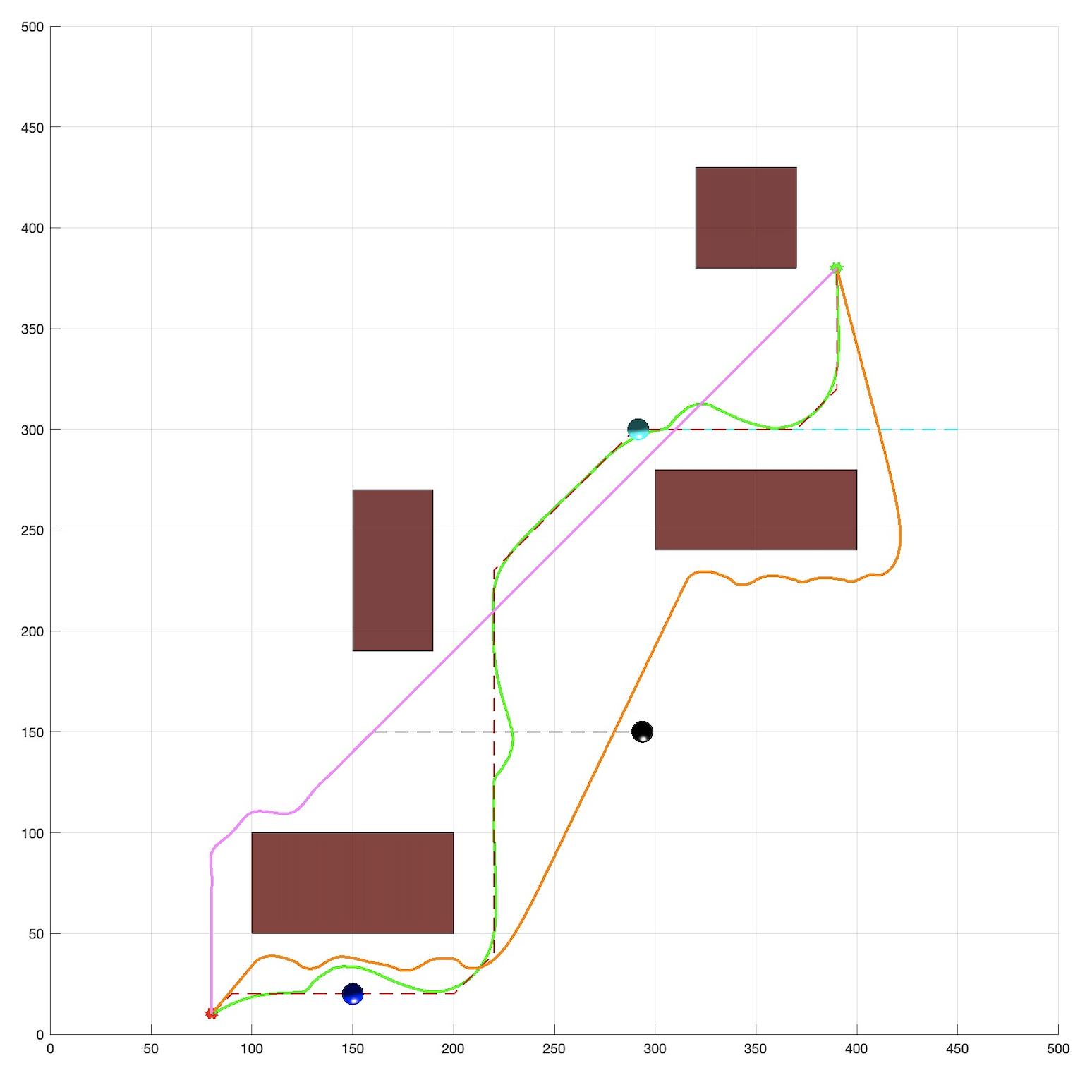}}
\caption{Top View of Comparison Result in 3D}
\label{ch5f6}
\end{figure}

Figure \ref{ch5f5} and figure \ref{ch5f6} shows the simulation results with three different navigation algorithms. The A-star algorithm and purely reactive control methods are benchmarks to compare with the proposed navigation algorithm. It is shown in the figures that the A-star algorithm takes the shortest path from the start node to the target node without considering any cost factor. In addition, the A-star algorithm relies on the full information of the map, which cannot handle unknown obstacles. Moreover, the pure obstacle avoidance (OA) method lacks information on the environment, and the trajectory is not ideal for either distance or energy cost. The total cost comparison is shown in Figure \ref{ch5f7}. However, the A-star algorithm has a shorter operation time; the proposed hybrid strategy can find a short path with a much lower energy cost, which provides better performance than the benchmark methods.

\subsection{Conclusion}
\label{conclusion}
In conclusion, a framework including a hybrid approach was proposed and applied for a solar-powered UAV in a dynamic urban environment. The computer simulation was performed, and the results show that the proposed hybrid strategy can finish the navigation mission without collision. Additionally, the proposed method reduces energy consumption compared with a purely reactive control method. The future extension of this method can be in a purely 3D environment instead of setting a fixed height for the SUAV. Furthermore, how to handle multiple obstacles at the same time can be considered as future work.

\section{Conclusion and Future Directions}
\label{ch6}

The report has addressed several challenging navigation strategies for an SUAV. This section concludes the whole report and provides some suggestions for future research.

\subsection{Conclusion}
In this report, the navigation strategies for an SUAV were studied. Both path planning and obstacle avoidance phases were investigated in detail. Furthermore, an urban environment was chosen as the background environment due to its greater relevance to practical applications.

Section \ref{ch2} presented an overview of the latest applications and navigation methods for single and multiple SUAVs. The previous navigation methods for SUAVs can be classified into three main categories: sample-based, optimisation-based, and coverage navigation. It can be seen that no scholar has employed search-based methods to solve the navigation problem of SUAVs. And this report filled the research gap.

Section \ref{ch3} presented a privacy-aware navigation strategy from a normal UAV. Different from regular path planning problems, UAVs need to be able to sense the environment to minimize the privacy violation risk. A dynamic programming-based approach was used to solve this specific problem. The cost-sensitive navigation feature, in other words, information-aware navigation, inspired the author to focus on the navigation of SUAVs.

Section \ref{ch4} studied the navigation problem of an SUAV in a static urban environment. Energy models were presented to evaluate the energy consumption and harvest. A modified informed graph search algorithm is used to solve the specific problem. In addition, by changing the objective cost function, the algorithm can find the minimum-time path while satisfying the energy constraint. Computer simulations were presented in two different cases to show the effectiveness of the proposed algorithm.

Section \ref{ch5} considered a more complicated problem based on the previous section. Instead of a static environment, some unknown dynamic obstacles were added to the urban environment. The complexity increased significantly because the path planning cannot be purely based on the map information. A hybrid approach was proposed to ensure the safe navigation in a dynamic environment. It started with a global energy-aware path-planning algorithm, followed by a path-tracking controller and a reactive obstacle avoidance algorithm. Computer simulation showed that the hybrid approach can generate a collision-free path when multiple unknown obstacles are presented in the environment.

\subsection{Future Directions}

In the light of the research results in the field of SUAV navigation and deployment, there exist multiple encouraging avenues for future investigation. The examination of deployment methods for SUAVs has been insufficient. It is possible to get comparable outcomes to the development of UAVs demonstrated in the papers \cite{huang2022deployment, savkin2022demand, huang2018method}, which can be regarded as a viable approach. More research directions are discussed in this section.

\subsubsection{Combining Reconfigurable Intelligent Surfaces Technologies}
Reconfigurable intelligent surfaces (RISs) are a novel technology that can make data communication more efficient \cite{liu2021reconfigurable}. Even if obstacles block the LoS between a UAV and sensors, data can still be transmitted to the UAV via an indirect path through an RIS. Therefore, multi-UAV communication systems can benefit a lot from using RISs technologies. Research on the multi-SUAV communication system with RIS technology is a novel research direction that can maximize the communication efficiency and coverage of the system. It can be benefited by using ground-based RISs \cite{savkin2023collision, savkin2022demand}, or using RIS-equipped SUAVs \cite{eskandari2023trajectory, eskandari2022model}. Some research can be conducted in this specific research area.

\subsubsection{SUAVs on Uneven Terrain}
One limitation of aerial mobile robots is that they are often unsuitable for missions that should be conducted very close to the ground. An ongoing topic of extensive debate is the navigation of UAVs in rough terrain. One proposed method in \cite{stodola2019cooperative} enhances simulated annealing technique that optimizes the placement of waypoints to expand the coverage area. This allows UAVs to conduct efficient surveillance on uneven terrain. In \cite{savkin2023autonomous}, the proposed method enhances data transmission efficiency over uneven terrain, reduces energy consumption and time delays, and takes into account time limits of ground sensors and data gathering concerns. Further research could be done considering SUAVs are deployed over uneven terrain, where RIS technologies might offer significant assistance.

\subsubsection{Multi-dimensional Robot Collaboration}
Inspired by the previous direction, research over the past decades has explored the collaboration between airborne and ground-based robots \cite{butzke2016planning, lee2014collaborative}. The use of teams of collaborating UAVs and Unmanned Ground Vehicles (UGVs) is quickly growing in many civilian applications. However, current research still does not fully consider the problem of collaborative navigation of aerial and ground robots on uneven terrain, which becomes a potential direction for future research.  Research can be further extended to scenarios with SUAVs collaborating with marine vehicles. For example, the navigation of UAVs collaborating with marine unmanned vehicles, including both unmanned surface vehicles and autonomous underwater vehicles for environmental data collection. However, UAVs have far less endurance than ground robots, which limits their widespread use. Consequently, SUAVs with extended endurance can fulfil the missions more effectively than traditional UAVs. Calibrating SUAVs with ground mobile robots and marine unmanned vehicles is another promising research direction.

\subsubsection{Natural Disaster Monitoring}
Monitoring natural disasters is an extremely important area of UAV applications. Among these disasters, earthquakes and forest fires \cite{savkin2020navigation} poses a serious threat to many countries due to their very high humanitarian, environmental and economic cost. Therefore, the application of the multi-UAVs technology especially plays a crucial role in addressing these challenges. SUAVs have the ability to swiftly evaluate the situation on-site following a disaster, as well as actively contribute to disaster relief efforts and environmental surveillance, thereby significantly enhancing the efficiency and effectiveness of responses to natural disasters. An application of an emergency relief distribution system after an earthquake using small-UAVs was proposed in \cite{nedjati2016post}. This scheme mainly focuses on the utilization of UAVs to carry out the distribution of relief supplies in the disaster area, gather critical data and provide recommendations for relief priorities depending on the conditions. This application enhances both the effectiveness of rescue efforts and the ability to precisely evaluate the consequence of disasters and execute efficient rescue missions. When it comes to volcano detection, SUAVs offer unparalleled flexibility and a wider field of view than traditional ground-based monitoring methods, allowing for the collection of large amounts of elevation data in a relatively short period of time. The application of this technology improves the accuracy of volcanic activity predictions, facilitates and improves volcano mapping, characterization, interpretation, monitoring and hazard assessment. There exist some research publications that demonstrate the superiority of UAVs in volcano detection \cite{amici2013volcanic, beni_uavs_2019, bonali_uav-based_2019}. Technically, the flight time of a UAV is limited by its power source. According to the study, the flight time of UAVs varies between 4 and 20 minutes \cite{jordan_collecting_2019}. In remote areas such as volcanoes, where charging grids are not readily available, the need to carry a large number of batteries has in the past been an added burden to volcano detection efforts. The use of SUAVs, on the other hand, can address the energy implications and change the past dilemma of non-continuous monitoring to obtain richer data. We see in this not only the great potential of UAVs for volcano exploration and data collection missions, but also the merits of SUAVs to further optimize and enhance mission efficiency and mission sustainability. In the future, the utilization of SUAVs in disaster relief is expected to become an essential and progressively prominent field of research and implementation.

\subsubsection{Wildlife Monitoring}
Similar to natural disaster monitoring, wildlife monitoring \cite{hodgson2016precision, linchant2015unmanned, gonzalez2016unmanned} plays the same important role. UAVs can provide higher precision monitoring than traditional methods. Moreover, the high mobility of animals makes it harder to monitor by stationary ground stations. The long endurance of SUAVs makes long-distance surveillance and control of wildlife more beneficial and reliable. Furthermore, an difficult challenge in UAV wildlife monitoring is making UAV surveillance covert to avoid scaring wild animals. In \cite{savkin2020bioinspired}, a biomimetic (biologically inspired) sliding mode control based navigation algorithm mimicking motion camouflage stealth behaviour observed in some attacking animals, such as hover flies, dragonflies and falcons, was proposed. Based on this, an interesting future research direction is to combine the biomimetic navigation algorithm for covert video surveillance of \cite{savkin2020bioinspired} with an efficient solar power harvesting method for implementation with SUAVs. This leads to a further branch in the field, SUAVs can be used for both ground and marine wildlife monitoring. The use of UAVs in fisheries has opened a new section in the sustainable utilization of marine resources. These UAVs are capable of flying for long periods of time over vast areas of the sea to monitor the dynamics of fish stocks and provide real-time data on marine ecosystems, while also helping fishermen to locate fish stocks more accurately, monitor marine pollution and illegal fishing activities, and make a contribution to marine policing, thereby supporting the development of sustainable fisheries \cite{yang_uav_2022}. Using UAVs in fish farming is becoming a popular option. In \cite{wang2021intelligent}, a fishery inspection equipment combining UAV and unmanned boat is mentioned, which can realize underwater daily management, video information collection and breeding ecological environment monitoring, and realize the all-round perception of environmental and ecological information and fish feeding information of smart fish farms \cite{anggraeni2019development, romeo2012very, yang2022hybrid}. In addition to this, some research have been conducted for shark monitoring \cite{colefax2020developing}, which can also be extended for SUAV applications. SUAVs have a bright future in this application thanks to the easier access to solar energy for longer flights over unobstructed seas. 

\subsubsection{Smart Agriculture}
Furthermore, adequate data information can be used in smart agriculture \cite{maddikunta2021unmanned}. Typical scenario can be described as a large number of ground sensors is deployed in a remote agricultural field for sensing and monitoring temperature, soil moisture and other environmental parameters. SUAVs can be used to collect the data from the ground sensors. In this way, the joint SUAV navigation and communication scheduling for sensor data collection in smart agriculture can be a future research direction. Similar work has been done on normal UAVs \cite{li2021continuous}, but SUAVs have a brighter future with longer endurance. Based on this, a special case can be introduced that with a vineyard in a mountainous area where LoS between the SUAVs and the ground sensors is often blocked by the terrain. The complexity of the problem is pushed to another level that is worth comprehensive and detailed research.

\subsubsection{Machine Learning Techniques}
Within the framework of the big-data era, the examination of vast amounts of data holds significant importance for further progress. In light of the development of AI technology, machine learning and other AI-based tools can be a powerful and robust solver for solving optimization problems \cite{bithas2019survey, liu2019trajectory, makineci2022ann}. To enhance the performance of SUAVs, it is imperative to create advanced AI-assisted optimization methods. By employing these instruments to greatly enhance the effectiveness of solar energy harvesting and minimize energy consumption, not only will the operational duration of UAVs be substantially prolonged, but the advancement of SUAVs technology as a whole will be facilitated. These advancements will expand the potential uses of UAVs, encompassing various significant sectors such as environmental monitoring, traffic control, and emergency response. Another interesting topic for future research is applying optimization methods based on the dynamic programming/Bellman optimality principle \cite{salgado2001control, savkin2002hybrid} for SUAV navigation. This approach is especially promising in various problems of joint navigation/sensor data transmission control that are similar to the problems studied in \cite{savkin2023autonomous, li2021continuous}.

\subsubsection{Energy Management Methods and Hybrid SUAVs}
Energy management methods for SUAVs can be a topic that deserve to be studied. Due to the capacity limitation of the onboard battery, properly controlling the time allocation of using battery energy and solar energy can extend the flight time of SUAV \cite{tao2019state}. In this case, another control variable is born that increases the difficulty of the problem. The next topic can thus be introduced from this, the Hybrid UAVs. There have been some studies on hybrid UAVs using solar energy and some other forms of energy sources. For example, fossil fuel and solar powered UAV \cite{panagiotou2016conceptual}, wind and solar energy powered UAV \cite{sekander2020statistical} and RF and solar energy powered UAV \cite{van2020advanced, xu2021resource}. The energy management method is worth being researched on hybrid UAVs.

\subsubsection{Multi-SUAVs Self-collision Avoidance}
Another important challenge in navigation of SUAV teams is that the developed navigation algorithm should also guarantee avoiding collisions between SUAVs. This make involved optimization problems far more difficult as the dimension of the optimization space increases in $n$ times, where $n$ is the number of SUAVs in the team. This greatly increases the computational complexity of navigation algorithms, which is similar to \cite{Solar1}. All in all, this research direction is still an immature field, with many novel applications and algorithms to be developed and investigated.

Overall, the development of unmanned systems and renewable energy has been rapid during the last few decades. Researching the navigation algorithms for SUAVs has the potential to help with the development of a sustainable future. There is no doubt that more research findings can accelerate the growth of autonomous UAVs, potentially reducing labour work in the near future.

\subsubsection{Energy Harvesting Model of SUAVs}
Proper energy harvesting model is an essential factor that build the foundation of the energy-aware path planning algorithm. Among all the aforementioned path planning algorithm of SUAVs, the assumption was made that SUAVs cannot harvest solar energy under the shadow area. However, this does not fully reflect reality. The energy absorbed should be directly related to the light intensity, the shadow is just a more intuitive dividing line. Since it is not possible to quantify and simulate the light intensity easily, the concept of shadow areas is introduced. SUAVs are still possible to harvest solar energy from the dispersed sunlight to some degree. A sigmoid function is a potential solution for mimicking their relation more accurately. More research is needed to find a better energy harvesting model.

\bibliographystyle{IEEEtran}
\bibliography{references.bib}

\end{document}